\documentclass[smallextended]{svjour3}       
\smartqed  
\usepackage{graphicx}
\usepackage{mathptmx}      
\usepackage{amssymb}
\usepackage{amsmath}
\usepackage{booktabs}
\usepackage{xcolor}
\usepackage[colorlinks=true,linkcolor=blue,citecolor=blue,urlcolor=blue]{hyperref}
\usepackage{natbib}
\usepackage{xspace}

\def \beq{\begin{equation}}
\def \eeq{\end{equation}}
\def \bear{\begin{eqnarray}}
\def \ear{\end{eqnarray}}

\newcommand{\AAA}{\vec{A}}
\newcommand{\BB}{\vec{B}}

\newcommand{\EE}{\vec{E}}

\newcommand{\Bnabla}{\vec{\nabla}}

\newcommand{\Eq}[1]{Eq.~(\ref{#1})}

\def\EM{E_{\rm M}}

\def\MATINS{\texttt{MATINS}\xspace}

\newcommand\psra{PSR~J0205$+$6449\xspace}
\newcommand\psrb{PSR~B2334$+$61\xspace}
\newcommand\velajr{CXOU~J0852$-$4617\xspace}

\newcommand{\dd}{\mathrm{d}}

\newcommand{\gcc}{\mbox{g~cm$^{-3}$}}

\newcommand{\dive} {\vec{\nabla}\cdot}

\newcommand{\derparn}[2] {\frac{\partial #2}{\partial #1}}

\newcommand{\de} {{\rm d}}

\journalname{Living Reviews in Computational Astrophysics}

\begin{document}

\title{Magnetic, thermal and rotational evolution of isolated neutron stars%
\thanks{This article is a revised version of \url{https://doi.org/10.1007/s41115-019-0006-7}\\
\textbf{Change summary} Major revision, updated and expanded.\\
\textbf{Change details} Section~2.2 reorganized and extended with Sect.~2.2.3 (new Fig.~4). New Sect.~3.3 and Sect.~3.6 rewritten. New Sect.~4.3 (new Figs.~11, 12). Removed former Sect.~5 ``Numerical tests''. Section~5 revised (new Fig.~14) with new Sect.~5.4.1. Section 6 completely rewritten and updated (new Figs.~16, 17, 19, 20, 21, 24, 25, 26, 27). The number of references has increased to 294. Added new co-author Clara Dehman.}
}



\author{Jos\'e A. Pons \and Clara Dehman \and
        Daniele Vigan\`o
}


\institute{J. A. Pons \at 
  Departament de F\'{\i}sica \at
  Universitat d'Alacant, Spain \\
  \email{jose.pons@ua.es}
  \and
  C. Dehman \at 
  Departament de F\'{\i}sica \at
  Universitat d'Alacant, Spain \\
  \email{clara.dehman@ua.es}
  \and
  D. Vigan\`o \at Institut de Ci\`encies de I'Espai (ICE-CSIC), Campus UAB, Carrer de Can Magrans s/n, 08193 Cerdanyola del Vallès, Catalonia, Spain\\
  Institut d’Estudis Espacials de Catalunya (IEEC), 08860 Castelldefels, Catalonia, Spain \\
  \email{vigano@ice.csic.es}
}

\date{Received: date / Accepted: date}

\maketitle

\begin{abstract}
The strong magnetic fields of neutron stars are closely linked to their observed thermal, spectral, and timing properties, such as the distribution of spin periods and their derivatives. To understand the evolution of astrophysical observables over time, it is essential to develop robust theoretical frameworks and numerical models that solve the coupled thermal and magnetic field evolution equations, incorporating detailed microphysics such as thermal and 
electrical conductivities and neutrino emission rates. These efforts are key to uncovering how the strength and geometry of magnetic fields change with age, ultimately shedding light on the diverse phenomenology of neutron stars. In this review, we outline the fundamental theory underlying magneto-thermal evolution models, with an emphasis on numerical methods and a comprehensive set of benchmark tests intended to guide current and future code development. 
We revisit established results from axisymmetric simulations, highlight recent progress in fully three-dimensional models, and offer a perspective on the anticipated developments in this rapidly evolving field.

\end{abstract}

\setcounter{tocdepth}{3}
\tableofcontents

\newpage
\section{Introduction}
\label{sec:1}

Neutron stars (NSs), the endpoints of the evolution of massive stars, are fascinating astrophysical sources that display a bewildering variety of manifestations. They are arguably the only stable environment in the present Universe where extreme physical conditions of density, temperature, gravity, and magnetic fields are realized simultaneously. Thus, they are ideal laboratories to study the properties of matter and the surrounding plasma under such extreme limits. 

For instance, their strong gravitational fields provide unique tests of general relativity in the strong-field regime, such as through the timing of pulsars in binary systems or the observation of gravitational wave signals from NS mergers. The ultra-dense interiors of NSs allow us to probe the behavior of nuclear and possibly quark matter at supranuclear densities, offering insights into the equation of state (EoS) of dense matter and the role of exotic particles like hyperons or quarks. Their intense magnetic fields, which can well exceed $10^{14}$ G in magnetars, provide a natural setting to explore quantum electrodynamics in the non-perturbative regime. Additionally, the magnetospheres and relativistic winds of pulsars serve as a natural laboratory for studying plasma physics under highly relativistic and magnetized conditions, including phenomena such as magnetic reconnection and particle acceleration. NSs also have implications for axion-like particles and dark matter candidates, making them increasingly relevant in the search for physics beyond the Standard Model.

Regarding observation, NSs were first discovered as rotation-powered radio {\it pulsars} (standing for pulsating stars, due to their periodic signal). These so-called {\it standard pulsars} constitute
the most numerous class of known NSs, approaching four thousand  identified members.\footnote{See the online Australia Telescope National Facility pulsar catalog, \url{http://www.atnf.csiro.au/research/pulsar/psrcat/}}
The number is continuously increasing thanks to the progress in wide-field low-frequency radio surveys and the use of the new generation of high-sensitivity radio interferometers, like LOFAR \citep{lofar13} and the soon-available Square Kilometre Array Observatory (SKAO). The latter is foreseen to be able to detect many thousands of regular pulsars. 

To a lesser extent, NSs have also been observed in $X$-rays (about one hundred NSs so far), as persistent or transient sources, and/or as $\gamma$-ray pulsars (340 in the recent third Fermi-LAT pulsar catalog, \citealt{fermi3pc}). The origin of this high-energy radiation is typically non-thermal, originated by particle acceleration (synchro-curvature emission, \citealt{zhang97,vigano15}) or Compton up-scattering of lower-energy photons by the particles composing the magnetospheric plasma \citep{lyutikov06}. An exception is the soft X-ray thermal emission from the surface, observed in only a few dozen, mostly young, neutron stars \citep{potekhin_rev15b}; despite its rarity, it is highly relevant to this review.

A particularly intriguing class of isolated NSs are the {\it magnetars} \citep{2015SSRv..191..315M,2015RPPh...78k6901T,2017ARA&A..55..261K,readegrandis25}, relatively slow rotators with typical spin periods of several seconds and ultra-strong timing-inferred magnetic fields ($10^{13}$--$10^{15}$ G). In most cases, they show a relatively high persistent (i.e., constant over many years) X-ray luminosity ($L_x \approx 10^{33}$--$10^{35}$ erg/s), well exceeding their rotational energy losses, in contrast with radio (standard) and $\gamma$-ray pulsars. This leads to the conclusion that the main source of energy is provided by the strong magnetic field, instead of rotational energy.
Magnetars are also identified for their complex transient phenomenology in high energy $X$-rays and $\gamma$-rays, including short (tenths of a second) bursts, occasional energetic outbursts with months-long afterglows \citep{2011ASSP...21..247R,coti18} and, much more rarely (only three observed so far), giant flares \citep{hurley99,palmer05}. During giant flares, the energy release is as large as $10^{46}$ erg in less than a second. The source of energy of such transient, violent behavior is also generally agreed to be of magnetic origin, as originally proposed in \citet{1995MNRAS.275..255T,1996ApJ...473..322T}. 

Although isolated NSs have been historically differentiated in sub-classes, mostly based on observational grounds (detectability in $X$ and/or radio,  
transient vs. persistent properties, and presence/absence of pulsations), there is arguably no sharp boundary between classes, and the distributions of their physical properties, such as the inferred magnetic field, partially overlap.
Indeed, the evidence accumulated in the last few decades has shown that the presence of a strong dipolar field, which can be reliably estimated from timing properties (see Sect.~\ref{sec:magnetosphere}), is not a sufficient condition to trigger observable magnetar-like events. In contrast, a growing number of NSs with relatively low inferred surface dipolar magnetic fields have been observed showing magnetar-like activity. These include some normal pulsars \citep{2010Sci...330..944R,2012ApJ...754...27R,2013ApJ...770...65R,2014ApJ...781L..17R,gavriil08,gogus16,lower2021,uzuner2023}, often referred to as {\it low-field magnetars}, but also other sources belonging to the sub-class of {\it central compact objects} (CCOs), a handful of young NSs surrounded by a supernova remnant, detectable due to a persistent, mostly non-pulsating X-ray emission \citep{deluca17}. It is now evident that the complex, non-linear dynamics of the internal magnetic field, coupled with its interaction with the magnetospheric plasma, is crucial for understanding the observed phenomena. 

Key theoretical challenges necessary to understand NS phenomenology include: the partitioning of magnetic energy between toroidal and poloidal components and across various spatial scales; the spatial distribution and long-term dissipation of electrical currents within the star; the mechanisms generating and transporting magnetic helicity outward to sustain magnetospheric currents (i.e., the processes twisting magnetic field lines); and the nature of instabilities that drive outbursts and flares. 
Although the underlying physics resembles that of other plasma environments, such as solar or laboratory plasmas, NS conditions are far more extreme, involving intense gravity, ultra-high densities, and potentially exotic states like superconductivity or superfluidity. 
Addressing these issues requires advanced 2D and 3D numerical simulations \footnote{Throughout this review, 2D indicates the use of 3D vectorial fields, but with no dependence of any quantity on the azimuthal coordinate; note that this is sometimes called 2.5D.} tailored specifically to NS physics.
The goal of this review is to provide a comprehensive yet accessible overview of NS evolution modeling, aimed not only at specialists but also at a broader astrophysical audience, including students entering the field. To this end, we will review the fundamental equations and numerical techniques employed in different aspects of the modeling, with particular emphasis on the unique physical features of NSs that set them apart from other stellar systems.

This review is organized as follows. In Sect.~\ref{sec:cooling}, the theory of the cooling of NSs is reviewed;  the magnetic field evolution is described in detail in Sect.~\ref{sec:magnetic_evolution}, where we discuss the physical processes in different parts of the star. In Sect.~\ref{sec:magnetic_methods} we review the different numerical methods and techniques used to model the magnetic evolution. In Sect.~\ref{sec:magnetosphere}, we explore the complex and dynamic coupling between the slowly evolving interior of the NS and the surrounding force-free magnetosphere, where plasma dynamics are dominated by the magnetic field. This interaction plays a crucial role in regulating the rotational evolution, as it governs the loss of angular momentum and thus the long-term behavior of the spin period and its derivative, which are generally the two most precisely measured observables in NSs. Sect.~\ref{sec:examples} presents selected examples of realistic evolution models from recent literature. Finally, in Sect.~\ref{sec:conclusions} we outline future directions and highlight key open questions in the field.


\section{Neutron star cooling}
\label{sec:cooling}

The evolution of the temperature in a NS was theoretically explored even before the first detections, in the 1960s \citep{tsuruta64}. Today, \emph{NS cooling} is the most widely accepted terminology for the research area studying how internal and surface temperature evolve as NSs age and their observable effects. We refer the interested reader to the introduction in the review by \citet{potekhin_rev15a} for a thorough historical overview of the foundations of the NS cooling theory. 

According to the standard NS theory, a proto-NS is born as extremely hot and liquid, with $T\gtrsim 10^{10}$ K, and a relatively large radius, $\sim 100$ km. Within a minute, it becomes transparent to neutrinos and shrinks to its final size, $R\sim 10-14$ km \citep{burrows86,keil1995,pons99}. Neutrino transparency marks the  starting point of the \emph{long-term cooling}. At the initially high temperatures, there is a copious production of thermal neutrinos that abandon the NS core draining energy from the interior. In a few minutes, the temperature drops by another order of magnitude to $T \sim 10^{9}$~K. The core of the star, forming its bulk, consists of a liquid mixture of neutrons, electrons, protons, and potentially exotic particles such as muons, hyperons, or deconfined quark matter, while the outermost layers ${\cal O}({\rm km})$ comprise heavy nuclei and relativistic electrons. In these outer regions, the matter has a melting temperature of $T\sim 10^9$ K, leading to rapid crystallization and the formation of the solid crust. Since the melting temperature depends on the local value of density, the gradual growth of the crust takes place from minutes to months after birth. This crystallization process does not extend to the entire outer region of the star up to its surface. The outermost layer, known as the envelope or sometimes the ocean, with a typical thickness of ${\cal O}(10^2~{\rm m})$ remains in a liquid state. Additionally, the star may be surrounded by a very thin ${\cal O({\rm cm})}$ gaseous atmosphere. 

The high thermal conductivity of the core rapidly leads to a nearly isothermal\footnote{We employ the term ``isothermal'' in the relativistic sense, including metric corrections, as we describe below.} state within the first year. After a few years, the gradual long-term cooling of residual heat proceeds slowly, on a timescale of $10^5$--$10^6$ years, with temperature gradients essentially limited to the crust and envelope. 
These temperature gradients are primarily radial, but, as discussed in Sect.~\ref{sec:heat_eq}, they can also develop in the angular directions in the presence of strong magnetic fields, leading to surface temperature variations that influence the observed X-ray spectra. Notably, the absence of strong temperature gradients in the bulk of the NS implies that, in contrast with main-sequence stars, no internal convection is triggered. Therefore, the common self-sustained convective dynamos, which operate ubiquitously in different astrophysical scenarios (planets, stars, brown dwarfs...), are not operative in NS and cannot be responsible for the observed long-term presence of the strong magnetic fields.

Thus, the primary goal of NS cooling studies is to develop realistic evolutionary models that, when compared with thermal emission observations from NSs of varying ages, yield valuable insights into the chemical composition, magnetic field strength and configuration of the emitting regions, or the properties of denser matter deeper within the star
\citep{Page2004,2004ARA&A..42..169Y,aguilera08a,aguilera08b, 2008AIPC..983..379Y,2009ASSL..357..247P,2009ASSL..357..289T,potekhin_rev15b,potekhin20}.

When comparing to observational data, it is important to distinguish that the emission in soft X-rays may be
reflecting either internal or external heating mechanisms. External heating can result from localized magnetospheric currents or particle bombardment, often producing a prominent non-thermal component in the X-ray spectrum, with an inferred emission area typically small ($\lesssim 100$ m$^2$). 
In this case, even if an additional thermal component is present in the spectrum, the high temperatures of such small spots are difficult to reconcile with standard cooling behavior, as they are likely driven by these external processes. 

In contrast, the internal process of heat transfer from the NS core manifests as a thermal component originating from a significant portion of the star’s surface. For a few dozen isolated neutron stars, the X-ray spectra are dominated by this thermal component, reflecting the residual cooling after the NS is born, probably compensated by internal heating processes connected to magnetic field decay. In this case, when an estimate of the star's age is available, one can examine the relationship between temperature and age, providing an indirect approach to investigate the physics of the NS interior through comparison of long-term cooling models with observations. Therefore, careful selection of sources is essential when testing these cooling models, focusing specifically on cases where internal processes govern the evolution.
The sample of sources suitable for these studies is gradually expanding, primarily consisting of young ($t\lesssim 10^4$ yr) NSs \citep{vigano13,potekhin_rev15b,potekhin20,Marino24},
along with a small number of slightly older NSs
($\sim10^5$--$10^6$ yr), detectable due to their proximity within a few hundred parsecs. These older sources are known as the magnificent seven, or X-ray dim isolated NSs \citep{haberl}.

NS cooling scenarios are divided into two types: 1) ``standard'' or ``minimal'' cooling, driven by modified Urca processes and possibly enhanced by core superconductivity or superfluidity, and 2) ``enhanced'' cooling, with faster cooling at early ages, triggered by neutrino production due to direct Urca processes \citep{lattimer01}, presence of hyperons \citep{Anzuini22a}, quark matter or meson condensates \citep{Umeda94}. The Vela pulsar, with its unusually low temperature and thermal luminosity for its age, is a key example suggesting rapid neutrino emission linked to high central density or exotic matter. This concept, first proposed by \citet{page1992}, has awaited further data and confirmation for over three decades.
Evidence of enhanced cooling existed for some isolated NSs, but uncertainties in spectral data, ages, distances, accretion history, and magnetic field dissipation in high-field stars ($B\gtrsim10^{14}$ G) hindered definitive confirmation. Very recently, a new analysis of three young, nearby, extremely cold NSs, whose properties require enhanced cooling to align models with observations \citep{Marino24}, allowing to establish some constraints on the EoS. Currently, there is compelling evidence for rapid cooling in at least some NSs, or possibly most of them, if one accounts for strong magnetic fields effects counteracting the fast cooling \citep[see e.g. Sect.~7.2 in ][]{aguilera08b}. In addition, similar conclusions about the need for fast cooling mechanisms have been reached by examining the cooling of NSs in Low Mass X-ray Binaries following an accretion phase \citep{Mendes22}.

In the same spirit, NS cooling has traditionally been used to establish constraints on the existence and properties of new particles beyond the Standard Model. By analyzing the thermal evolution of NSs, researchers can probe the parameter space of hypothetical particles that could enhance cooling via additional energy loss mechanisms.
Specific examples include \citet{Hamaguchi18}, who derived upper limits on the axion decay constant, while \citet{Buschmann2022} used the cooling behavior of young NSs to constrain the axion mass, and \citet{gomez24} tightened these constraints by incorporating structural effects in the envelope. 

To illustrate a particularly intriguing case, we briefly examine the debated observational claims regarding the rapid cooling of the NS linked to the Cassiopeia A (Cas A) supernova remnant. Initial detailed studies of its soft X-ray thermal spectrum, spanning multiple years, indicated a surface temperature drop of approximately 4\% per decade \citep{2010ApJ...719L.167H}. 
This was interpreted as evidence of a superfluid transition in the NS core, leading to enhanced neutrino emission, 
which sparked significant interest in modeling and motivated further in-depth X-ray observations \citep{Page2011,2011MNRAS.412L.108S,2015PhRvC..91a5806H}. 
However, the inferred cooling rate has gradually decreased to 2--3\% as more data and improved instrumental calibrations became available \citep{posselt18}. The most recent reanalysis, using data up to 2020 and incorporating updated background and detector models, places the cooling rate between 1.6\% and 2.2\% per decade, depending on the treatment of the hydrogen column density \citep{wijngaarden19}. While this confirms that the NS in Cas A is cooling, the revised (slower) rate challenges some of the earlier interpretations that required very strong neutrino emissivity. More recently, a thorough and critical revision has raised some doubts about possible systematic biases or imperfect calibrations which further challenge the reliability of the previously inferred cooling rate values \citep{posselt22}. 
This highlights the intricate nature of both theoretical modeling and data analysis challenges, demonstrating the dynamic and evolving character of this research field.

We now revisit the theory of NS cooling, beginning with a brief revision of the stellar structure equations and by introducing notation for the rest of the review.

\subsection{Neutron star structure}

Initial and most recent NS cooling studies typically modeled a spherically symmetric 1D background star, both for simplicity and because the extreme gravity results in minimal deviations from symmetry.
The matter distribution can be considered spherically symmetric to a very good approximation, except in extreme, unobserved cases involving structural deformations from near-breakup spin rates ($P \lesssim 1$ ms) or ultra-strong magnetic fields ($B \gtrsim 10^{18}$~G), which are unlikely to occur in nature. 
Therefore, using spherical coordinates $(r,\theta,\varphi)$, the space-time structure is accurately described by the interior Schwarzschild metric:
\begin{equation}
\dd s^2 = - \mathrm{e}^{2\nu(r)} c^2 \dd t^2 + \mathrm{e}^{2\lambda(r)} \dd r^2 
+ r^2 (\dd \theta^2 + \sin^2\theta \dd \varphi^2),
\label{Schw}
\end{equation}
where  $ \lambda(r) = - \frac{1}{2} \ln \left[ 1-  \frac{2 G}{c^2} \frac{m(r)}{r^2}\right]$ accounts for the space-time curvature, $$m(r)=4\pi\int_0^r\rho(\tilde{r})\tilde{r}^2\dd \tilde{r}$$ 
is the enclosed gravitational mass within a sphere of radius $r$, $\rho$ is the mass-energy density, $G$ is the gravitational constant, and $c$ is the speed of light. The \emph{lapse} function $e^{2\nu(r)}$ is determined by the equation
\begin{equation}
   \frac{\dd \nu(r)}{dr} =  \frac{G}{c^2} \frac{m(r)}{r^2}\,
      \left( 1 + \frac{4\pi r^3 P}{c^2 m(r)} \right)\,
      \left( 1 - \frac{2 G}{c^2} \frac{m(r)}{r} \right)^{-1}, 
\end{equation}
with the boundary condition $\mathrm{e}^{2\nu(R)}=1-2GM/c^2R$
at the stellar radius $r=R$.  Here, $M\equiv m(R)$ is the total gravitational mass of the star.
The pressure profile, $P(r)$, is determined by the Tolman-Oppenheimer-Volkoff equation
\begin{equation}
\frac{\dd P(r)}{dr} =   - \left(\rho +\frac{P}{c^2}\right)   \frac{\dd \nu(r)}{dr}. 
\label{Phi}
\end{equation}
Throughout the text, we will keep track of the metric factors for consistency, unless indicated. The Newtonian limit can easily be recovered by setting $\mathrm{e}^{\nu}=\mathrm{e}^{\lambda}=1$ in all equations.

To close the system of equations, one must provide the EoS, i.e., the dependence of the pressure on the other variables $P=P(\rho,T, Y_i)$ ($Y_i$ indicating the particle fraction of each species).
Since the Fermi energy of all particles is much higher than the thermal energy (except in the outermost layers) the dominant contribution is given by degeneracy pressure. The thermal and magnetic contributions to the pressure, for typical conditions, are negligible in most of the star volume. Besides, the assumptions of charge neutrality and $\beta$-equilibrium uniquely determine the composition at a given density. Thus, one can assume an effective barotropic EoS, $P=P(\rho)$, to calculate the background mechanical structure. Therefore, the radial profiles describing the energy-mass density and chemical composition can be calculated once and kept fixed as a background star model for the thermal evolution simulations. 

\begin{figure}
	\centering
	\includegraphics[width=0.75\textwidth]{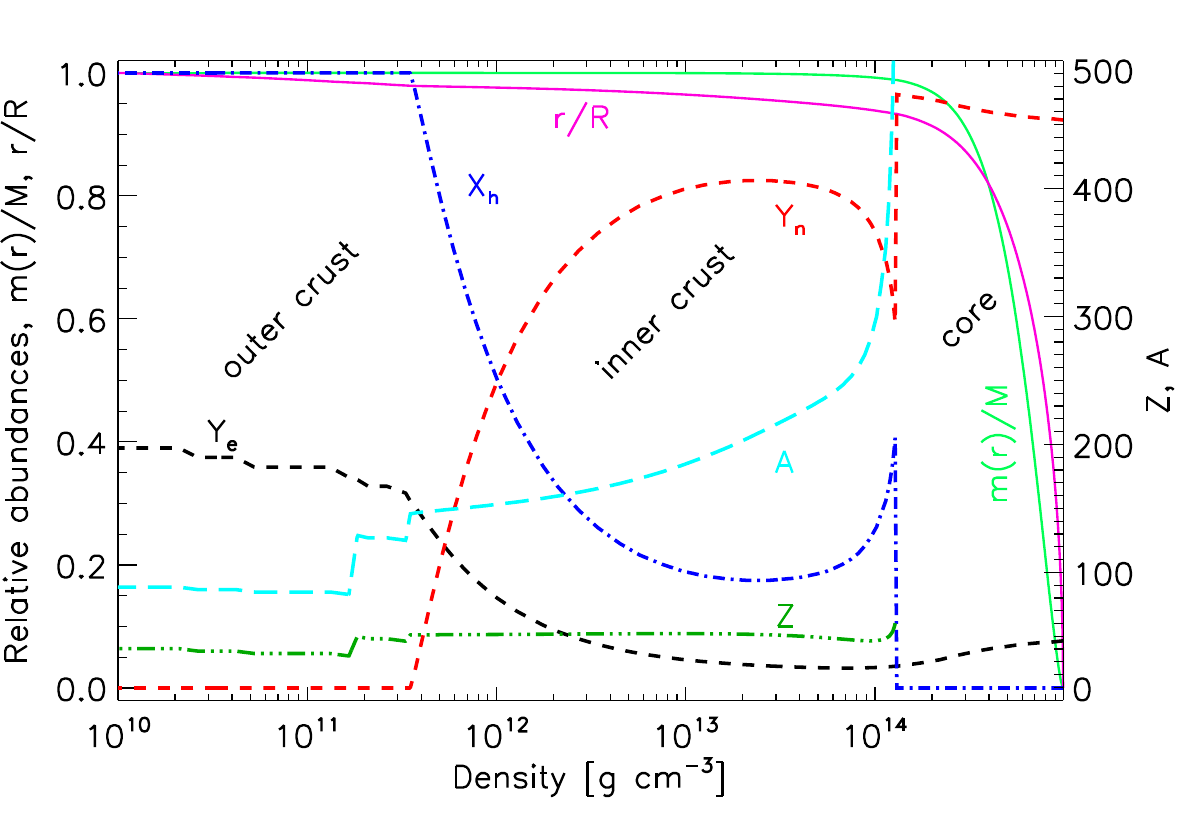}
	\caption{Structure and composition of a $1.4\,M_\odot$ NS, with SLy EoS. The plot shows, as a function of density from the outer crust to the core, the following quantities: mass fraction in the form of nuclei $X_h$ (blue dot-dashed line), the fraction of electrons per baryon $Y_e$ (black dashes), the fraction of free neutrons per baryon $Y_n$ (red dashes), the atomic number $Z$ (dark green triple dot-dashed), the mass number $A$ (cyan long dashes), radius normalized to $R$ (pink solid), and the corresponding enclosed mass normalized to the star mass (green solid). Note that the total neutron fraction (not shown here) varies continuously across the NS interior, unlike the fraction of free neutrons per baryon, $Y_n$.
}
	\label{fig:ns_profile}
\end{figure} 


In Fig.~\ref{fig:ns_profile} we show a typical profile of a NS, obtained with the EoS SLy4 \citep{douchin01}, which is among the realistic EoS supporting a maximum mass compatible with the observations, $M_{\max}\sim 2.0$--$2.2\,M_\odot$ \citep{demorest10,antoniadis13,margalit17,ruiz18,radice18,cromartie19}. We show the enclosed radius and mass, and the fractions of the different components, as a function of density, from the outer crust to the core. For densities $\rho\gtrsim 4\times 10^{11}~\gcc$, neutrons drip out of the nuclei and, for low enough temperatures, they would become superfluid. Note that the core contains about 99\% of the mass and comprises 70--90\% of the star volume (depending on the total mass and EoS). Envelope and atmosphere are not represented here. For a more detailed discussion, see e.g., \citet{2007ASSL..326.....H,potekhin_rev15a}.

\subsection{Heat transfer equation}\label{sec:heat_eq}

Spherical symmetry was also assumed in most NS cooling studies during the 1980s and 1990s. However, in the 21st century, the unprecedented amount of data collected by soft X-ray observatories such as Chandra and 
XMM-Newton supports that most nearby NSs whose thermal emission is visible in the X-ray band of the electromagnetic spectrum show some anisotropic temperature distribution \citep{haberl,2007Ap&SS.308..171P,2011ApJ...736..117K}. 
This observational evidence made clear the need to build multi-dimensional models and gave a new impulse to the development of the cooling theory including multidimensional effects \citep{geppert04,geppert06,2007Ap&SS.308..403P,aguilera08a,aguilera08b,vigano13,Beznogov23}. The cooling theory builds upon the heat transfer equation, which includes both flux transport and source/sink terms.

\begin{figure}[t]
	\centering
	\includegraphics[width=0.49\textwidth]{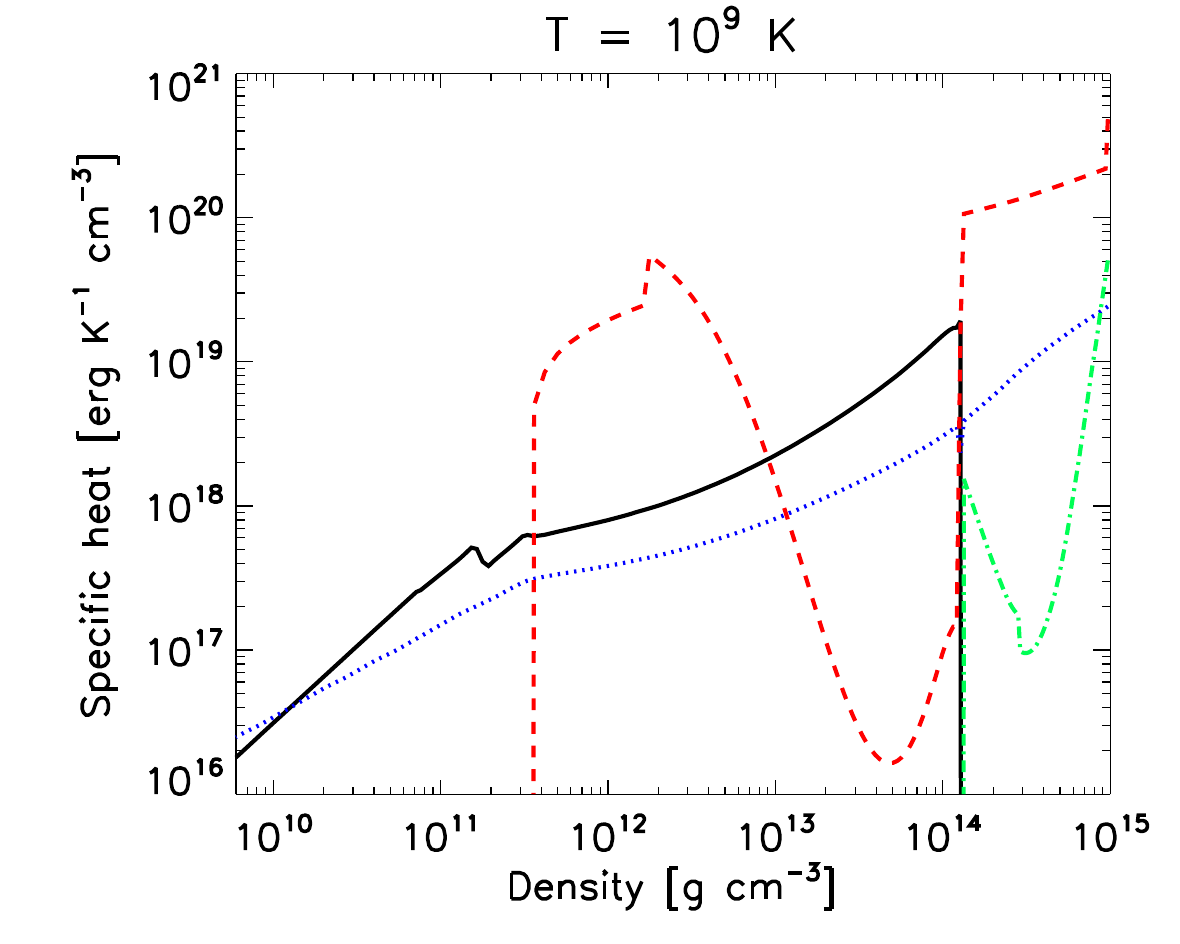}
	\includegraphics[width=0.49\textwidth]{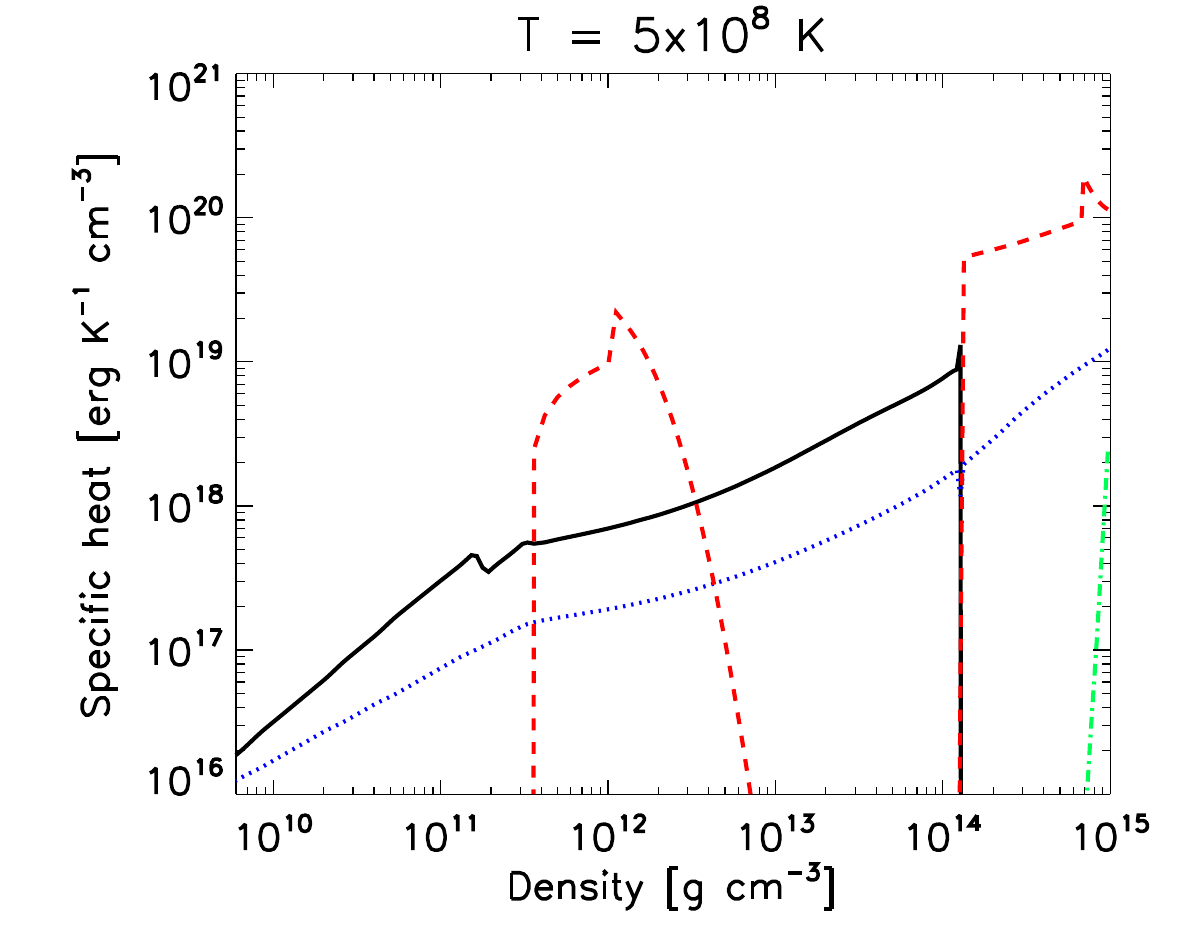}\\
	\includegraphics[width=0.49\textwidth]{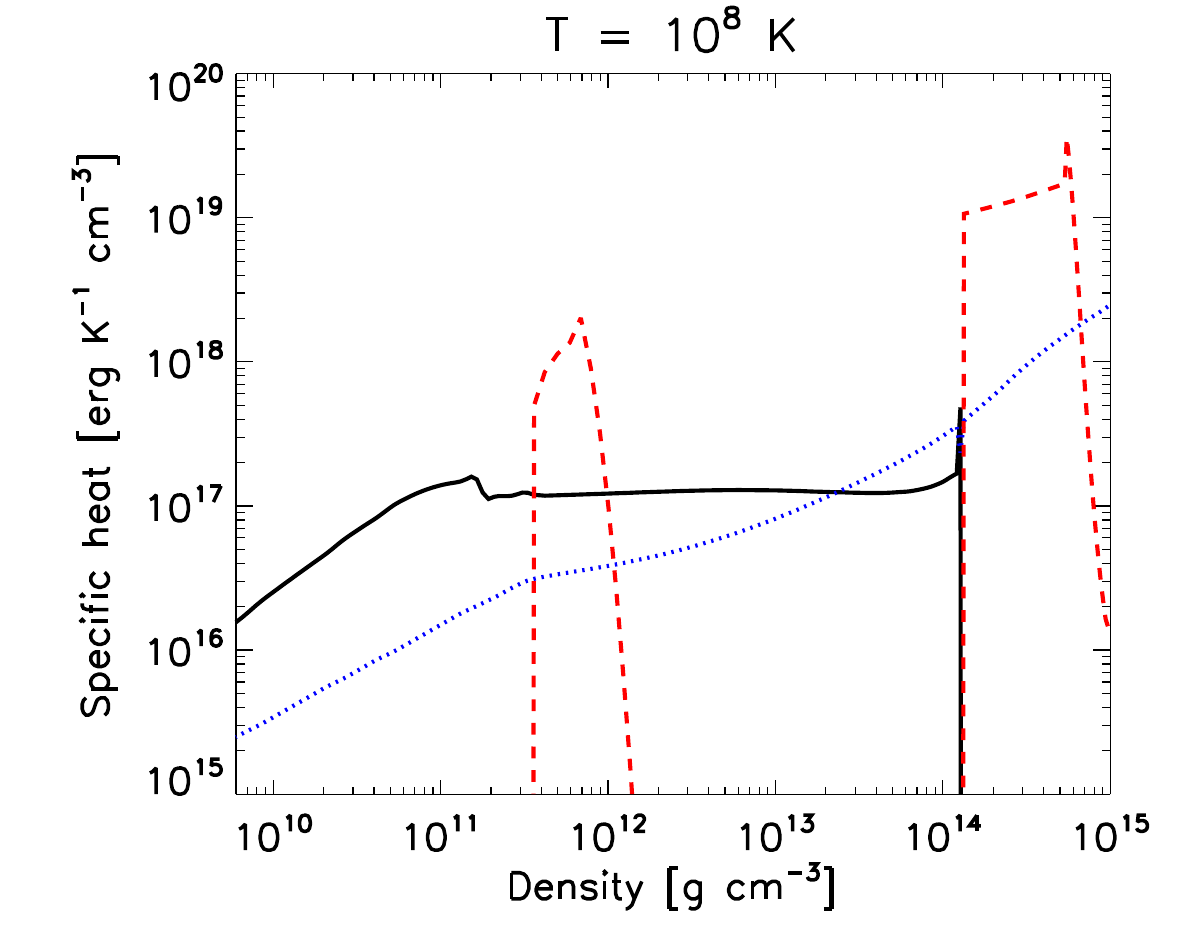}
	\includegraphics[width=0.49\textwidth]{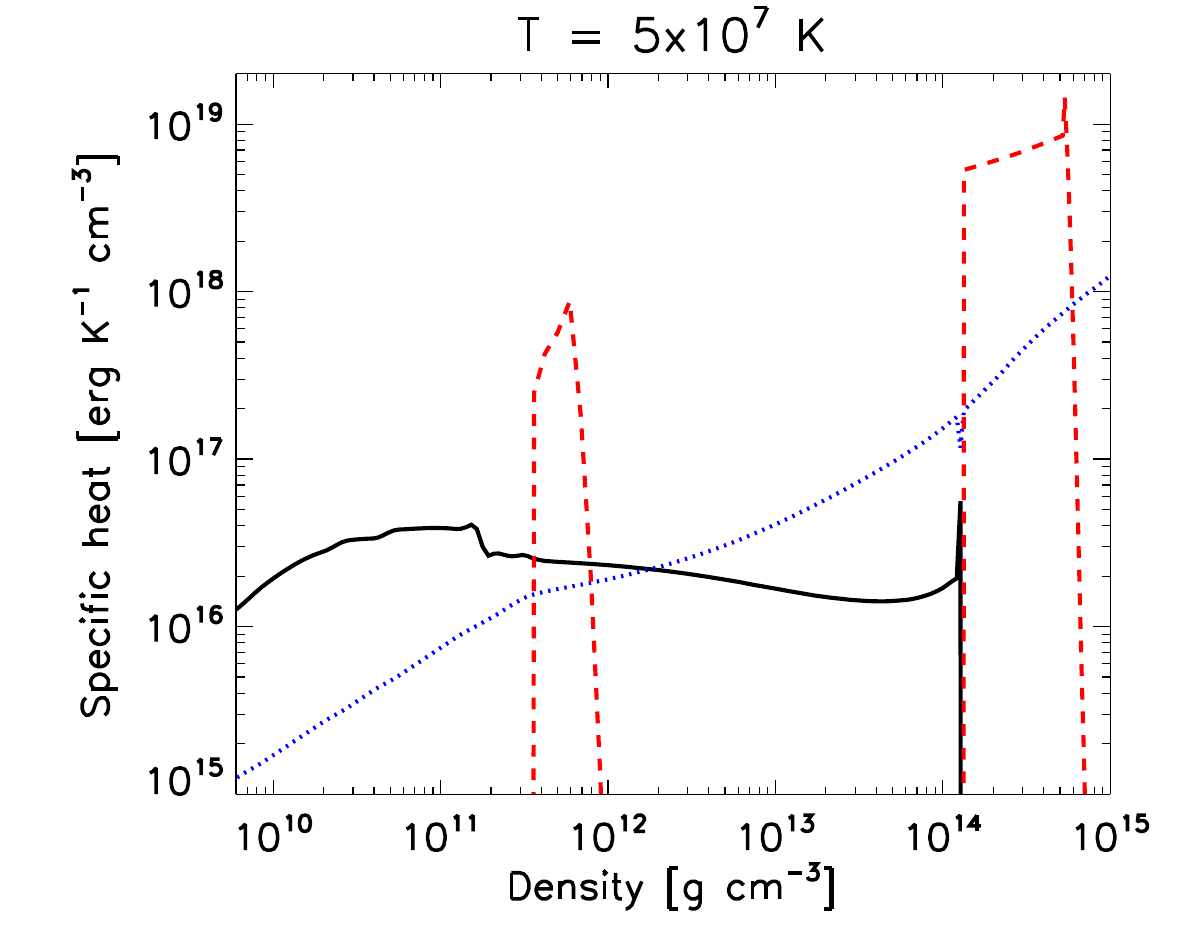}
	\caption{Contributions to the specific heat from neutrons (red dashes), protons (green dot-dashed), electrons (blue dots), and ions (black solid line) as a function of density, from the outer crust to the core, and for different temperatures in each panel (as indicated). The superfluid models employed here are the same as in \citet{Ho2012}. The plots refer to the representative NS shown in Fig.~\ref{fig:ns_profile}.}
	\label{fig:cv}
\end{figure}

The equation governing the temperature evolution at each point of the star's interior reads:
\begin{equation}
  c_\mathrm{v}\,\frac{\partial (T \mathrm{e}^\nu)}{\partial t}
      + \vec{\nabla}\cdot(\mathrm{e}^{2\nu} \vec{F}) = 
           \mathrm{e}^{2\nu} (H - Q)\, ,
\label{Tbalance}
\end{equation}
where $c_\mathrm{v}$ is specific heat, and the heat flux $\vec{F}$ is given by 
\begin{equation}
 \vec{F} = - \mathrm{e}^{-\nu} \hat{\kappa}\cdot\vec{\nabla}(\mathrm{e}^\nu T)\, ,
\end{equation}
with $\hat{\kappa}$ being the thermal conductivity tensor.
Throughout the text, we will use the $\vec{\nabla}$ operator for conciseness, but we note that it must include the metric factors of Eq.~(\ref{Schw}), so that its components in spherical coordinates are $\vec{\nabla}\equiv \left( \mathrm{e}^{-\lambda} \frac{\partial}{\partial r},\, \frac{1}{r} \frac{\partial}{\partial  \theta},\,  \frac{1}{r \sin\theta} \frac{\partial}{\partial \varphi} \right).$  
The source term has contributions from the neutrino emissivity $Q$ (accounting for energy losses by neutrino emission), and the heating power per unit volume $H$, both functions of temperature, in general. The latter may include contributions from, for example, accretion and—more relevant for this paper-Joule heating due to magnetic field dissipation. All these quantities (including the temperature) vary in space and are measured in the local frame, with the metric (redshift) corrections accounting for the change to the observer's frame at infinity.

For weak enough magnetic fields, the conductivity can be safely considered isotropic, so that the tensor reduces to a scalar value multiplied by the identity matrix. In this case, given the approximately spherically symmetric background, temperature gradients are primarily radial across most of the star, making 1D models sufficiently accurate for the core and inner crust of weakly magnetised NSs.

However, in strong magnetic fields, such as those in magnetars, the electron thermal conductivity tensor in the crust becomes anisotropic. The thermal conductivity is significantly reduced in the direction perpendicular to the local magnetic field, limiting heat flow across the magnetic field lines.
In this case, in the relaxation time approximation, the ratio of conductivities parallel ($\kappa^\parallel$) and orthogonal ($\kappa^\perp$) to the magnetic field can be written as
\begin{equation}\label{eq:omegatau}
\frac{\kappa^\parallel}{\kappa^\perp}  \approx 1 + (\omega_B^e \tau_e)^2\, ,
\end{equation}
where we have introduced the so-called \emph{magnetization parameter} \citep{1980SvA....24..425U}, $\omega_B^e \tau_e$, where $\tau_e$ is the electron relaxation time and $\omega_B^e = eB/m^*_ec$ is the gyro-frequency of electrons with charge $-e$ and  effective mass $m^*_e$ moving in a magnetic field with intensity $B$.
Eq.~(\ref{eq:omegatau}) is only strictly valid in the classical approximation (see \citealt{potekhin2018} for a recent discussion of quantizing effects), but this dimensionless quantity is always a good indicator of the suppression of the thermal conductivity in the transverse direction. We will see later that this is also the relevant parameter to discriminate between different regimes for the magnetic field evolution.

To understand the role of anisotropy in strong magnetic fields, we can examine electron conductivity while neglecting quantizing effects. The heat flux, as derived by \cite{perez06},
is expressed in the compact form:
\begin{equation}
\vec{F} = -  e^{-\nu} \kappa^\perp \left[ \vec{\nabla} (e^\nu T) + (\omega_B^e \tau_e)^2 (\vec{b}\cdot  \vec{\nabla} (e^\nu T)) \vec{b} + \omega_B^e \tau_e (\vec{b} \times  \vec{\nabla} (e^\nu T)) \right]\, ,
\label{hflux}
\end{equation}
where $\vec{b} \equiv \vec{B}/B$ represents the unit vector aligned with the local magnetic field. This expression breaks the heat flux into three components: heat flow along the redshifted temperature gradient $\vec{\nabla} (e^\nu T)$, heat flow parallel to the magnetic field lines (along $\vec{b}$), and heat flow perpendicular to both the gradient and the field.

In axial symmetry, the $\varphi$-component of the heat flux is typically non-zero but does not need to be calculated, as it is independent of $\varphi$, resulting in a zero contribution to the flux divergence. For instance, with a purely poloidal magnetic field (only $r,\theta$ components), the last term in the heat flux equation (Eq.~\ref{hflux}) can be neglected, as it does not affect the temporal evolution of temperature. However, when a significant toroidal component $B_\varphi$ is present, this term contributes to the heat flux in the direction perpendicular to $\vec{\nabla} (e^\nu T)$.

Next, we provide a more detailed description of the relevant microphysics inputs, the source terms, and a key ingredient in the NS cooling models: the heat blanketing envelope.

\subsubsection{Heat capacity}

In Fig.~\ref{fig:cv} we show the different contributions to the specific heat (per unit volume) by ions, electrons, protons, and neutrons, for the same TOV solution as in Fig.~\ref{fig:ns_profile}, and considering four uniform temperature profiles across the star, $T=\{10,5,1,0.5\} \times 10^8$ K. For the superfluid/superconducting corrections we use the phenomenological formula for the momentum dependence of the energy gap at zero temperature employed in \citet{Ho2012}, in particular their \emph{deep neutron triplet} model.

The bulk of the total heat capacity of a NS is given by the core, where most of the mass is contained. The regions with superfluid nucleons are visible as deep drops in the specific heat. The proton contribution is always negligible. For this particular choice (other options are possible, given the large uncertainties in the superfluid gaps), neutrons in the outer core are not superfluid, thus their contribution is dominant.
The crustal specific heat is given by the dripped neutrons, the degenerate electron gas and the nuclear lattice \citep{vanriper91}. The specific heat of the lattice is generally the main contribution, except in parts of the inner crust where neutrons are not superfluid, or for temperatures $T\lesssim 10^8$ K, when the electron contribution becomes dominant. In any case, the small volume of the crust implies that its heat capacity is small in comparison to the core contribution. For a detailed computation of the specific heat and other transport properties, we recommend the codes publicly available at \url{http://www.ioffe.ru/astro/EIP/}, describing the EoS for a strongly magnetized, fully ionized electron-ion plasma \citep{potekhin10}.

\subsubsection{Thermal conductivity}

\begin{figure}[ht]
	\centering
	\includegraphics[width=0.49\textwidth]{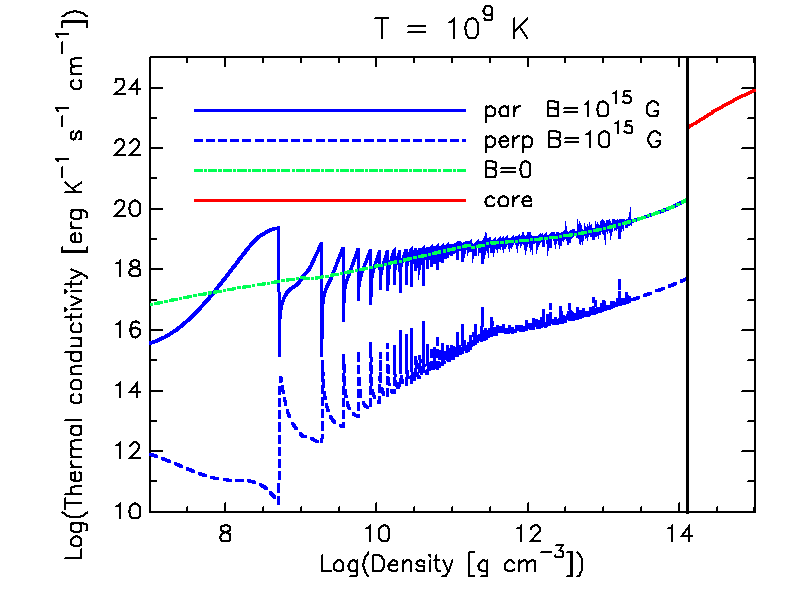}
	\includegraphics[width=0.49\textwidth]{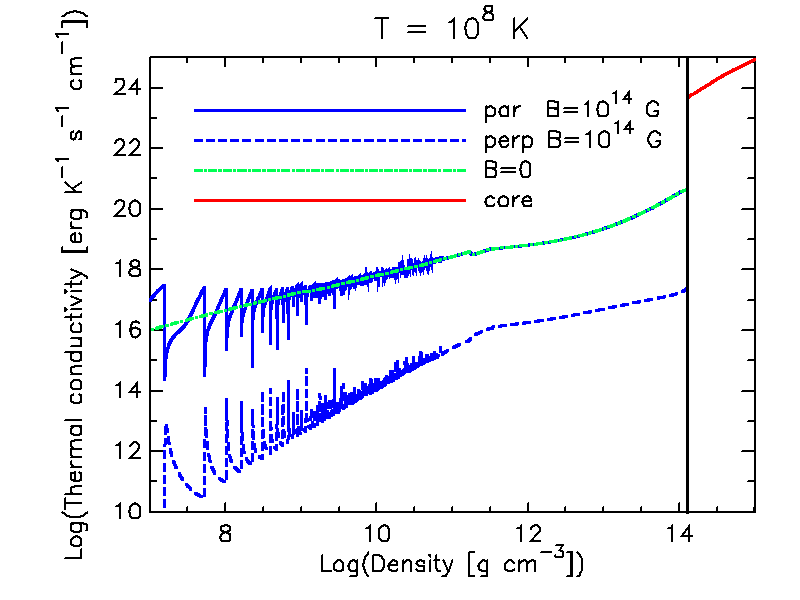}
	\caption{Thermal conductivity in the directions parallel (solid lines) and perpendicular (dashes) to the magnetic field, including quantizing effects. We show the cases $T=10^9$ K, $B=10^{15}$ G (left panel) and $T=10^8$ K, $B=10^{14}$ G (right panel). For comparison, the $B=0$ values are shown with green lines in both figures. The plots refer to the representative NS shown in Fig.~\ref{fig:ns_profile}.}
	\label{fig:cond_class_quant}
\end{figure}

Figure~\ref{fig:cond_class_quant} shows the thermal conductivity including the contributions of all relevant carriers, for two different combinations of constant temperatures and magnetic field: $T=10^9$ K, $B=10^{15}$ G (left panel) and $T=10^8$ K, $B=10^{14}$ G (right), for the same fiducial NS of Fig.~\ref{fig:ns_profile}. 
For simplicity, we present profiles using idealized, uniform values of $T$ and $B$, which are roughly representative of a newly formed magnetar and one that has evolved over approximately
($\sim 10^4$ yr), respectively. 
Note that the thermal conductivity of the core, far exceeding that of the crust by several orders of magnitude, quickly leads to a nearly isothermal core, irrespective of the initial thermodynamic conditions, as previously discussed.

Thus, the precise value of the core thermal conductivity becomes unimportant, and thermal gradients can only be developed and maintained in the crust and the envelope. In the crust, the dissipative processes responsible for the finite thermal conductivity include all the mutual interactions between electrons, lattice phonons (collective motion of ions in the solid phase), impurities (defects in the lattice), superfluid phonons (collective motion of superfluid neutrons) or normal neutrons. The mean free path of free neutrons, which is limited by the interactions with the lattice, is expected to be much shorter than for the electrons, but a fully consistent calculation is yet to be done \citep{chamel08}.
Quantizing effects due to the presence of a strong magnetic field become important only in the envelope, or in the outer crust for very large magnetic fields ($B \gtrsim 10^{15}$ G). For comparison, we also plot the $B=0$ values. The quantizing effects are visible as oscillations around the classical (non-magnetic) values, corresponding to the gradual filling of Landau levels. More details about the calculation of the microphysics input ($\hat{\kappa}, c_v, \, Q$) can be found in Sect.~2 of \citet{potekhin_rev15a}. 

\subsubsection{Neutrino emissivity}
\begin{figure}
    \centering
    \includegraphics[width = \textwidth]{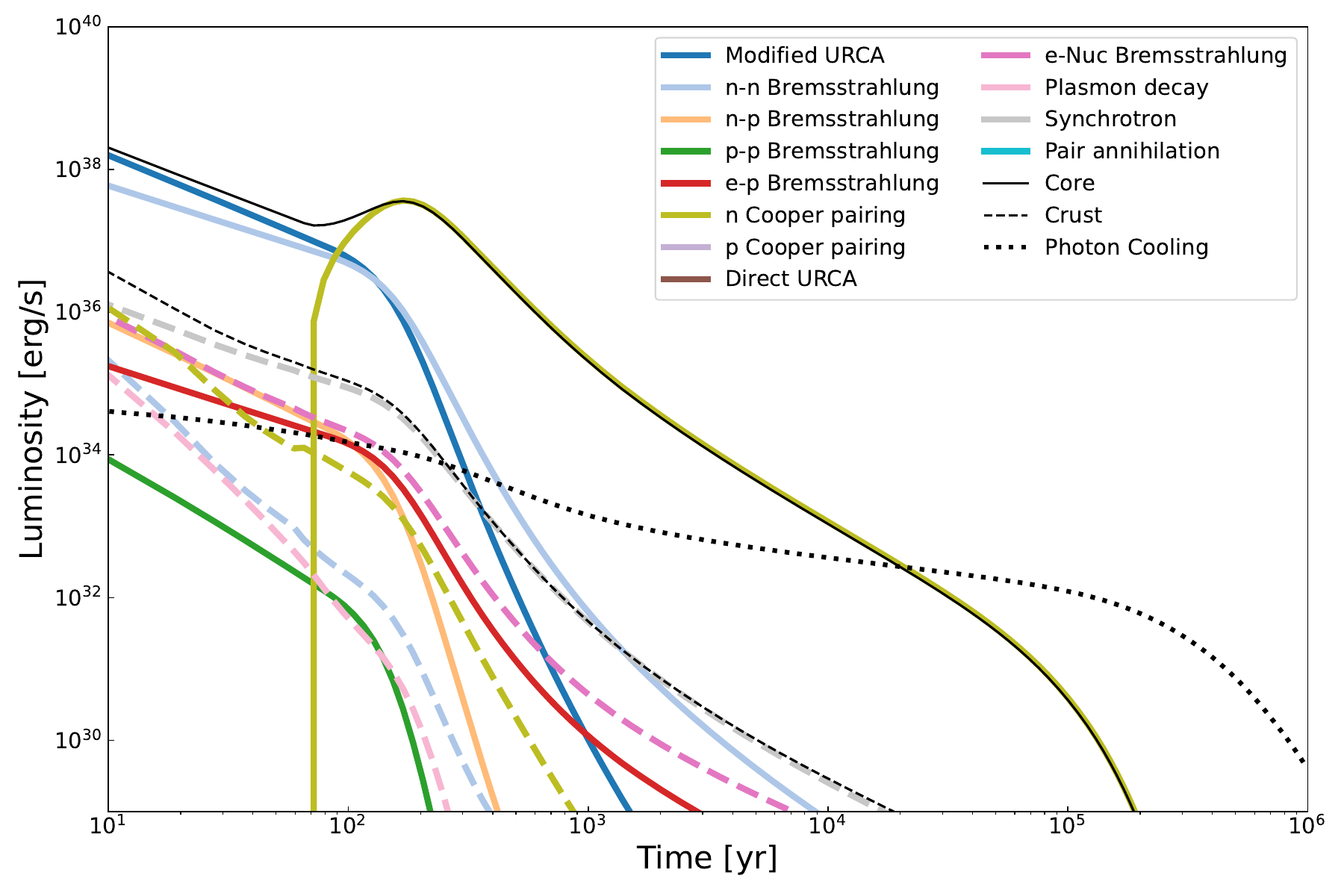}
    \caption{Evolution of neutrino and photon luminosities from the different emission processes throughout the NS history, computed with the SLy4 EoS for a mass of $M = 1.6 \, M_\odot$. Solid-colored lines represent different neutrino emission processes occurring in the core. Dashed lines represent different neutrino emission processes occurring in the crust. The same process (marked with a given color) may involve both the core and the crust. Black lines represent the total neutrino luminosity for processes involving the core (continuous line) and the crust (dashed line), respectively. The black-dots report the surface photon luminosity. This figure assumes the gap model of \citet{Ho2015}.
    It is worth noticing that while the legend includes all the possible neutrino processes included in the simulation, some of them are not effectively active in this particular simulation, as such, they do not appear in the figure. Moreover, in case magnetic fields are present in the core, additional processes can become relevant \citep{Kantor2021}. Image reproduced with permission from \citet{MATINS2}, copyright by the author(s)}
    \label{fig:cooling_curve_benchmark}
\end{figure}

A third (and crucial) component in NS cooling studies is the neutrino emissivity. For a detailed overview, we direct readers to the comprehensive review by \citet{2001PhR...354....1Y}, the summary of key processes with references in Table~3 of \citet{aguilera08b}. Recent advancements not incorporated in the above cited include:
in-medium enhancement of the modified URCA rates \citep{Shternin18,Alford24}, the role of non-equilibrium reactions \citep{Yanagi20}, the effect of short-range correlations \citep{Sedrakian24}, a critical reassessment of Bremsstrahlung and modified URCA rates \citep{Bottaro24}, or the always important role of magnetic fields \citep{Tambe25}. Fig.~\ref{fig:cooling_curve_benchmark} summarizes the evolution of neutrino and photon luminosities over one million years, from the different emission processes throughout the NS history, computed with the SLy4 EoS for a mass of $M = 1.6 \, M_\odot$. Note that in Fig.~\ref{fig:cooling_curve_benchmark} the gap models used for nuclear superfluidity and proton superconductivity are those adopted by \citet{Ho2015}: SFB \citep{Schwenk2003} for neutrons in the crust, TToa \citep{Takatsuka2004} for neutrons in the core, and CCDK \citep{Chen1993, Elgaroy1996} for protons in the core. The choice of gap models used here differs from that adopted in Fig.~\ref{fig:cv}, which employs the superfluid models indicated in \citet{Ho2012}.

\subsubsection{Heating sources}
The other important contribution in the source term in Eq.~(\ref{Tbalance}) accounts for possible heating mechanisms.
Various internal heating mechanisms, such as magnetic field Ohmic dissipation, dark matter accretion, crust cracking, and vortex creep have been proposed in the literature \citep[see e.g.][for a comparative study]{GR2010,beloborodov16}. 
In this review, we will discuss later Ohmic dissipation, since it is arguably the dominant effect for young and middle-aged pulsars.

Regarding other mechanisms, note that rotochemical heating \citep{Reis1995,PR2010,GR2015} and vortex creep have been proposed to produce detectable thermal emission in old NSs. In particular, the rotochemical mechanism is sourced by the loss of angular momentum and rotational energy, which makes the cores slightly contract, increasing the internal density and driving the matter out of $\beta$-equilibrium. This imbalance leads to the accumulation of chemical energy, which can be released through weak interactions. This mechanism enhances reaction rates and neutrino emission, and when the chemical potential imbalance is sufficiently large, it can result in net heating of the star. Rotochemical heating is sensitive to the spin-down history and can be particularly relevant in old millisecond pulsars with low magnetic fields (NSs which have been span-up by the long-term accretion from a companion), where it may dominate the thermal evolution. The presence of superfluid nucleons further alters this picture by suppressing standard neutrino processes while enabling additional reactions via Cooper pair formation.

Another potential heating mechanism arises from dark matter particles accumulating inside the neutron star, releasing energy via annihilation. This could offset surface thermal losses, leading to a stabilized surface temperature evolution in stars older than 1–10 million years \citep{Hamaguchi19}.

\subsubsection{The heat blanketing envelope}
\label{subsection:envelope}

In the low-density region (envelope and atmosphere), radiative equilibrium will be established much faster than the interior evolves. The difference by many orders of magnitude of the thermal relaxation timescales between the envelope and the interior (crust and core) makes it computationally unpractical to perform cooling simulations in a numerical grid including all layers up to the star surface.
Therefore, the outer layer is effectively treated as a boundary condition. It relies on a separate calculation of stationary envelope models to obtain a functional fit giving a relation between the surface temperature $T_s$, which determines the radiation flux, and the temperature $T_b$ at the crust/envelope boundary. This $T_s - T_b$ relation provides the outer boundary condition to the heat transfer equation. The radiation from the surface is usually assumed to be blackbody radiation, although the alternative possibility of more elaborated atmosphere models, or anisotropic radiation from a condensed surface, has also been studied \citep{turolla04,vanadelsberg05,perez05,potekhin12}.

Section~5 of \citet{potekhin_rev15a} provides a historical overview and contemporary examples of NS envelope models. For an up-to-date and thorough review of advanced envelope models, we recommend \citet{Beznogov2021}.
This study explores various heat blanket models, analyzing the effects of layered compositions, with or without diffusion equilibrium, the influence of strong magnetic fields, and the role of high temperatures in driving significant neutrino emission. It also examines how these properties shape the thermal evolution of NSs, offering insights into their internal structure. Extending this line of work, \citet{Dehman23b} used a 2D magneto-thermal evolution model to investigate envelope properties and magnetic field topology, showing that different envelope models can lead to radically different predictions for the surface temperature and its evolution with age.

In the remainder of this section, we focus on the main aspects of the numerical methods employed to solve Eq.~(\ref{Tbalance}) alone. We will return to the specific problems arising from the coupling with the magnetic evolution in the following sections.

\subsection{Numerical methods for multidimensional cooling}
\label{sect:cooling 2D}

Classically, there are two broad strategies to solve the heat equation: spectral methods and finite-difference schemes. Spectral methods are well known to be elegant, accurate and efficient for solving partial differential equations with parabolic and elliptic terms, where Laplacian (or similar) operators are present. However, they are much more tedious to implement and to be modified, and usually require some strong previous mathematical understanding. On the contrary, finite-difference schemes are very easy to implement and do not require any complex theoretical background before they can be applied. On the negative side, finite-difference schemes are less efficient and accurate when compared to spectral methods using the same amount of computational resources. The choice of one over the other is mostly a matter of taste. However, in realistic problems with ``dirty'' microphysics (irregular or discontinuous coefficients, stiff source-terms, quantities varying many orders of magnitude, etc), simpler finite-difference schemes are usually more robust and more flexible than the heavy mathematical machinery normally carried along with spectral methods, which are often derived for constant microphysical parameters. 

A third novel strategy has appeared in the last few years: the so-called Physics Informed Neural Networks (PINNs), introduced by \citet{PINNs2019}, which utilize deep learning to approximate solutions to linear and non-linear partial differential equations. Enabled by recent advances in computational power, graph-based automatic differentiation, and frameworks like TensorFlow and PyTorch, PINNs integrate physical laws into the neural network's loss function, minimizing PDE residuals during training. Unlike traditional deep learning, PINNs require minimal or no data. They have been applied in fields like fluid dynamics, nuclear reactor dynamics, radiative transfer, black-hole spectroscopy and, as we will discuss later in this review, NS magnetospheres \citep{Urban23, Stefanou23b}. The heat equation, particularly in 2D or 3D, is an optimal problem for PINNs because the solutions are expected to be smooth, and PINNs scale better than classical methods with increasing dimensionality. Although these methods are believed to be less efficient and precise than classical finite-difference or finite-element methods, the gap is closing very fast \citep{Urban2025}. At present, PINNs offer flexibility as general-purpose PDE solvers, handling arbitrary, unstructured meshes without high-resolution grids. Once trained, PINNs provide fast solutions via a forward pass, offering potential speed advantages over traditional methods.

Most studies on NS cooling in the literature have utilized standard finite difference methods; therefore, we briefly review the key references and discuss the primary technical challenges. The first 2D models (in axial symmetry) of the stationary thermal structure in a realistic context (including the comparison to 
observational data) were obtained by \citet{geppert04,geppert06} and \citet{perez06}, paving the road for subsequent 2D simulations of the time evolution of temperature in strongly magnetized NS \citep{aguilera08a,aguilera08b,2014MNRAS.442.3484K}.
In all these works, the magnetic field was held fixed, as a background, exploring different possibilities, including superstrong ($B\sim10^{15}$\,--\,$10^{16}$~G) toroidal magnetic fields in the crust to explain the strongly non-uniform distribution of the surface temperature. 

\begin{figure}[h]
 \centering
\includegraphics[width=.5\textwidth]{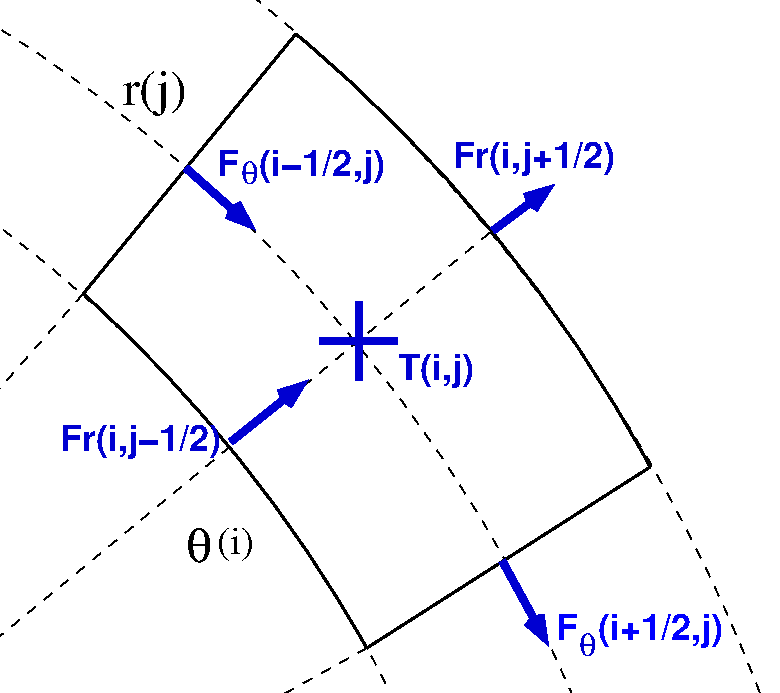}
\caption{Schematic illustration of the allocation of temperatures (cell centers) and fluxes (cell interfaces) in a typical grid in polar coordinates.}
\label{fig_staggered_T}
\end{figure}

In \citet{aguilera08a,aguilera08b,vigano13,vigano21} and related works, values of temperature are defined at the center of each cell, where also the heating rate and the neutrino losses are evaluated, while fluxes are calculated at each cell-edge, as illustrated in Fig.~\ref{fig_staggered_T}. 
The boundary conditions at the center ($r=0$) are simply $\vec{F}=0$, while on the axis the non-radial components of the flux must vanish. As an outer boundary, they consider the crust/envelope interface, $r=R_b$, where the outgoing radial flux, $F_{\rm out}$, is given by a formula depending on the values of $T_b$ and $\vec{B}$ in the last numerical cell. For example, assuming blackbody emission from the surface, for each outermost numerical cell, characterized by an outer surface $\Sigma_r$ and a given value of $T_b$ and $\vec{B}$, one has $F_{\rm out}=\sigma_B \Sigma_r T_s^4$ where $\sigma_B$ is the Stefan-Boltzmann constant, and $T_s$ is given by the $T_s - T_b$ relation (dependent on $\vec{B}$), as discussed in the previous subsection on envelope models.

To overcome the strong limitation on the time step in the heat equation, $\Delta t \propto (\Delta x)^2$, the diffusion equation can be discretized in time in a semi-implicit or fully implicit way, which results in a linear system of equations described by a block tridiagonal matrix \citep{bookRM67}.
The ``unknowns'' vector, formed by the temperatures in each cell, is advanced by inverting the matrix with standard numerical techniques for linear algebra problems, like the lower-upper (LU) decomposition, a common Gauss elimination based method for general matrices, available in open source packages like \textsc{LAPACK}. However, this is not the most efficient method for large matrices. A particular adaptation of the Gauss elimination to the block-tridiagonal systems, known as Thomas algorithm \citep{Thomas1949} or matrix-sweeping algorithm, is much more efficient, but its parallelization is limited to the operations within each of the block matrices. A new idea that has been proposed to overcome parallelization restrictions is to combine the Thomas method with a different decomposition of the block tridiagonal matrix \citep{Belov2017}.

A word of caution is in order regarding the treatment of the source term. The thermal evolution during the first Myr is strongly dominated by neutrino emission processes, which enter the evolution equation through a very stiff source term, typically a power-law of the temperature with a high index ($T^8$ for modified URCA processes, $T^6$ for direct URCA processes). These source terms cannot be handled explicitly without reducing the time step to unacceptable small values but, since they are local rates, linearization followed by a fully implicit discretization is straightforward and results in the redefinition of the source vector and the diagonal terms of the matrix. 
A very basic description to deal with stiff source terms can be found in Sect.~17.5 of \citet{NumRec}.
This procedure is stable, at the cost of losing some precision, but it can be improved by using more elaborated implicit-explicit Runge--Kutta algorithms \citep{IMEX}.

Typically, effects of rotation are neglected in magnetar studies due to their characteristically slow rotation rates, which minimally impact their thermal evolution. However, the role of rapid rotation, which can significantly enhance temperature anisotropy has received attention 
in recent research \citep{Beznogov23}. 
This study explores the long-term thermal evolution of axisymmetric rotating NSs using a fully general relativistic framework. To achieve this, they introduce NSCool 2D Rot, a substantial upgrade to the one-dimensional NSCool code \citep{Page2016} developed by Dany Page. 

The transition to 3D models with realistic microphysics has only recently occurred \citep{Degrandis21, Igoshev21, MATINS_MT, MATINS2}, primarily because radial gradients dominate in most scenarios, and angular anisotropies in the outer layers (crust and envelope) become significant only in the presence of ultra-strong magnetic fields. In \citet{MATINS2}, the authors introduce the thermal evolution module of a new three-dimensional magnetothermal code, 
\MATINS (MAgneto-Thermal evolution of Isolated Neutron Stars; \citealt{MATINS1,MATINS_MT,MATINS2}). \MATINS utilizes a finite volume approach and incorporates a realistic background structure, alongside advanced microphysical models for conductivities, neutrino emissivities, heat capacity, and superfluid gap calculations.

\subsection{Temperature anisotropy in a magnetized NS}\label{sec:T_anis}

\begin{figure}[ht]
 \centering
\includegraphics[width=.9\textwidth]{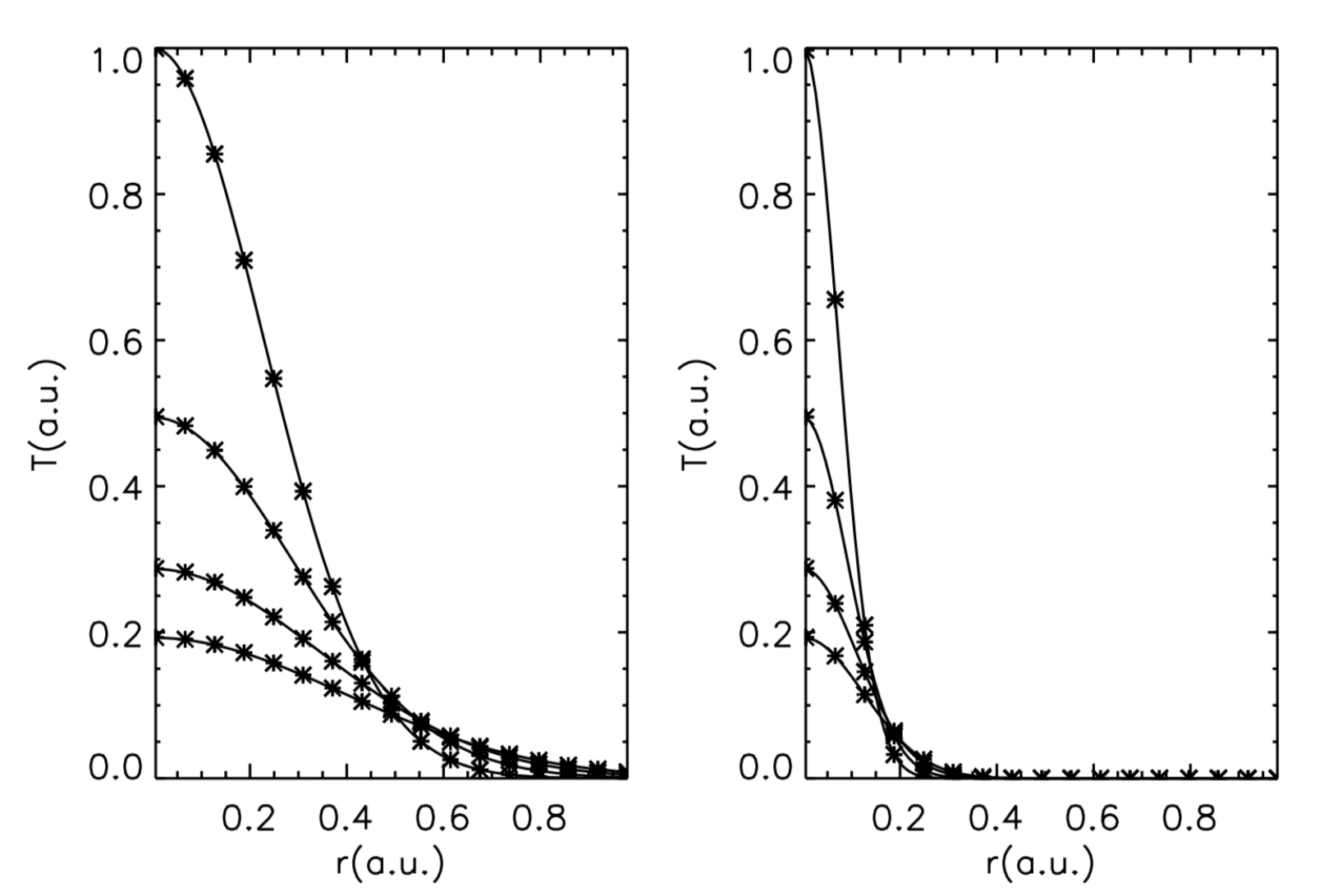}
\caption{Temperature profiles at different times comparing the analytic solution (solid) and the numerical evolution (stars) of a thermal pulse in a medium embedded in a homogeneous magnetic field. The left (right) panel shows four different times during the evolution of polar (equatorial) profiles in arbitrary units. The simulation has been done with a fully implicit scheme and the linear system is solved with the Thomas algorithm.
Image reproduced with permission from \citet{perez06}, copyright by ESO.} 
 \label{fig_test_cool}
\end{figure}

\begin{figure}[ht]
 \centering
\includegraphics[width=.9\textwidth]{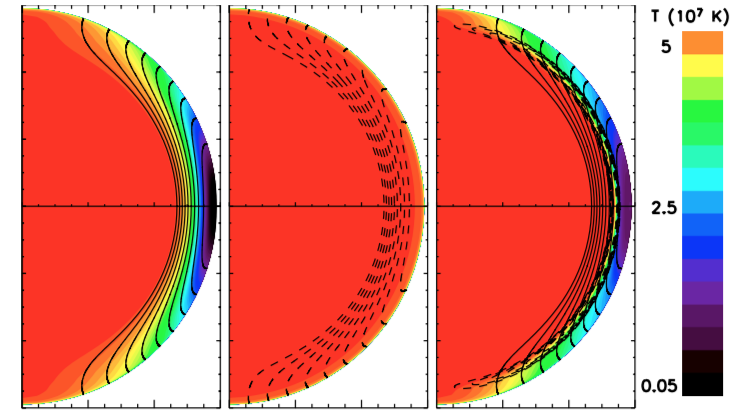}
\caption{Temperature anisotropy induced in the NS crust by the presence of a strong magnetic field confined into the crust. The projections of the poloidal field lines are shown with solid lines in the left and right panels, and dashed lines in the central panel. The left panel corresponds to a model without toroidal field, the central panel to a force-free configuration (toroidal magnetic flux contours and poloidal magnetic field lines are aligned), and the right panel shows a model with a toroidal component confined to a narrow region of the crust represented by dashed lines. Image reproduced with permission from  \citet{perez06}, copyright by ESO.}
\label{tanis}
\end{figure}
An analytical solution that can be used to test numerical codes in multi-dimensions is the evolution of a thermal pulse in an infinite medium, embedded in a homogeneous magnetic field oriented along the $z$-axis, which causes the anisotropic diffusion of heat. Assuming constant conductivities, and neglecting relativistic effects, the following analytical solution for the temperature profile can be obtained for $t>t_0$:
\begin{equation}
T(t, r, \theta) = T_0 \left(\frac{t_0}{t}\right)^{3/2} \exp\left[ - \frac{r^2}{4  t {\kappa^\perp} } \left( {\sin^2\theta}+ 
\frac{\cos^2\theta}{1 + (\omega_B^e \tau_e)^2} \right) \right]\, ,
\end{equation}
where $T_0$ is the central temperature at the initial time $t_0$. In Fig.~\ref{fig_test_cool} we show the comparison between
the  analytical (solid) and numerical (stars) solution for a model with $t_0=10^{-4}$, $T_0=1$, $\kappa^\perp = 10^2$ and $\omega_B^e \tau_e = 3$. The boundary conditions employed are $F=0$ at the center and the temperature  corresponding to the analytical solution at the surface ($r=1$). \citet{perez06} found deviations from the analytical solution to be less than 0.1\% in any particular cell within the entire domain, even with a relatively low grid resolution of 100 radial zones and 40 angular zones. The same numerical test has been used to test the new 3D code \MATINS \citep{MATINS2}.

To conclude this section, the induced anisotropy in a realistic NS reported by \citet{perez06} is shown in Fig.~\ref{tanis}.
The figure shows equilibrium thermal solutions, in the absence of heat sources and sinks. The core temperature is kept at $5\times 10^7$ K, and the surface boundary condition is given by the $T_s-T_b$ relation, assuming blackbody radiation. The poloidal component is the same in all models ($B_p = 10^{13}$ G).
The effect of the magnetic field on the temperature distribution can be easily understood by examining the expression of the heat flux (\ref{hflux}). When $\omega_B^e \tau_e \gg 1$, the dominant contribution to the flux is parallel to the magnetic field and proportional to  $\vec{b}\cdot \vec{\nabla} (e^\nu T)$. Thus, in the stationary regime (i.e., $\vec{\nabla}\cdot(e^{2\nu}\vec{F})=0$ if no sources are present), the temperature distribution must be such that $\vec{b} \perp \vec{\nabla} (e^\nu T)$: magnetic field lines are tangent to surfaces of constant temperature. This is explicitly visible in the left panel, which corresponds to the stationary solution for a purely poloidal configuration with a core temperature of $5 \times 10^7$ K. Only near the surface, the large temperature gradient can result in a significant heat flux across the magnetic field lines. When we add a strong toroidal component, the Hall term (the one proportional to $\omega_B^e \tau_e$ in Eq.~(\ref{hflux})) results in meridional heat fluxes which lead to a nearly isothermal crust. 
The central panel shows the temperature distribution for a force-free magnetic field with a global toroidal component, present in both the crust and the envelope. The right panel shows a third model with a strong toroidal component confined to a thin crustal region (dashed lines). It acts as an insulator maintaining a temperature gradient between both sides of the toroidal field.

\section{Magnetic field evolution in the interior of neutron stars: theory review}
\label{sec:magnetic_evolution}

The interior of a NS is a complex multifluid system, where different species coexist and may have different average hydrodynamical velocities. 
In most of the crust, nuclei have very restricted mobility and form a solid lattice. Only the ``electron fluid'' can flow, providing the currents that sustain the magnetic field. In the inner crust superfluid neutrons are partially decoupled from the heavy nuclei, providing a third neutral component. In the core, the coexistence of superfluid neutrons and superconducting protons makes the situation even less clear. Since a full multifluid, MHD-like description of the system is far from being affordable, one must rely on different levels of approximation that gradually incorporate the relevant physics. In this section we give an overview of the theory, trying to capture the most relevant processes governing the magnetic field evolution in a relatively simple mathematical form. 

The evolution of the magnetic field is given by Faraday's induction law:
\begin{equation}
\frac{\partial \vec{B}}{\partial t} = - c \vec{\nabla}\times ( \mathrm{e}^{\nu} \vec{E} )\, ,
\label{induction0}
\end{equation}
which needs to be closed by the prescription of the electric field $\vec{E}$ in terms of the other variables (constituent component velocities and the magnetic field itself), either using simplifying assumptions (e.g., Ohm's law) or solving additional equations. Very often, this prescription involves the electrical current density, which is typically obtained from Amp\'ere's law, neglecting the displacement currents due to the high electrical conductivity  
(the usual MHD approximation):
\begin{equation}\label{eq:current_mhd}
\vec{j} = \mathrm{e}^{-\nu} \frac{c}{4 \pi} \vec{\nabla}\times \left(\mathrm{e}^{\nu} \vec{B}  \right)~. 
\end{equation}
To maintain consistency with the previous section, we adopted the same spherically symmetric background metric and retained relativistic corrections in this introductory subsection. However, to avoid cluttering the text with unnecessary $\mathrm{e}^{\nu}$ factors (which are not essential for our present purposes), we will omit them in the reminder of this review for clarity, with a few exceptions in which they play a relevant role and it will be explicitly indicated in the text.

In a complete multi-fluid description of plasmas, the set of hydrodynamic equations complements Faraday's law. From the multi-fluid hydrodynamics equations, a generalized Ohm's law --- in which the electrical conductivity is a tensor --- can be derived \citep{1990SvAL...16...86Y,1995MNRAS.273..643S} 
$$ \vec{j} = \hat{\sigma}  \vec{E}.$$
Expressing the tensor components in a basis referred to the magnetic field orientation, one can identify longitudinal, perpendicular and Hall components, that give rise to a complex structure when the equation is inverted to express $\vec{E}$ as a function of $\vec{j}$, $\vec{B}$, and possibly other terms independent of the magnetic field (gradients of temperature and chemical potential). 

In some regimes, one can make simplifications to make the problem more affordable \citep{1980SvA....24..425U,jones1988,goldreich92}, although one should incorporate as much physics as possible.
The three main processes are Ohmic dissipation, Hall drift (mostly relevant in the crust) and ambipolar diffusion (mostly relevant in the core, \citealt{goldreich92,1995MNRAS.273..643S,cumming04}), although additional terms could in principle be included in the induction equation. For instance, there are theoretical arguments proposing additional slow-motion dynamical terms, such as plastic flow \citep{beloborodov14,lander16,lander19}, magnetically induced superfluid flows \citep{ofengeim18} or vortex buoyancy \citep{muslimov1985,konenkov2000,Elfritz2016,dommes}. Typically, all these effects are introduced as advective terms, of the type $\vec{E} = -\vec{\mathrm v}\times \vec{B}$, with $\vec{v}$ being some effective velocity. The thermoelectric effect (with a contribution to the electric field of the form
$\vec{E} =  - s \vec{\nabla} T$) has also been proposed to become significant in regions with large temperature gradients \citep{geppert91,wiebicke91,wiebicke92,wiebicke95,geppert95,wiebicke96}, and has been recently revisited in \citet{gourgouliatos22,gakis2024}.
These additional terms are typically not included in most of the existing literature. However, some of them may play a more important role than expected and should be carefully reconsidered. 

We now individually address the most significant and well-understood contributions to the electric field, discussing the physical processes at their origin.

\subsection{Ohmic dissipation}
\label{sec:ohmic}

In the simplest case, the electric field in the reference frame co-moving with matter is simply related to the 
electrical current density, $\vec{j}$, by:
\begin{equation}\label{eq:ohm}
\vec{E} = \frac{\vec{j}}{\sigma} \, ,
\end{equation}
where the conductivity $\sigma$, dominated by electrons, must take into account all the (usually temperature-dependent) collision processes of the charge carriers. 
Here, $\sigma$ actually represents the longitudinal (to the magnetic field) component of the general conductivity tensor $\hat{\sigma}$. In the weak field limit, the tensor becomes a scalar ($\sigma \equiv \sigma_\parallel $) times 
the identity, and anisotropic effects are absent.

The induction equation, when we have only Ohmic dissipation, conforms a {\it vector diffusion equation}:
\begin{equation}
	\frac{\partial \vec{B}}{\partial t} +  \vec{\nabla}\times \left( \eta \vec{\nabla}\times  
    \vec{B} \right) = 0\, ,
	\label{ohm0}
\end{equation}
where we have defined the magnetic diffusivity $\eta \equiv \frac{c^2}{4 \pi \sigma}$. 

\begin{figure}[t]
	\centering
	\includegraphics[width=0.49\textwidth]{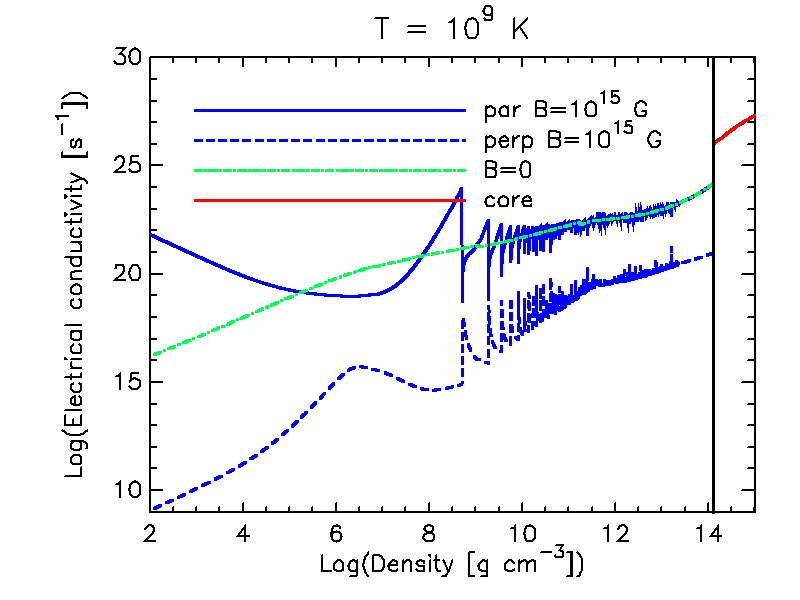}
	\includegraphics[width=0.49\textwidth]{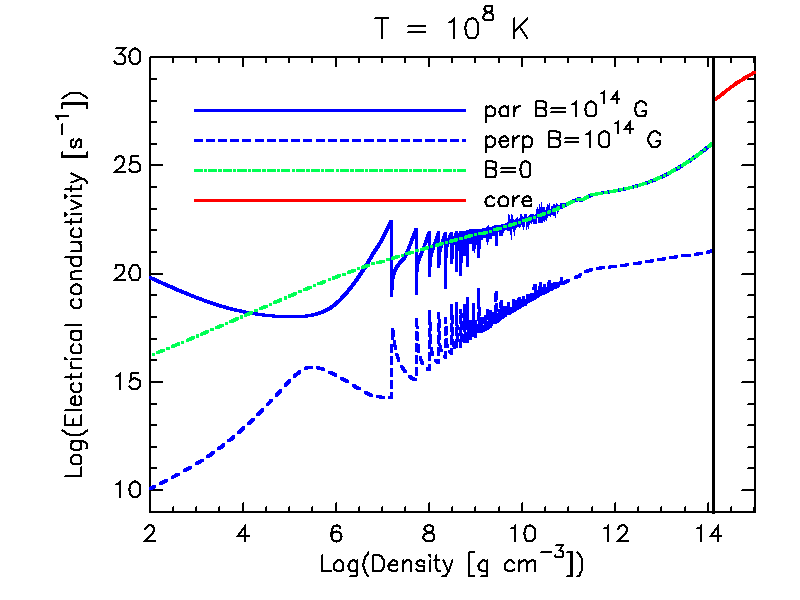}
	\caption{Electrical conductivity in the directions parallel (solid lines) and perpendicular (dashes) to the magnetic field, including quantizing effects. We show the cases $T=10^9$ K, $B=10^{15}$ G (left panel) and $T=10^8$ K, $B=10^{14}$ G (right panel). For comparison, the $B=0$ values are shown with green lines in both figures. Plots refer to the same representative NS shown in Fig.~\ref{fig:ns_profile}.}
	\label{fig:cond_el}
\end{figure}

In the relaxation time approximation, the electrical conductivity parallel to the magnetic field, $\sigma=e^2 n_e \tau_e/m^*_e$, with $n_e$ being the electron number density.
Typical values of the electrical conductivity in the crust are $\sigma \sim 10^{22}$--$10^{25}$ s$^{-1}$, several orders of magnitude larger than in the most conductive terrestrial metals described by the band theory in solid state physics. In the core, the even larger electrical conductivity ($\sigma \sim 10^{26}$--$10^{29}$ s$^{-1}$) results in much longer Ohmic timescales, thus potentially affecting the magnetic field evolution only at a very late stage ($t \gtrsim 10^8$ yr), when isolated NSs are too cold to be observed. In Fig.~\ref{fig:cond_el} we show typical profiles of the electrical conductivity, for the same combinations of $T$ and $B$ shown for the thermal conductivity in Fig.~\ref{fig:cond_class_quant}. Since, neglecting inelastic scattering, both thermal and electrical conductivities are proportional to the collision time $\tau_e$, they share some trends: the suppression of the conduction in the direction orthogonal to a strong magnetic field, and the quantizing effects visible as oscillations around the classical value \citep{potekhin_rev15a,potekhin2018}. We note that, if inelastic scattering contributes significantly, $\tau_e$ can be different for thermal and electrical conductivities. 

\subsection{The Hall drift}
\label{sec: Hall drift}

At the next level of approximation, it is necessary to account not only for Ohmic dissipation but also for the advection of magnetic field lines by the charged component of the fluid, predominantly the electrons, moving with velocity $\vec{\mathrm v_e}$. The electric field has the following form 
\begin{equation}\label{eq:induction}
\vec{E} =  \frac{\vec{j}}{\sigma} - \frac{\vec{\mathrm v_e}}{c} \times \vec{B}   ~.
\end{equation}

In the crust, where electrons are the only mobile charge carriers, their velocity is directly proportional to the electric current
\begin{equation}
\vec{\mathrm v}_e = - \frac{ \vec{j} }{e n_e} \, ,
\end{equation}
and the Hall--MHD (or electron--MHD) induction equation reads
\begin{equation}\label{eq:hall_induction}
\frac{\partial \vec{B}}{\partial t} =
- \vec{\nabla}\times \left\{ \eta \vec{\nabla}\times \vec{B} 
+ \frac{c }{4\pi e n_e} (\vec{\nabla}\times \vec{B}) 
\times \vec{B}\right\}~.
\end{equation}

Here, the first term on the right-hand side is the same as in Eq.~(\ref{ohm0}) and accounts for Ohmic dissipation, while the second term is the nonlinear Hall term. Note that the coefficient of the second does not depend on the temperature, but it varies by orders of magnitude in the crust due to the inverse dependence with density. We can factor out the magnetic diffusivity and express the Hall induction equation in the form
\begin{equation} 
\frac{\partial\vec B}{\partial t}= - {\Bnabla \times}\left(\eta 
\left\{\Bnabla \times \vec{B} + \omega_B^e \tau_e [(\Bnabla \times \vec{B} ) \times \hat{b} ]\right\} \right).
\label{Hallind} 
\end{equation} 
where $\hat{b} = \BB/B$. This makes explicit that the magnetization parameter $\omega_B^e \tau_e$
(the same that determined the degree of anisotropy of heat transfer, Eq.~(\ref{eq:omegatau}), plays the role of a magnetic Reynolds number: it gives the relative weight of the Hall and Ohmic dissipation terms. Generally speaking, as we approach the surface from the interior, $\omega_B^e \tau_e$ increases. It is important to note that, in light of these considerations, analytical estimates of the Ohmic or Hall timescales must be interpreted with caution, as they can vary by many orders of magnitude depending on the local physical conditions.

Most previous studies of magnetic field evolution in NS crusts
\citep{HR2002,HR2004,PonsGeppert2007,Reise2007,Pons2009,Kondi2011,vigano12a,vigano13,gourgouliatos13,marchant14,gourgouliatos14a,gourgouliatos15a,gourgouliatos15b,wood15}
have been restricted to 2D simulations. The few recent 3D models
\citep{vigano19,gourgouliatos19,Degrandis21,Degrandis22,Igoshev23a,Igoshev23b,MATINS1,MATINS_MT} suggest that most 2D features persist, most notably, the role of the Hall term in driving a direct cascade, transferring magnetic energy from large to small scales and thereby enhancing Ohmic dissipation (see Sect.~\ref{sec:ohmic}). Fully 3D simulations, however, reveal distinct behaviors, such as the emergence of long-lived, Hall-driven small-scale azimuthal magnetic structures (see Sect.~\ref{sec:examples} for details) and the occurrence of an inverse cascade \citep{brandenburg2020,DehmanBrandenburg25}. The latter transfers magnetic energy from small-scale turbulence to larger scales, a process of key relevance for the amplification of large-scale fields in astrophysical contexts.


The inverse cascade is rooted in the conservation of magnetic helicity,
\begin{equation}
    \chi_M = \int_V \AAA \cdot \BB \, dV
    \label{eq: A.B}
\end{equation}
where $\AAA$ is the magnetic vector potential, $\BB = \Bnabla \times \AAA$, and the integration is performed over a control volume $V$. The maximum helicity that a magnetic field can contain is constrained by the realizability condition, derived from the Schwarz inequality \citep{Mof78}:
\begin{equation}
    k |\chi_M (k)| /2 \leq \EM(k),
    \label{eq: realisability cond}
\end{equation}
where $k$ is the wavenumber, and $\chi_M(k)$ and $\EM(k)$ are the spectra of magnetic helicity and magnetic energy, respectively. Since $\chi_M(k)$ may take positive or negative values, the inequality involves its absolute value. Saturation of \Eq{eq: realisability cond} at a given $k$ corresponds to maximal helicity at that scale. If this holds for all $k$, the system is said to be maximally helical \citep{frisch1975}.\footnote{A single-mode helical field is also force-free, with $(\Bnabla \times \BB) \parallel \BB$ and vanishing Lorentz force. However, a superposition of helical modes with different wavenumbers generally breaks the force-free condition.}

For a fully helical field (e.g., $\chi_M(k) = 2\EM(k)/k$ for positive helicity), \citet{frisch1975} showed that neither magnetic energy nor helicity can cascade directly to smaller scales. Instead, nonlinear interactions of modes with wavenumbers $\vec{p}$ and $\vec{q}$ generate new modes with $\vec{k} = \vec{p} + \vec{q}$, constrained such that $|\vec{k}| \leq \max(|\vec{p}|,|\vec{q}|)$ \citep{brandenburg2005}. This restriction drives magnetic helicity and energy toward progressively larger length scales in an approximately self-similar fashion, giving rise to the inverse cascade.
The role of magnetic helicity in enabling the inverse cascade was first recognized by \citet{frisch1975} and later explored in the context of NSs, initially with box simulations of the Hall term \citep{brandenburg2020} and more recently with global NS models \citep{DehmanBrandenburg25}.

\subsection{The chiral magnetic effect}
\label{sec: CME}

Beyond macroscopic physics such as Ohmic dissipation, the Hall drift, or dynamos and turbulence during NS formation, some quantum effects can also play a significant role in the magnetic evolution, linking magnetic helicity to fermionic chirality---the handedness of particles (left vs right). One particularly interesting effect is the so-called chiral anomaly, which enables bidirectional transfer between chiral imbalance and magnetic helicity, facilitating the transfer of energy to larger-scale structures in a manner reminiscent of the inverse cascade driven by the non-linear terms in the induction equation \citep{boyarsky2012}. In this subsection, we briefly review the topic and some recent relevant results. 

A small imbalance in the chemical potentials between left- and right-handed electrons, denoted by $\mu_5\equiv \mu_R-\mu_L$\footnote{Note that definitions of $\mu_5$ vary in the literature, sometimes differing by a sign \citep{sigl2016} or a factor of 2 \citep{kaplan2017}, depending on the source.}, generates an electric current parallel to the magnetic field, an effect known as the Adler--Bell--Jackiw anomaly \citep{adler1969,bell1969}. When $\mu_5 \neq 0$, Maxwell's equations acquire an additional current term \citep{vilenkin1980}:
\begin{equation}
    \vec{j}_5 = \frac{\alpha \mu_5}{\pi \hslash } \BB, 
    \label{eq: J5}
\end{equation}
where $\alpha = e^2 / (\hslash  c)$ is the fine structure constant, $e$ is the fundamental charge, $\hslash $ is the reduced Planck constant, and $c$ is the speed of light. We use Gaussian units throughout the section. 

Thus, the modified induction equation accounting for Ohmic dissipation, Hall drift, and the new chiral magnetic contribution, can be written as follows:
\begin{eqnarray}
\frac{\partial \vec{B}}{\partial t} =  - \Bnabla \times \left[  \eta \left( \Bnabla \times \BB    
    + \omega_B^e \tau_e \left(\Bnabla \times \BB \right)\times\hat{b}
    - k_5  \BB \right) \right],
    \label{Eq:CME}
\end{eqnarray}
where $k_5 = 4 \alpha \mu_5/\hslash  c$ is the chiral wavenumber. 

The chiral term, which plays a similar mathematical role to turbulent dynamos, has led to studies of stellar core collapse or the proto-NS phase where large-scale fields are generated via chiral asymmetries \citep{masada2018,matsumoto2022}. A major obstacle in these scenarios is the existence of spin-flip processes, driven by the finite electron mass, which quickly suppress the chiral imbalance, reducing the efficiency of the Chiral Magnetic Effect (CME) and inhibiting the Chiral Magnetic Instability (CMI) \citep{grabowska2015,sigl2016,kamada2023}. This raises questions about the CME’s relevance in such short-lived environments. A recent study by \citet{DehmanPons25} found that in NSs, chiral asymmetries can persist for centuries after birth in the presence of small tangled magnetic structures, enabling a continuous transfer of energy from small to large scales, despite strong suppression by spin-flip processes.

The induction equation must be coupled to the evolution equation for the chiral number density $n_5 \equiv n_R - n_L$ \citep{adler1969,kamada2023,DehmanPons25}, which includes both source and sink terms:
\begin{equation}
    \frac{\partial n_5}{\partial t} =  \frac{2 \alpha }{\pi \hslash} \EE \cdot \BB + {n_e} \Gamma_w^\mathrm{eff} - n_5 \Gamma_f.
    \label{eq: n5}
\end{equation}
Here, the reaction rate $\Gamma_f$ accounts for spin-flip interactions, arising from electromagnetic interactions and the finite mass of electrons, and acts as a sink term \citep{grabowska2015,kaplan2017}, while $\Gamma_w^\mathrm{eff}$ represents the effective weak reaction rate \citep{epstein1981} and acts as a source term.
The $\EE \cdot \BB$ term governs the coupling between the chiral density and the electromagnetic field: twisting or untwisting magnetic field lines alters the net chirality in the system, acting as either a source or a sink depending on its sign.

\Eq{eq: n5} should be viewed alongside the time evolution of the magnetic helicity, which can be written in the form \citep{biskamp1997,boyarsky2012}:
 \begin{eqnarray}
\frac{\partial (\AAA \cdot \BB)}{\partial t} 
&=& - 2 c  \EE \cdot \BB -  c \Bnabla \cdot \left( \EE \times \AAA\right) .
 \label{eq: chirality and magnetic helcity}
\end{eqnarray}
The two equations, when combined and integrated over a volume, yield a generalized helicity balance law:
 \begin{equation}
    \frac{d}{dt} \left( Q_5 + \frac{\alpha }{\pi \hslash c} \chi_m  \right) + {\Gamma_5} = 0.
    \label{eq: modified helicity conservation}
\end{equation}
Here, $Q_5 = \int n_5 \, dV$ denotes the total axial charge, which quantifies the imbalance between left- and right-handed fermions in the system, $\chi_m$ is the total magnetic helicity (\ref{eq: A.B}), and ${\Gamma_5} = \int n_5 \Gamma_f dV $ is the total spin-flip rate. Note that the total helicity is not strictly conserved, due to the sink term $\Gamma_5$. 

Given that all reaction rates are much faster than typical astrophysical timescales, the system can be considered in a quasi-equilibrium state. In this limit, an explicit expression of $k_5$ in terms of the magnetic field can be derived: 
\begin{equation}
    k_5\left(\boldsymbol{x},t \right) =
    \frac{ \left(\Bnabla\times \BB \right) \cdot \BB  }
    {\left(\dfrac{2 \mu_e^2 }{m_e^2 c^4}\right) \dfrac{B^2_\mathrm{QED}}{3 \pi}
    + B^2  } \, ,
    \label{eq: k5}
\end{equation} 
and $B_\mathrm{QED} \equiv m_e^2 c^3 / (e \,\hslash) = 4.41 \times 10^{13}$\,G is the Schwinger quantum electrodynamic critical field. 

The chiral asymmetry can be maintained as long as $\BB \cdot (\Bnabla\times \BB) \neq 0$ (non-vanishing current helicity); see \Eq{eq: k5}. This allows a small but sustained asymmetry to persist, despite the action of spin-flip processes. In this regime, one can insert \Eq{eq: k5} into \Eq{Eq:CME}, explicitly factoring out the chiral anomaly from the induction equation. In simpler terms, the CME facilitates energy transfer across scales by introducing a new nonlinear term in the magnetic field evolution equation, linked to the current component parallel to the magnetic field $\BB$.
We note that this non-linearity is mathematically different from the Hall term, that appears as a quadratic ($B^2$) term, driven by the current component perpendicular to the field.
In \citet{DehmanPons25} the saturation of the field amplification due to the CME is estimated to be:
\begin{equation}
B_\mathrm{sat} \approx \sqrt{\frac{2}{3 \pi}} \frac{\mu_e }{m_e c^2} B_\mathrm{QED}.
    \label{eq: Bmax}
\end{equation}
Under typical NS conditions, this ranges from 10 $B_\mathrm{QED}$ in the outer crust to 200 $B_\mathrm{QED}$ in the inner crust, yielding $B_\mathrm{sat} \sim 10^{14}$\,G near the surface and up to $\sim 5 \times 10^{15}$\,G in deeper layers, consistent with typical magnetar field strengths. 


\subsection{Elasticity, crustal failures and plastic flow.}

The main idea for the Hall--MHD description of the crust is that ions are locked in the crustal lattice and only electrons are mobile. However, molecular dynamics simulations \citep{horowitz09} indicate that the matter behaves elastically up to a certain threshold stress. Beyond this point, the magnetic stresses exceed the capacity of the elastic response, leading to mechanical failure. This failure can occur through brittle fracture, where the material breaks abruptly with little prior deformation, or through plastic flow, in which the material undergoes irreversible deformation without fracturing. The dominant failure mode depends on factors such as temperature, composition, and the local stress environment.

In the first detailed simulations of this process, crustal failures were treated in the most simplified manner as {\it star-quakes}. By evaluating the accumulated stress, \citet{pons11,perna11} simulated the frequency and energetics of the internal magnetic rearrangements, which were proposed to be at the origin of magnetar outbursts. 
This model mimics earthquakes since, under terrestrial conditions, the low densities of the material allow for the propagation of sudden fractures. The Earth mantle, in this respect, can be thought of as brittle.
However, materials subject to very slow shearing forces could behave differently and enter a slow plastic flow regime instead.
Although the dynamics of failure may differ, the energetic considerations underlying the release of energy due to accumulated magnetic stresses remain broadly similar. Failure or yielding is expected to occur when components of the elastic strain tensor, $\sigma_{ij}$, approach the critical strain threshold, $\sigma_{\mathrm{max}}$, typically in the range of 0.001 to 0.1
\citep{horowitz09}. 
Assuming that the system evolves slowly and remains close to quasi-equilibrium, the elastic strain can be related to the magnetic stress, ${\cal M}_{ij}$, and the shear modulus, $\mu$, via the relation $\mu \sigma_{ij} = \Delta {\cal M}_{ij}$ (but see discussion at the end of the subsection), where $\Delta$ denotes the change in magnetic stress between the current configuration and the previous equilibrium configuration, established at the time of the previous failure.
Making use of this simple algebraic relation, \citet{lander15} apply the von Mises criterion to establish that the crust would yield when
\begin{equation}
\sqrt{\sigma_{ij} \sigma^{ij}} \geq \sigma_{\mathrm max}~.
\end{equation} 
As anticipated, this happens at magnetic field strengths around $10^{15}$ G, 
consistent with magnetar estimates, though the outer crust is weaker than the inner crust and may yield at lower strengths.
Notably, the prior estimate aligns with the field strength needed to trigger plastic flow \citep{lander16}, depending on crustal depth and the viscosity of the plastic phase, indicating that both scenarios arise under comparable conditions.

Using a plane-parallel model, \citet{lander19} investigated the features of such plastic flow under the assumption of Stokes flow, where a viscous term balances magnetic and elastic stresses. They study the crustal response under Ohmic and Hall evolution and find that there can be significant plastic-like motions in the external layers of the star.
Similar arguments have also been proposed to account for the deposition of heat by the visco-plastic flow and the propagation of thermo-plastic waves \citep{beloborodov14}. In a later work, \citet{gourgouliatos21} conducted global axisymmetric simulations to investigate various failure mechanisms impacting a specific crustal region. 
They find that flow does not merely dampen the Hall effect, even when plastic viscosity is low; instead, it drives complex evolutionary patterns, sometimes amplifying the Hall effect’s influence. They concluded that plastic flow influences both the characteristics of magnetar bursts and their spin-down behavior. 

However, as pointed out in recent works \citep{kojima24,bransgrove25}, in magneto-elastic equilibrium, it is the divergence of stress (i.e., the force) that is balanced, not the stress itself. Kojima's methodology for calculating the elastic tensor is needed to ensure the model accuracy. In numerical simulations of magnetic field evolution, $\sigma_{ij}$ must be tracked by solving additional differential equations, at the cost of a significant increase in computation time compared to previous simple algebraic relations. \citet{bransgrove25} indicate that, in configurations close to magneto-elastic equilibrium, the elastic stress is typically several orders of magnitude smaller than the local Maxwell stress. 
Consequently, criteria based only on stress balance may significantly overestimate the frequency of crustal failures, and previous conclusions should be approached with caution. 
Nevertheless, the results in \citet{bransgrove25} also suggested that Hall waves, initiated following the superconducting transition in the core, may be sufficiently intense to fracture the crust, potentially leading to starquakes that induce rotational glitches and alterations in the radio pulse profile. Their simulations incorporated the time evolution of the elastic deformation of the lattice following previous work \citep{bransgrove18}, which enables them to monitor time-dependent shear stresses throughout the crust. Further investigations along these lines are worthwhile for future research.

\subsection{Ambipolar diffusion in neutron star cores}

The number of works concerning mechanisms operating in NS cores is sensibly smaller, and most contain far less detail than the studies of the crust. Ambipolar diffusion involves the coordinated movement of magnetic field lines and charged particles (protons and electrons) relative to neutrals (neutrons). There are several astrophysical scenarios, involving environments with partially ionized plasma like the Solar chromosphere, protostellar disks, and parts of the interstellar medium, where the ambipolar diffusion is regarded as a dominant mechanism. The seminal works by 
\citet{goldreich92} and \citet{1995MNRAS.273..643S}, already proposed ambipolar diffusion as a viable mechanism for the
dissipation of magnetic energy in regions of NSs where the charged particle fluid is chemically homogeneous.
Owing to its cubic dependence on $B$, ambipolar diffusion could be the dominant process driving the evolution of magnetars during the first $10^3-10^5$ yr, although there is some controversy.
In particular, we refer the reader interested in the role of chemical potential gradients, which is out of the scope of this review, to the literature, for instance the original arguments in \citet{goldreich92} or the discussion in \citet{passamonti17b}. However, \citet{gusakov17} questioned the validity of the approach followed by previous works
in stratified matter, and obtained a different equation from the momentum equation (implicitly assuming magnetostatic equilibrium), in which the small deviations of the chemical potentials from their equilibrium values do not depend on temperature and are determined by the Lorentz force. With the same methodology, \citet{ofengeim18} estimated the
instantaneous particle velocities and other parameters of interest, determined by specifying the magnetic field configuration, and found that the evolution timescales could be shorter than expected.

A simple way to incorporate ambipolar diffusion is to introduce the ``ambipolar velocity'' $\vec{\mathrm v}_a$ and consider an advective term $\propto (-\vec{\mathrm v}_a \times \vec{B})$ in the electric field. As discussed in \citet{goldreich92, passamonti17b},
the simplest case is realized in the regime where the system attains $\beta-$equilibrium faster than it evolves, and the ambipolar velocity is proportional to the Lorentz force 
\begin{equation}
\label{ambipolar1}
{\vec{\mathrm v}_a}  \propto (  \vec{j}  \times \vec{B} )~.
\end{equation}
We refer to
the original work by \citet{goldreich92} for a detailed description of the origin of the proportionality coefficients in terms of microphysical relaxation times. In general, the velocity field can be decomposed into solenoidal and irrotational components. The solenoidal component preserves chemical equilibrium among neutrons, protons, and electrons, encountering resistance only from friction with neutrons. Conversely, the irrotational component is hindered by pressure gradients that arise due to deviations from chemical equilibrium it induces. At low temperatures, weak interactions that restore chemical equilibrium are slow, causing these pressure gradients to significantly impede the irrotational modes.

We note that, assuming Eq.~(\ref{ambipolar1}), the ambipolar contribution to the electric field can also be written as:
\begin{equation}\label{Eq:ambipolar}
  - {\vec{\mathrm v}_a} \times \vec{B} \propto  [B^2 \vec{j} 
  - (\vec{j}\cdot\vec{B})\vec{B}] =  B^2\vec{j}_\perp\, ,
\end{equation} 
which explicitly manifests as an enhanced resistive-like term, featuring a dissipative term, proportional to $B^2$, that acts exclusively on currents perpendicular to the magnetic field ($\vec{j}_\perp$). 
Consequently, this term does not fully dissipate the magnetic field but instead aligns it with the sustaining electrical currents, driving the system towards a force-free configuration, i.e. $\vec{j} \times \vec{B}=0$.
Notably, the impact of this term is highly sensitive to the magnetic geometry, in addition to its strength, and it does not affect currents flowing parallel to the magnetic field lines.

Most previous studies of ambipolar diffusion have focused on estimating timescales, with only a few exceptions that include simulations. These simulations have primarily been limited to simplified one-dimensional models \citep{2008A&A...487..789H,2010MNRAS.408.1730H,Tsuruta23}.
However, given the important considerations outlined above regarding the role of geometry, simplified one-dimensional results should be interpreted with caution, as they require confirmation through multidimensional studies.
Two-dimensional \citep{castillo17,passamonti17b,bransgrove18,Skiathas24,vigano21}, 
and the first three-dimensional simulations \citep{Igoshev23a}, have become possible only very recently, although typically assuming constant coefficients and with some simplifying assumptions. Despite current numerical limitations, they already point to intriguing possibilities. A more consistent strategy based on a multifluid approach has recently gained attention, yielding promising results \citep{castillo2025,moraga2025}. Although still restricted to axisymmetric models, these initial findings support earlier estimates, indicating an enhancement in dissipation rates. At constant temperature, they recover the expected outcome: neutrons reach diffusive equilibrium, the Lorentz force is balanced by the chemical potential gradients of the charged particles, and the magnetic field configuration is governed by a non-linear Grad–Shafranov equation.

In addition, in a realistic scenario, there is a further complication that usually is omitted. The NS core cools very fast (less than a year) below the critical temperatures for neutron superfluidity and proton superconductivity, which has important implications, sometimes controversial. \citet{goldreich92} argued that ambipolar diffusion would still be a significant process, but \citet{glampedakis11b} studied in detail the ambipolar diffusion in superfluid and superconducting stars and concluded that its role on the magnetic field evolution would be negligible. Other works \citep{Graber2015} also showed that ambipolar diffusion with superconducting protons is very slow. Moreover, \citet{kantor18} argued that, in finite-temperature superfluid NS matter, magnetic field dissipates exclusively due to Ohmic losses and non-equilibrium beta-processes, limiting the effects to the case where muons are present.
 
In summary, the role of ambipolar diffusion remains an active area of research, with recent progress marked by the development of new multi-dimensional simulations. These advances may offer valuable insights into the mechanisms behind the high luminosity observed in magnetars.

\subsection{Mathematical structure of the generalized induction equation}\label{sec:induction_eq_char}

In order to understand the dynamical evolution of the system and to design a successful numerical algorithm, 
it is important to identify the mathematical character of the equations and the wave modes. The magnitude of $\omega_B^e \tau_e$  defines the transition from a purely parabolic equation ($\omega_B^e \tau_e \ll 1$) to a hyperbolic regime ($\omega_B^e \tau_e \gg 1$). The Hall term introduces two wave modes into the system. \citet{huba03} has shown that,
in a constant density medium, the only modes of the Hall--MHD equation are the {\it whistler or helicon waves}. They are
transverse field perturbations propagating along the field lines. In presence of a charge density gradient, 
additional {\it Hall drift waves} appear. These are transverse modes that propagate in the $\vec{B} \times \vec{\nabla} n_e$ direction. We also note that the presence of charge density gradients results in a Burgers-like
term \citep{Vai2000}. Furthermore, even in the constant density case but without planar symmetry, 
the evolution of the toroidal component also contains a quadratic term that resembles the Burgers equation \citep{PonsGeppert2007} with a coefficient dependent on the distance to the axis. This term leads to the formation of discontinuous solutions (current sheets) that require proper treatment. It is fundamental for a numerical Hall--MHD code to reproduce these modes and features, which are easily testable.

Consider the generalized induction equation (\ref{Eq:CME}), extended with an ambipolar
diffusion term of the form of Eq.~(\ref{Eq:ambipolar}). One can factor out the magnetic diffusivity $\eta$ of all terms to
obtain the following equation
\begin{eqnarray}\label{ind_eq_final}
\frac{\partial \vec{B}}{\partial t} =  - \Bnabla \times \left[\eta \left( \Bnabla \times \BB -  k_5 \vec{B}   
+  f_H \left(\Bnabla \times \BB \right)\times \hat{b} 
- f_a \left( (\Bnabla \times \vec{B}) \times \hat{b} \right) \times  \hat{b}\right) \right].  
\end{eqnarray}
Here, we remind that $\hat{b}$ is the unit vector along the magnetic field direction and we have introduced the notation
$f_H = \omega_B^e \tau_e$. The ambipolar coefficient ($f_a$) in the simplest case where proton collisions are dominated by proton-neutron collisions with a relaxation time $\tau_{p}$, can be written as $f_a = (\omega_B^e \tau_e)(\omega_B^p \tau_p)$, with $\omega_B^p$ being the proton gyrofrequency. Note that, since the gyrofrequencies scale with $B$, $f_H\propto B$ and $f_a\propto B^2$. Generally speaking, $\vec{B}$, $(\vec{\nabla} \times \vec{B})$, and $(\vec{\nabla} \times \vec{B} ) \times \vec{B}$ form a complete basis, provided that $\vec{B}$ and $ \vec{\nabla} \times \vec{B}$ are not parallel, so that the components considered in the generalized relation above can formally absorb any further contribution to the electric field (with coefficients having different physical meaning).

By assuming a generic, small perturbation $\vec{\delta B}$ over a fixed constant background field $\vec{B}_o$:
\begin{equation}
\vec{B} = \vec{B}_o + \vec{\delta B} \, e^{i(\vec{k}\cdot \vec{x} - \omega t)}\, ,
\end{equation}
where $\vec{k}$ is the wavenumber of the perturbation and $\omega$ its angular frequency, and considering the high-frequency limit, for which the gradients of the pre-factors $\eta,k_5,f_H,f_a$ are negligible, the linearized Eq.~(\ref{ind_eq_final}) reads:
\begin{eqnarray}
\omega \vec{\delta B} &=&  - i k^2 \eta \vec{\delta B} 
- \eta k_5 (\vec{k} \times \vec{\delta B})
+ i  \eta f_H (\vec{k} \cdot \hat{b}_0) (\vec{k} \times \vec{\delta B})
\nonumber \\
&-& i \eta f_a \left[ (\vec{k} \cdot \hat{b}_0)^2 \vec{\delta B}
       + (\vec{\delta B} \cdot \hat{b}_0) \left( k^2 \hat{b}_0  - (\vec{k} \cdot \hat{b}_0) \vec{k} \right) \right]~.
\end{eqnarray}
Introducing the notation $b_{\parallel} = \hat{k}\cdot\hat{b}_o$, and $\vec{b_{\perp}} = \hat{b}_o - {b}_{\parallel} \hat{k}$, 
with $\hat{k} = \vec{k}/k$, we can write
\begin{eqnarray}
\omega \vec{\delta B} &=&  \eta \left( - i k^2 \vec{\delta B} 
-(k_5 - i k f_H b_{\parallel}) (\vec{k} \times \vec{\delta B})
- i f_a k^2 \left[ b_{\parallel}^2 \vec{\delta B}
       + (\vec{\delta B} \cdot \hat{b}_0) \vec{b_{\perp}} 
       \right]~\right)
\end{eqnarray}
and, following the standard procedure (see \citealt{vigano19}, who did not include the CME term), the dispersion relation is given by
\begin{equation}
{\omega^{\pm}} = - i k^2 \eta \left( 1 + f_a b_{\parallel}^2 
+ \frac{1}{2} f_a b_{\perp}^2 \right)  
\pm \frac{i k}{2} \eta \sqrt{ k^2 \left( {f_a b_{\perp}^2}\right)^2  
- 4 \left(f_H k {b_{\parallel}} + i k_5\right)^{2}}
~.
\end{equation}
It is also useful to examine the modes that propagate only along the direction of the magnetic field $\vec{B_0}$ (with no magnetic component perpendicular to $k$, that is $b_{\perp}=0$) and which follow the dispersion relation:
\begin{eqnarray}
{\omega^{\pm}} = - i k^2 \eta (1 + f_a)
\mp \eta \left( f_H k^2 + i k k_5\right)~.
\end{eqnarray}
This relation explicitly confirms that the Hall term is the only contribution to the real part of the frequency, and the only one that could be associated with waves, although the $k^2$ dependence shows the dispersive character of the Hall whistler waves (Hall drift waves can appear if the high-frequency assumption is relaxed). On the contrary, both Ohmic and ambipolar terms are intrinsically dissipative. 

Separating real and imaginary parts explicitly, one has (still considering the simplest case $b_\perp=0$, but the following qualitative considerations hold for the general case)
\begin{equation}\label{eigenvalues}
{\omega^{\pm}} = - i k^2 \eta (1 + f_a \mp k_5/k) \mp \eta \left( f_H k^2\right)~.
\end{equation}
In this form, we identify the ambipolar diffusion as a purely dissipative term, with a field-dependent diffusivity through $f_a\propto B_0^2$, but the CME term has opposite signs in each of the two modes, indicating that 
short-wavelength ($k<k_5/(1+f_a)$) unstable modes are possible, if $k_5 >0$ (see Eq.~\ref{eq: k5}). Indeed, the mathematical form of the CME term, $\propto k_5\vec{B}$, is the same as the so-called $\alpha$-term in dynamo theory, although the CME pre-coefficient has a different physical origin and can have either sign.

\section{Magnetic field evolution in the interior of neutron stars: numerical methods}\label{sec:magnetic_methods}

In this section, we explore some of the key aspects of numerical methods. One of the first crucial decisions is the choice of formalism. There are two main approaches:
(i) working directly with the magnetic field components, which avoids additional mathematical transformations but requires careful handling of the divergence-free condition; and 
(ii) using the solenoidal constraint to reduce the problem to two scalar functions that represent the true degrees of freedom
(the so-called poloidal-toroidal decomposition, see Appendix~\ref{app:formalism}).
In the context of NS evolution, finite-difference schemes have been developed for both approaches, whereas pseudo-spectral methods are more commonly based on the poloidal-toroidal decomposition.

Within spectral methods, another important choice is whether to apply a spectral decomposition in the radial direction (typically using Chebyshev polynomials) or to adopt a hybrid approach, combining spectral methods in angular directions with finite-difference schemes radially, which can more effectively resolve the steep density gradients in the NS crust.
Examples of a fully spectral approach are the first 2D simulations of the evolution of the crustal magnetic field assumed a constant density shell \citep{HR2002} and were later extended to include density gradients \citep{HR2004}. They used an adapted version of the spherical harmonic code described in \citet{Hollerbach2000}, including modes up to $\ell =100$, and 25 Chebyshev polynomials in the radial direction, but they were restricted to $\omega_B \tau_e<200$ by numerical issues.

\citet{PonsGeppert2007,Pons2009} used a hybrid code (spectral in angles but finite-differences in the radial direction) to perform 2D simulations in realistic profiles of NSs over relevant timescales (typically, Myr). This approach allowed us to reach higher values of the magnetization parameter ($\omega_B \tau_e \approx 10^3$), and to study the Hall instability \citep{Pons2010}. Similarly, a number of recent 3D simulations \citep{gourgouliatos18,gourgouliatos19,Degrandis20,Degrandis22,Igoshev21,Igoshev23b}, use a modified version of the PARODY code \citep{dormy98,aubert08}. An earlier version of this code, excluding thermomagnetic coupling, was already 
used in \citet{wood15} in the context of NSs. The code also employs a pseudo-spectral approach, with a radial grid and spherical harmonic expansions for the angular components. Time-stepping uses the Crank-Nicholson method for ohmic diffusion, backward Euler for the isotropic thermal diffusion, and Adams-Bashforth for all additional terms.

\citet{wood15,gourgouliatos16} were limited to magnetization parameters of the order of $\simeq 100$. The main problem arises from the presence of non-linear Burgers-like terms, which naturally lead to discontinuities \citep{Vai2000,vigano12a}, which are notoriously problematic for spectral codes. For this reason, subsequent works aiming at extending the simulations to more general cases have been gradually shifting towards the use of finite-difference schemes. An example of a finite difference approach, while still employing poloidal-toroidal decomposition, is seen in the axially symmetric simulations
\citep{gourgouliatos14a,gourgouliatos14b,gourgouliatos15a}. A more elaborated scheme using finite volume methods applied directly to the induction equation in terms of field components was presented in \citet{vigano12a} and used for different 2D applications \citep{vigano12b,vigano13,Dehman20}. They utilized Stokes' theorem and the conservative form of the equations to apply techniques from high-resolution shock-capturing schemes, enabling handling of higher magnetization parameters. Extending these methods to 3D with a new code (\MATINS) has only recently become feasible \citep{MATINS1,MATINS_MT,MATINS2,DehmanPons25}. 

It is worth noting that, despite differences in methods and coordinate choices, all existing codes handle the radial direction
differently from the angular directions due to the stronger gradients along the radial axis. 
Alternatively, Cartesian grids can be used, but they present two main challenges: first, resolution cannot be enhanced solely in the radial direction, leading to a significant increase in computational cost compared to spherical coordinate-based codes (scaling as $\propto N^3$ instead of $N$, where $N$ is the number of radial points or dual basis components in spectral methods). Second, Cartesian discretization introduces numerical noise and spurious modes due to imperfect mapping of spherical boundaries. While these difficulties have recently been addressed in various star-in-a-box simulations in other contexts, the only existing preliminary application to neutron star evolution is presented in \citet{vigano19}, to which we refer for further details.

We continue in this section with a brief overview of spectral methods, before turning to some key aspects of finite-difference schemes.

\subsection{Spectral methods with the toroidal-poloidal decomposition}\label{methods_spectral}

Using the notation of \citet{geppert91}, the basic idea is to expand the poloidal ($\Phi$) and toroidal ($\Psi$) scalar functions (see Appendix ~\ref{app:formalism}) in a series of spherical harmonics\footnote{We note that other authors use different notations, for example \citet{gourgouliatos14a} denote by $\Psi$ and $I$ the poloidal and toroidal functions, respectively. See Appendix \ref{app:formalism} for more details.}

\bear
\Phi = \frac{1}{r} \sum_{\ell,m} \Phi_{\ell m}(r,t) Y_{\ell m}(\theta,\varphi)\, , 
\nonumber \\ 
\Psi = \frac{1}{r} \sum_{\ell,m} \Psi_{\ell m}(r,t) Y_{\ell m}(\theta,\varphi)\, ,
\label{expans} 
\ear
where $\ell=1,\ldots,\ell_{\rm max}$ is the degree and $m=-\ell,\ldots,+\ell$ is the order of the harmonics.

Assuming a radial dependent diffusivity, $\eta=\eta(r)$, it can be shown that the Ohmic term for each multipole effectively decouples, and the set of coupled evolution equations for the radial parts ($\Phi_{\ell m}$ and $\Psi_{\ell m}$) can be readily obtained. Omitting relativistic factors (see \citealt{Pons2009} for the relativistic expressions in the Schwarzschild metric) we have:
 %

\bear
\frac{\partial \Phi_{\ell m}(r)}{\partial t} &=& \eta (r)  \left[
\frac{\partial^2 \Phi_{\ell m}}{\partial r^2} 
- \frac{\ell (\ell+1)}{r^2}\Phi_{\ell m} \right] + D_{\ell m} 
+ k_5 \, \eta (r) \Psi_{\ell m}
\label{Phi_diff}
\\
\frac{\partial \Psi_{\ell m}(r)}{\partial t}&=&
\frac{\partial}{\partial r} 
\left( \eta(r) \frac{\partial  \Psi_{\ell m}}{\partial r}\right)
- \eta(r) \frac{\ell (\ell+1)}{r^2} \Psi_{\ell m} + C_{\ell m}
\nonumber \\
&& -  k_5\,\eta (r)  \left[\frac{\partial^2 \Phi_{\ell m}}{\partial r^2} - \frac{\ell (\ell +1)}{r^2}\Phi_{\ell m} \right].  \nonumber\\
\label{Psi_diff}
\ear
Terms proportional to $k_5$ encode the chiral magnetic effect (Appendix A of \citealt{DehmanPons25}). For the quasi-analytical illustration we take $k_5$ constant; in general $k_5=k_5(\vec{r},t)$ is a spatially and temporally varying pseudo-scalar field. We use $D_{\ell m}$ and $C_{\ell m}$ as a shorthand for the nonlinear Hall terms (the full expressions can also be found in \citealt{geppert91}). These include sums over running indices and coupling constants related to Clebsch--Gordan coefficients (the sum rules to combine angular momentum operators are used to determine which multipoles are coupled to each other). All these coefficients can be evaluated once at the beginning of the evolution and stored in a memory-saving form since only specific combinations of indices are non-zero.

In the general case, however, the magnetic diffusivity also depends on the angular coordinates, for example through the temperature dependence of $\eta$ when the temperature is non-uniform. In this case we can also expand the magnetic diffusivity in spherical harmonics 
\begin{equation}
\eta = \sum_{\ell,m} \eta_{\ell m}(r,t) Y_{\ell m}(\theta,\varphi)\, , 
\end{equation}
where the sum must include the monopole term, $\ell =0,\ldots,\ell_{\rm max}$.
These new terms couple different multipoles of the same component (poloidal or toroidal). 
The inclusion of additional terms in the electric field (e.g. ambipolar diffusion) would introduce even more complicated non-linear couplings (the theory has not yet been developed). In general, we end up with a system of the order of $\approx 2 \ell_{\rm max}^2$, strongly coupled, differential equations. Partly for this reason, recent 3D studies have favored the use of simpler finite-difference schemes, which facilitate the incorporation of additional terms.

\subsection{Finite-difference and finite-volume schemes}

To capture the magnetar scenario in detail, numerical codes need to tackle a substantially more challenging regime.
In \citet{vigano12a}, a novel approach making use of the well-know High-Resolution Shock-Capturing (HRSC) techniques \citep{toro97}, designed to handle shocks in hydrodynamics and MHD, was proposed. These techniques have been successfully applied to a range of problems, from a simple 1D Burgers equation to complex ideal MHD problems \citep{anton06,giacomazzo07,cerdaduran08}, 
avoiding the appearance of spurious oscillations near discontinuities. We refer to \citet{marti2015} for a general review on grid-based methods and to \citet{balsara17} for a review on finite-volume methods, applied to other astrophysical scenarios.
Let us review some of the main characteristics of these methods, of particular interest in our problem.

\subsubsection{Conservation form and staggered grids}

In hydrodynamics and MHD, the system of partial differential equations (PDEs) involves the divergence operator acting on vector or tensor fields. Thus, Gauss' theorem is usually employed in the design of the algorithms, exploiting the formulation of the equations in conservation form. Analogously, for problems involving the induction equation, the presence of the curl operator makes it natural to apply Stokes' theorem to the equation.
Considering a numerical cell and its surface $\Sigma_\alpha$ normal to the $\alpha$ direction, delimited by the curve $C_\Sigma$, we have a discretized version of Eq.~(\ref{induction0}):
\begin{eqnarray}
\frac{1}{c}\frac{\partial }{\partial t}\left[\int_{\Sigma_\alpha} B_\alpha d\Sigma_\alpha\right]  =  - \oint_{C_\Sigma} \vec{E} \cdot d\vec{l} ~.
\end{eqnarray}
The space-discretized evolution equation for the average of the magnetic field component normal to the surface over the cell surface  is then
\begin{eqnarray}\label{phialpha}
\frac{\partial \overline{B}_\alpha}{\partial t}  =  - \frac{c \sum_k E_k l_k}{\Sigma_\alpha} ~. \label{eq:induction_discretized}
\end{eqnarray}
Here, the circulation of the electric field is approximated by the sum $\sum_k E_k l_k$, where $E_k$ is the average value of the electric field over each cell of length $l_k$, and $k$ identifies each of the four edges of the face. 
For clarity, in this section, we omit relativistic metric factors that must be consistently 
incorporated in the definitions of lengths, areas, and volumes.

\begin{figure}[ht]
 \centering
\includegraphics[width=.5\textwidth]{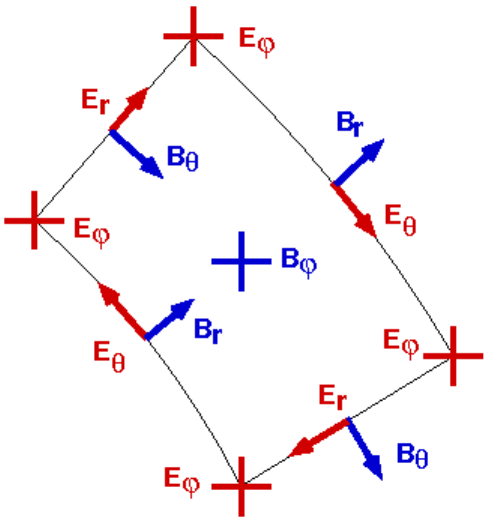}
\caption{Location of the variables on a staggered grid in spherical coordinates for the axisymmetric case.
Solid lines delimit the edges of the surface $\Sigma_\varphi$.} 
 \label{fig_staggered}
\end{figure}

The problem is then reduced to designing an accurate and stable discretization method to calculate the $E_k$ components at each edge. A natural choice is to use staggered grids, for which in each numerical cell the locations of the different field components are conveniently displaced, instead of being all located at the same position (typically, the center), 
as in standard centered schemes. In our case, we allocate the normal magnetic field components at each face center and electric field components along cell edges. 
Fig.~\ref{fig_staggered} shows an example of the location of the variables in a numerical cell in spherical coordinates $(r, \theta, \varphi)$, considering  axial symmetry (in the general 3D case, there would be a displacement of $B_\varphi, E_\theta, E_r$ in the direction orthogonal to the plane of the figure). 

Making use of Gauss' theorem, the numerical divergence can be evaluated, for each cell with volume $\Delta V$, as follows: 
\begin{equation}\label{divb_staggered}
  \vec{\nabla}\cdot\vec{B}=\frac{1}{\Delta V}\sum_\alpha{\overline{B}_\alpha \Sigma_\alpha}~.
\end{equation}
With this definition, the divergence-preserving character of the methods using the conservation form to advance $\overline{B}_\alpha$ in time becomes evident: taking the time derivative of Eq.~(\ref{divb_staggered}), and using Eq.~(\ref{phialpha}), every edge contributes twice (one per each face) with opposite signs, so that all discrete contributions to the divergence time evolution exactly cancel out. Thus, by construction, the divergence condition is preserved to machine error for any divergence-free initial data. Examples of applications of such methods can be found, among many others, in \citet{toth00,vigano12a,balsara15}.

\subsubsection{Divergence cleaning methods in finite-difference schemes}

An algorithm built on a staggered grid can be designed to preserve the divergence constraint by construction, 
but the different allocation of variables makes its implementation relatively complex, particularly in 3D problems and with the inclusion of quadratic and cubic terms in the electric field. Among alternative formulations that have recently gained popularity, and can also handle many MHD-like problems, a relatively simple option is the family of {\it divergence-cleaning} schemes built on standard grids (all components of the fields are defined and evolved at every grid node).
A popular divergence-cleaning method \citep{dedner02}, extensively used in MHD, consists in the extension of the system of equations as follows: 
\begin{eqnarray}
  &&  \frac{1}{c} \frac{\partial \vec{B}}{\partial t} +  \vec{\nabla}\times \vec{E} + \vec{\nabla}\chi = 0 \, ,\nonumber \\
  && \frac{\partial \chi}{\partial t} + c_h^2 \vec{\nabla}\cdot\vec{B} = -\gamma \chi \, ,
\end{eqnarray}
where $\chi$ is a scalar field that allows the propagation and damping of divergence errors, and $c_h$ and $\gamma$ are two parameters to be tuned: $c_h$ is the propagation speed of the constraint-violating modes, which decay exponentially on a timescale $1/\gamma$. In principle, a large value of $\gamma$ will damp and reduce divergence errors very quickly, but in practice the optimal cleaning is reached for {$c_h \approx \gamma \sim {\cal O}(1)$ because, if $\gamma$ is too large, the source term becomes stiff and more difficult to handle with explicit numerical schemes.

\subsubsection{Evaluation of the current and the electric field}

\begin{figure}
 \centering
 \includegraphics[width=.45\textwidth]{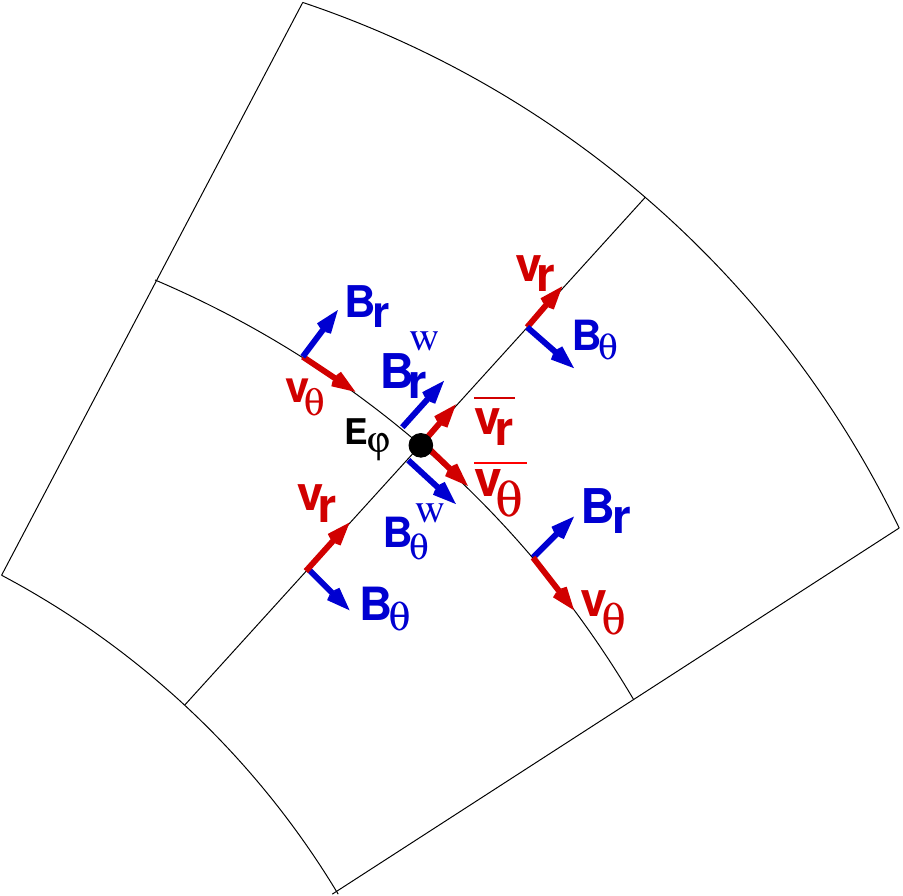}
\caption{Illustration of the procedure to calculate the electric field in a staggered grid: location of the 
components of velocity (red arrows) and magnetic field (blue) involved in the definition of contribution to  $E_\varphi$ (black dot) from the Hall term.}
 \label{fig_ewind}
\end{figure}

As a practical example, let us consider an electric field of the form:
\begin{eqnarray}\label{efield_v}
\vec{E} = \frac{\vec{j}}{\sigma} 
- \frac{\vec{\mathrm v}}{c} \times \vec{B} , 
\end{eqnarray}
where nonlinear (Hall and/or ambipolar) dependencies on the magnetic field are implicitly contained in the expression of $\vec{\mathrm v}$. The current density $\vec{j}$ is 
\begin{equation}
\vec{j} = \frac{c}{4\pi}\Bnabla \times \BB - \vec{j}_5 , 
\end{equation}
where $\vec{j}_5$ denotes the chiral current.

By considering the allocation of the components in the staggered grid (Fig.~\ref{fig_staggered}), the components of the current density can be naturally defined along the edges of the cells, in the same positions as the electric field components, exploiting the discretized version of the Stokes' theorem applied to $\vec{j}\propto  \vec{\nabla}\times \vec{B} $. Therefore, the ohmic term in the electric field can be directly evaluated, but the other terms involving vector products require special care since they involve products of field components that are not defined at the same place as the desired electric field component. The simplest option is a direct interpolation of both $\vec{\mathrm v}$ and $\vec{B}$ using the first neighbors, but this often results in numerical instabilities.

In the spirit of HRSC methods, we can instead think of the interpolated value of $\vec{\mathrm v}$ as the advective velocity acting at that point (although it depends on $\vec{B}$ itself), and consistently take the {\it upwind} components $\overline{B}_\alpha^w$ of the magnetic field at each interface. For example, in the axisymmetric case and considering the evolution of the poloidal components ($B_r, B_\theta$), the contributions of $E_r$ and $E_\theta$ to the circulation cancel out and we only need to evaluate the contribution of $E_\varphi$, which is given by
\begin{eqnarray}\label{efield_wind}
{E_\varphi} =\frac{1}{\sigma}  {J_\varphi}  - \frac{1}{c} \left( \overline{\mathrm v}_r {\overline{B}_\theta}^w -  \overline{\mathrm v}_\theta {\overline{B}_r}^w \right) 
\end{eqnarray}
In Fig.~\ref{fig_ewind} we explicitly show the location of $E_\varphi$ (black point) and the location on the staggered grid of the quantities needed for its evaluation. First, $\overline{\mathrm v}_r$ and $\overline{\mathrm v}_\theta$ are calculated by taking the average of the two closest neighbors; in the example, they point outward and to the right, respectively. Second, one considers the upwind values of  $\overline{B}_r^w$ and $\overline{B}_\theta^w$; in the example, they are taken from the bottom and left sides.

\subsubsection{Cell reconstruction and high-order accuracy}

The original upwind (Godunov's) method is well known for its ability to capture discontinuous solutions, but it is only first-order accurate: the variables are assumed to be constant on each cell. This method can be easily extended to give second-order spatial accuracy on smooth solutions, but still avoiding non-physical oscillations near discontinuities, by using
a reconstruction procedure that improves the piecewise constant approximation. 

A very popular choice for the slopes of the linear reconstructed function is the {\it monotonized central-difference limiter}, proposed by \citet{vanleer77}. 
Given three consecutive points $x_{i-1},x_i,x_{i+1}$ on a numerical grid, and the numerical values of the function $f_{i-1},f_i,f_{i+1}$, the reconstructed function within the cell $i$ is given by $f(x)=f(x_i)+ \alpha (x-x_i)$, where the slope is
$$\alpha = {\rm minmod}\left( \frac{f_{i+1}-f_{i-1}}{x_{i+1}-x_{i-1}},2\frac{f_{i+1}-f_{i}}{x_{i+1}-x_{i}},
2\frac{f_{i}-f_{i-1}}{x_{i}-x_{i-1}}\right).$$
The ${\rm minmod}$ function of three arguments is defined by
\begin{equation}
{\rm minmod}(a,b,c) = \left\{ 
\begin{array}{cc}
{\rm min}(a,b,c) & {\rm if} ~a,b,c>0 ;\\
{\rm max}(a,b,c) & {\rm if} ~a,b,c<0 ;\\
0 & {\rm otherwise} .
\end{array}
\right.\nonumber
\end{equation}
Other popular higher order reconstruction, are PPM \citep{colella84}, PHM \citep{PHM}, MP5 \citep{suresh97}, the FDOC families \citep{bona09}, or the Weighted-Essentially-Non-Oscillatory (WENO) reconstructions \citep{jiang96,shu98,yamaleev09,balsara17}.
In \citet{vigano19} they presented and thoroughly tested a two-step method consisting of the reconstruction with WENO methods of a combination of fluxes and fields at each node, known as flux-splitting \citep{shu98}.
This reconstruction scheme does not require the characteristic decomposition of the system of equations (i.e., the full spectrum of characteristic velocities) and, at the lowest order of reconstruction, their flux formula reduces to the popular and robust Local-Lax--Friedrichs flux \citep{toro97}.

\subsection{Axis Singularities in 3D and Cubed-Sphere Grid}

\begin{figure}
	\centering
	\includegraphics[width=\textwidth]{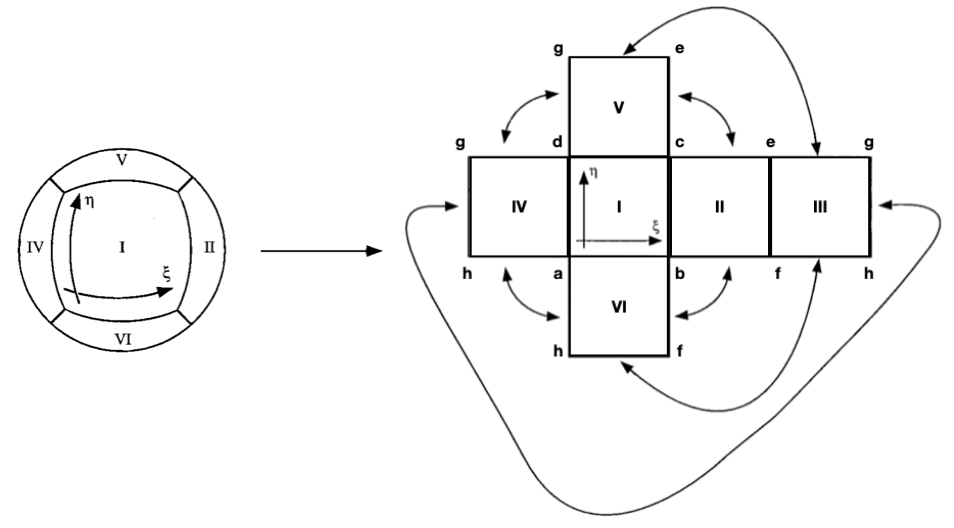}
	\includegraphics[width=\textwidth]{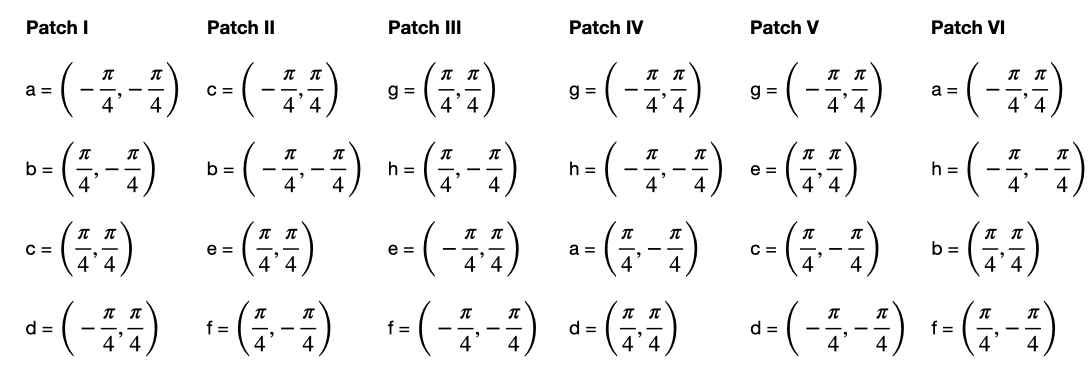}
	\caption{Exploded, cubed view of the patches \citep{ronchi96}. Each patch is identical and is described by the coordinates $\xi$ and $\eta$, both spanning the range $[-\pi/4;\pi/4]$. In the exploded view $\xi$ and $\eta$ grow to the right and upward, respectively, for all patches (only patch I is explicitly drawn here). Arrows identify the 12 edges between patches. The coordinate values $(\xi,\eta)$ of the corners for each of these patches are written in the bottom part as well. Image reproduced with permission from \citet{MATINS1}, copyright by the author(s). }
	\label{fig:full_grid}
\end{figure}

3D finite-difference and finite-volume schemes in spherical coordinates encounter the axis singularity problem. At the axis, the azimuthal direction becomes degenerate, making the $\varphi$-component of vector fields, such as magnetic or velocity fields, numerically ill-defined. Additionally, metric terms (e.g., those proportional to $\sin\theta$) vanish, leading to divisions by near-zero values that amplify round-off and truncation errors close to the axis. These issues 
often lead to severe numerical instabilities.

Common strategies to mitigate this problem include employing pseudo-spectral methods in the angular directions (e.g., {\tt MaGIc}, \citealt{wicht2002}; {\tt Parody}, \citealt{gourgouliatos16}), excising the axis from the computational domain (e.g., {\tt the Pencil Code}\footnote{\url{https://github.com/pencil-code}} in spherical geometry, \citealt{pencil2021}), or adopting alternative coordinate systems such as Yin–Yang grids \citep{yinyang2004}.
In the newly developed \MATINS\footnote{\url{https://github.com/ice-csic-astroexotic/MATINS}} code \citep{MATINS1,MATINS_MT,MATINS2,DehmanPons25}, the cubed-sphere coordinate system was implemented to overcome the axis singularity problem. Originally introduced by \citet{ronchi96}, this formalism 
keeps the radial direction as one coordinate, similar to spherical coordinates, while decomposing the volume into a stack of concentric radial layers. As illustrated in Fig.~\ref{fig:full_grid}, each layer is covered by six non-overlapping patches. These patches can be visualized as the inflated faces of a cube expanded to a spherical surface. Unlike the Yin–Yang grid \citep{yinyang2004}, the cubed-sphere has no overlap: its six patches meet edge-to-edge along great-circle arcs \citep{ronchi96}.

In \MATINS, the cubed-sphere coordinates are prescribed as in the original work by \citet{ronchi96}, with the only difference given by the metric factors in the radial direction, according to Schwarzschild interior metric, obtained from a TOV solution. In this framework, each patch has the same functional form for the metric in terms of two coordinates mapping the spherical surface portion ($\xi, \eta$). The radial unit vector $\hat{e}_r$ is orthogonal to each spherical surface, but $\hat{e}_\xi$ and $\hat{e}_\eta$ are generally not orthogonal to each other, with the degree of skew varying across the patch. This non-orthogonality is evident in the spatial part of the metric tensor with coordinates ($r,\xi,\eta$): it contains non-zero off-diagonal terms in the ($\xi,\eta$) sub-block: 
\begin{equation} 
\begin{pmatrix}
1 &0  & 0\\
0 & 1 &  - \frac{XY}{CD} \\
0 &  - \frac{XY}{CD} & 1
\end{pmatrix}
\label{eq: metric tensor}
\end{equation}
Here, $X$, $Y,$ $C,$ and $D$ are auxiliary variables of the cubed-sphere grid. Further details or the explicit transformations between cubed-sphere, spherical, and Cartesian coordinates can be found in \citet{ronchi96} or Appendix~A of \citet{MATINS1}.

The non-orthogonality of the cubed-sphere grid requires a clear distinction between the covariant and contravariant components of vector fields. This is essential for ensuring that differential operators, such as the curl used to compute the current density and advance $\BB$ in time, are evaluated consistently in the metric.

\begin{figure}
     \centering
     \includegraphics[width=0.9\textwidth]{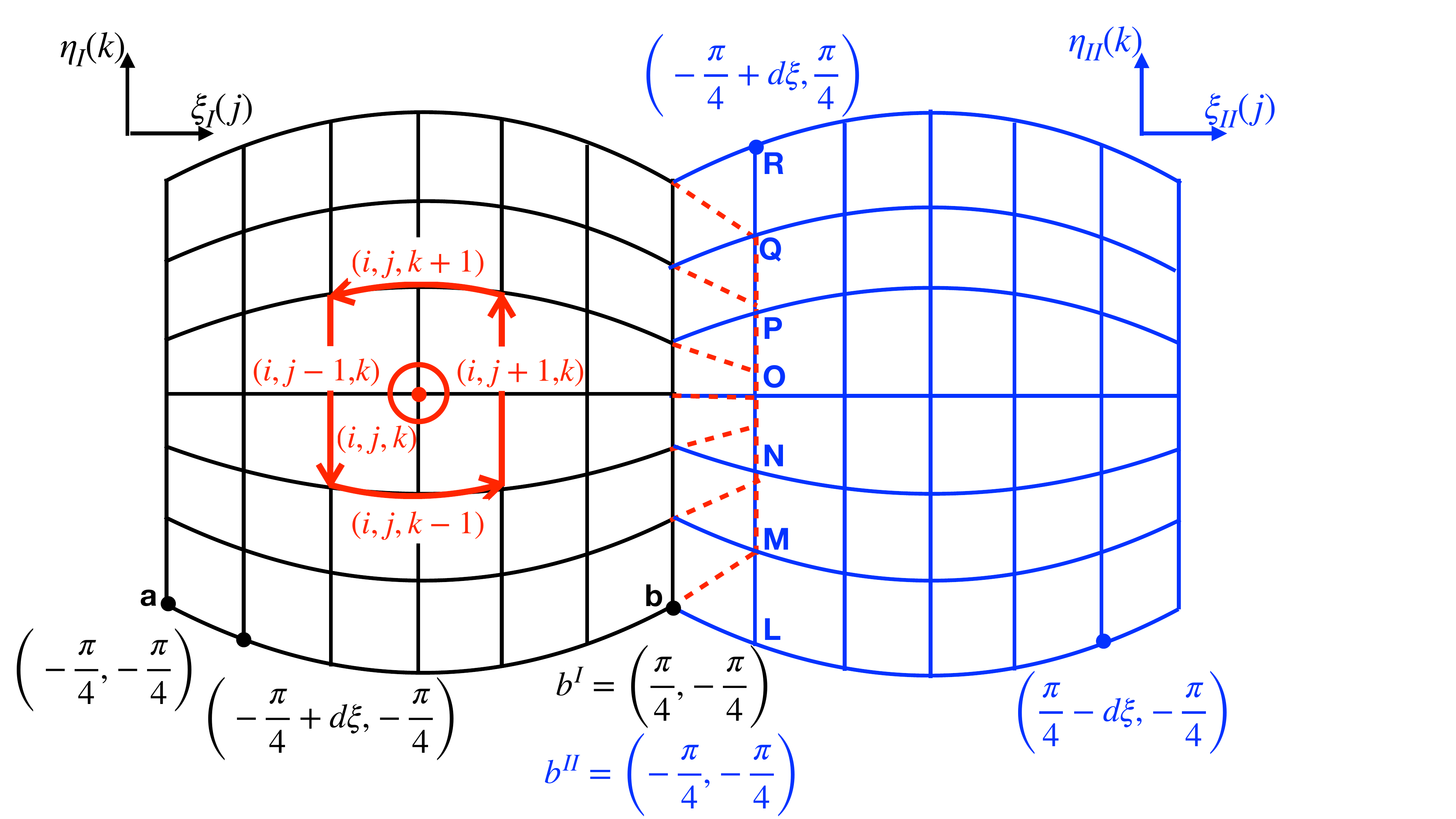}
     \caption{Schematic view of two contiguous equatorial blocks, e.g., patch I (black) and patch II (blue), and the ghosts cells of patch I (endpoints of the red dashes). The view is centered on the common vertical boundary line. The pseudo-horizontal coordinates $\xi$ of the ghost points of one grid, i.e., patch I, coincide with the second points along the $\xi$ coordinates of the last one interior grid points of the contiguous block, i.e., patch II. The ghost points are traced by the red line, and the values of the fields along the pseudo-vertical coordinate, $\eta$, are obtained by interpolations among the adjacent patch points (blue letters). Note that for other pairs of patches, the correspondence of coordinates may be less trivial (see Table~1 of \citealt{MATINS1}). A sketch of a centered discretized circulation which extends twice the size of the cell (once per each side around a central point $(i,j,k)$) is displayed on the left hand side of this plot, in red. The circulation shown here is applied to calculate the radial component of the curl operator for a given vector A, such as $(\vec{\nabla} \times \vec{A})^r$. Image reproduced with permission from \citet{MATINS1}, copyright by the author(s). }
     \label{fig: overlap region}
 \end{figure}

When evaluating derivatives near patch edges or corners, field values are needed at locations that lie in the coordinate system of adjacent patches (see Fig.\ref{fig: overlap region}). In \MATINS, this is handled by adding a single layer of ghost cells around each patch, which is sufficient for a second-order, centered finite-volume scheme. The field components in these ghost cells are obtained by interpolating neighbouring-patch data expressed in the local coordinates of the patch. Once these components are retrieved, Jacobian transformations are applied to convert them from the neighbour coordinate system to that of the patch where the derivative is actually being evaluated. As shown in Fig.~\ref{fig: overlap region}, a ghost grid line in one patch coincides with an interior grid line in the neighbouring patch, requiring only a one-dimensional interpolation along the relevant angular coordinate. Because $\xi$ and $\eta$ share the same spacing ($d\xi = d\eta = \Delta$) and range $[-\pi/4, \pi/4]$, this procedure applies in both vertical and horizontal directions. Further details on patch-interface handling can be found in \citet{MATINS1} and Appendix A of \citet{Dehmanthesis}.

\subsection{Courant condition and time advance}

In explicit algorithms to solve PDEs involving propagating waves, the time step is limited by the Courant condition, which essentially states that numerical stability requires waves not to travel more than one cell length on each time step.
Since we want to evolve our system on long (Ohmic) timescales, the Courant condition makes the simulation computationally expensive for Hall-dominated 
regimes, $\omega_B^e \tau_e \gg 1$.
For each cell, we can estimate the Courant time related to the Hall term by 
\begin{equation}\label{estimate_courant_hall}
\Delta t_h \approx \frac{4\pi e n_c L ~\Delta l }{c B}\, ,  
\end{equation}
where $L$ is a typical distance in which the magnetic field varies (e.g., the curvature radius of the lines), 
$\Delta l$ is the minimum length of the cell edges in any direction, i.e. the radial one in the case of thin NS crusts. In the case of a spectral code, $\Delta l \simeq L_{dom}/\ell_{max}$, i.e., the ratio between the length  of the dominion and the maximum number of multipoles calculated. 

The analogous stability condition for the CME term is
\begin{equation}
    \Delta t_5 \approx \frac{\Delta l}{\eta |k_5|},
\end{equation}
where the absolute value accounts for the fact that $k_5$ may be positive or negative, and for the ambipolar diffusion term we have
\begin{equation}\label{estimate_courant_ambip}
\Delta t_a  \approx  \frac{4\pi  L ~\Delta l }{c f_a B^2}\, ,  
\end{equation}
which becomes more restrictive than the Hall term when $e n_c f_a B \gg 1$. The time step must then be chosen according to
\begin{equation}
\Delta t = k_c \, \mbox{min} \left[ \Delta t_h , \Delta t_5, \Delta t_a \right]\, ,
\end{equation}
where the minimum is calculated among all the numerical cells and the Courant factor $k_c<1$ is empirically determined to ensure stability. For test-bed problems in Cartesian coordinates, taking $k_c=0.1-0.3$ is usually sufficient. In realistic models, however, often numerical instabilities caused by the quadratic dispersion relation of the Hall waves arise (or other nonlinearities), and more restrictive values of $k_c$ are required.

Other stabilizing techniques introduced in \citet{osullivan06} for the time advance of the non-linear terms are used in \citet{gonzalez18}. These methods, namely the Super Time-Stepping and the Hall Diffusion Schemes, allow the code to maintain stability and efficiently speed up the time evolution when the ambipolar or the Hall term dominates. 
Another common technique is the use of high-order dissipation (also called hyper-resistivity; \citealt{huba03}), or a predictor-corrector step advancing alternatively different field components.

\citet{vigano12a} used a particularly simple method that significantly improves the stability of the scheme in spherical coordinates. Their procedure to advance the solution from $t_n$ to $t_{n+1}=t_n+\Delta t$  can be summarized as follows:
\begin{itemize}
	\item
	starting from $\vec{B}^n$, all currents and electric field components are calculated \\ 
	$\vec{B}^n\rightarrow \vec{J}^n\rightarrow \vec{E}^n$;
	\item
	the toroidal field $\vec{B}_t^{n}$ is updated: $\vec{E}^n \rightarrow \vec{B}_t^{n+1}$;
	\item
	the new values $\vec{B}_t^{n+1}$ are used to calculate the modified current components and the toroidal part of the electric field $\vec{E}_t$: 
	$\vec{B}_t^{n+1} \rightarrow \vec{J}_p^\star \rightarrow \vec{E}_t^\star$;
	\item
	finally, we use the values of $\vec{E}_t^\star$ to update the poloidal components 
	$\vec{E}_t^\star \rightarrow \vec{B}_p^{n+1}$.
\end{itemize}
In \citet{Toth2008}, the authors discussed that such a two-stage formulation is equivalent to introducing a fourth-order hyper-resistivity. Since the toroidal component is advanced first, it follows that the hyper-resistive correction only acts on the evolution of the poloidal components. In \citet{vigano12a} it was also shown that the additional correction given by $\vec{E}_t^\star$ contains higher-order spatial derivatives and scales with $(\Delta t)^2$, which is characteristic of hyper-resistive terms. They found a significant improvement in the stability of the method when comparing a fully explicit algorithm with the two-steps method, allowing to work with $k_c \approx 10^{-2}-10^{-1}$.

In the finite-difference schemes of \citet{vigano19}, the authors used a fourth-order Runge--Kutta scheme and found that the instabilities are especially significant when using fifth-order-accurate methods for the flux reconstruction (i.e.\ WENO5), which needed to be combined with the application of artificial Kreiss--Oliger dissipation along each coordinate direction \citep{Calabrese2004}. A sixth-order derivative dissipation operator has a similar stabilizing effect, filtering the high-frequency modes which cannot be accurately resolved by the numerical grid, at the cost of a potential loss of accuracy \citep{vigano19}. For this reason, they recommend using third-order schemes that do not require any additional artificial Kreiss--Oliger dissipation. The typical Courant factors used were again quite low, $k_c \approx 10^{-2}-10^{-1}$.

\section{Magnetosphere-interior coupling and rotational evolution}
\label{sec:magnetosphere}

A central challenge in realistic simulations of magnetic field evolution lies in handling the outer boundary of the numerical domain. Just as in cooling calculations (see Sect.~\ref{sec:cooling}), direct modeling of the magnetic field in the thin ($\sim 100$ m) envelope is numerically prohibitive, since physical timescales (particularly the resistive one) are much shorter there than in the crust. The difficulty is even more severe in the magnetosphere, where the density (and therefore the relevant timescales) is more than twenty orders of magnitude smaller than in the outer crust, where numerical grids usually terminate. Because magnetospheric dynamical timescales are far shorter than those of the neutron star interior, it is generally assumed that, on the evolutionary timescales of the interior, the exterior relaxes almost instantaneously (on light-crossing timescales typical of MHD waves in such dilute plasmas) to a stationary state determined by the magnetic field and currents at the stellar surface. In this picture, the magnetosphere behaves as a perfect conductor, with currents rapidly canceling electromagnetic forces. Consequently, the interior evolution fixes the surface field that sets the external configuration, which in turn must be fed back into the interior evolution as a boundary condition at each time step. 
This creates an interdependence between the two regions and demands a consistent coupling.

The dynamics of the magnetosphere is essentially governed by the electro-magnetic field, since the plasma pressure and inertia are negligible. Therefore, a suitable approximation is that the large-scale magnetospheric structure follows force-free configurations, where electric and magnetic forces on the plasma are perfectly balanced, as in a perfect conductor (see \citealt{cerutti17,philippov22} for comprehensive reviews on electrodynamics of pulsar magnetospheres).

The force-free condition is expected to hold throughout most of the magnetosphere, with the exception of certain regions, 
such as the separatrix (the boundary between open and closed field lines), the zone immediately above the polar cap, and the current sheet forming near the light cylinder, where continuous magnetic reconnection is expected. This departure from the force-free condition (essentially induced by rotation) is central to pulsar phenomenology. It enables the component of the electric field parallel to the magnetic field to accelerate charged particles, either extracted from the stellar surface or produced via pair creation, to ultra-relativistic speeds. The motion of these particles then generates non-thermal emission across the electromagnetic spectrum, thereby converting part of the angular momentum losses (spin-down; see below) into the observed radio, $X$, $\gamma$-ray emission.

For magnetar conditions, rotational effects in the magnetospheric region close to the star can be safely ignored and the force-free condition reduces to $\vec{j}\times \vec{B}=0$: the electric currents flow parallel to the magnetic field lines that they sustain.\footnote{Since $\vec{j} \propto \Bnabla\times\vec{B}$, a force-free magnetic field is a also Beltrami vector field.} Among possible solutions, the simplest and most commonly used is the current-free or potential solution, $\vec{j}= 0$, which also applies in vacuum. Matching the interior magnetic field to a magnetospheric potential field effectively prevents current from escaping or entering the star. However, a non-zero Poynting flux across the boundary enables magnetic energy exchange between the two regions, though magnetic helicity remains conserved.

Although this simple current-free solution (adopted by the majority of works) serves as a reasonable first approximation,  developing more realistic models requires more general solutions. In particular, stable electrical currents can flow within regions of closed magnetic field lines, much like the coronal loops observed on the Sun. These current systems can persist for relatively long timescales, from months to decades \citep{beloborodov09}, likely sustained by the interior dynamics. 
There is indirect observational evidence of such currents in some magnetars, where the presence of a plasma much denser than the Goldreich-Julian value has been inferred. Soft X-ray photons emitted from the star surface are up-scattered to higher energy \citep{lyutikov06,rea08,beloborodov13} through resonant Compton processes, resulting in the observed spectra.  

Equilibrium solutions for force-free twisted magnetospheres in the context of magnetars have been investigated in several studies \citep{fuji14, glampedakis14,pili15,akgun16,kojima17}. However, the interior evolution can sometimes lead to configurations that cannot be smoothly matched to a force-free exterior. This mismatch results in discontinuities in the tangential magnetic field components at the surface, corresponding to current sheets, which may introduce numerical instabilities.

Moreover, solving a fully consistent 2D or 3D problem at each timestep in global simulations is computationally expensive, primarily due to the demands of elliptic solvers required to solve exactly the exterior force-free condition.
To address these challenges, a novel approach has recently been proposed \citep{Urban23,Stefanou23b}, exploring the use of PINNs for modeling the magnetic field evolution inside a NS coupled to a force-free magnetosphere. This method offers a promising alternative to traditional techniques. Initial results show that PINN-based solutions are accurate, robust, and numerically stable. Notably, \citet{Stefanou23b} found the computational cost to be over an order of magnitude lower than that of comparable simulations using conventional methods, and there is plenty of room for improvement \citep{Urban2025}.
These findings opened the door to extensions to fully 3D problems \citep{Stefanou25}, where implementing generalized boundary conditions becomes even more computationally demanding.

A final remark is in order. While rotation has a negligible impact on magnetic field evolution, the reverse is not true: the spin period evolves under the influence of electromagnetic torques dictated by the magnetospheric configuration. Although the equations governing rotational evolution are simpler than those for magnetic and thermal evolution, they are crucial for predicting the observable timing properties of isolated NSs.

In the remainder of this section, we outline the methodology for prescribing boundary conditions on the magnetic field when solving the induction equation using different types of numerical codes, discussing practical challenges that may arise. To conclude the section, we present the procedure for modeling the rotational evolution, including some important remarks about relativistic effects.

\subsection{Current-free boundary conditions}\label{app:vacuum_spectral} 

\paragraph{Finite-difference schemes.} 

When using a finite difference or finite volume method to advance magnetic field components in time, rather than a spectral method, boundary conditions must be applied directly to the magnetic field components instead of individual multipoles. The radial component of the magnetic field, $B_r$, is provided by the interior evolution and is known at the star's surface at each timestep. Applying the boundary conditions involves specifying the angular components consistent with the physical assumptions in one or more ghost cells outside the physical grid, based on the values of $B_r$ at the boundary.

Here, we outline the approach adopted in the 3D code \MATINS \citep{MATINS1}. As the external potential solution is, by definition, both solenoidal and irrotational, the magnetic field can be expressed as the gradient of a magneto-static potential $\chi$ which obeys the Laplace equation (see Appendix \ref{app:potential_solutions} for more details). One can then expand the scalar function $\chi$ in spherical harmonics as follows:
\begin{equation}
\chi = - B_0 R \sum_{\ell=1}^{\infty} \sum_{m=-\ell}^{m=+\ell} Y_{\ell m}(\theta,\varphi) \bigg( b_{\ell m} \bigg(\frac{R}{r} \bigg)^{\ell+1} + c_{\ell m}\bigg(\frac{r}{R} \bigg)^\ell \bigg) 
\label{eq: magnetostatic potential}
\end{equation}
where $B_0$ is a normalization factor and the dimensionless coefficients $b_{\ell m}$ and $c_{\ell m}$ correspond to the two branches of solutions. The second branch, proportional to $\big(r/R \big)^\ell$, diverges in a domain extending to infinity, such as the magnetosphere, so we must set all $c_{\ell m}=0$. 

Continuity of the radial component across the surface enables us to express it in terms of the magneto-static potential as:
\begin{equation}
     B_r= \frac{\partial \chi}{\partial r} = {B_0} \sum_{\ell=1}^{\infty} \sum_{m=-\ell}^{m=+\ell} (\ell+1) Y_{\ell m}(\theta,\varphi)  b_{\ell m} \bigg(\frac{R}{r} \bigg)^{\ell+2},
     \label{eq: spectral decomposition of Br}
\end{equation}
so that, we can evaluate the coefficients by applying the orthogonality properties of spherical harmonics to Eq. (\ref{eq: spectral decomposition of Br}). Integrating over the star surface one can obtain: 
\begin{equation}
b_{\ell m} = \frac{1}{B_0(\ell+1)} \int d\Omega ~ Y_{\ell m}(\theta,\varphi) ~ B_r(r=R),
\end{equation}
where $d\Omega = \sin\theta d\theta d\varphi$. Once the $b_{\ell m}$'s are known, the angular components of the magnetic field for $r > R$ can be readily reconstructed: 
 \begin{eqnarray}
    B_{\theta} = - B_0 \sum_{\ell=1}^{\infty} \sum_{m=-\ell}^{\ell} b_{\ell m} \bigg(\frac{R}{r}\bigg)^{\ell+2} \frac{\partial Y_{\ell m}(\theta,\varphi)}{\partial \theta},
 \nonumber \\
    B_{\varphi} =   - \frac{B_0}{\sin \theta} \sum_{\ell=1}^{\infty} \sum_{m=-\ell}^{\ell} b_{\ell m} \bigg(\frac{R}{r}\bigg)^{\ell+2} \frac{\partial Y_{\ell m}{(\theta,\varphi)}}{\partial \varphi}.
    \label{bvacuum}
 \end{eqnarray}

In summary, the procedure is the following: 
\begin{itemize}
\item
First, at each time step, obtain the $b_{\ell m}$ coefficients from the values the radial component of the radial magnetic field over the surface $B_r(r=R)$. We note that, in a discretised grid, values of $b_{\ell m}$ can be calculated only up to a maximum multipole $\ell_{\max}=n_\theta/2$, where $n_\theta$ is the number of angular points of the grid.
\item
Second, from the coefficient we reconstruct the values of $B_r$ and $B_\theta$ in the external ghost cells, by using Eq.~(\ref{bvacuum}).
\end{itemize}

This method is very accurate for smooth functions. In the case of sharp features in $B_r$, which may be created by the Hall term, the largest multipoles acquire a non-negligible weight, and, since $\ell_{\max}$ is limited,  fake oscillations in the reconstructed $B_\theta$ may appear (Gibbs phenomenon). An alternative method to impose potential boundary conditions is based on the Green's representation formula, a formalism often used in electrostatic problems
able to correctly handle the angular discontinuities in the normal components. Details about the derivation of the Green's integral relation between $B_r$ and $B_\theta$ at the surface are given in Appendix~\ref{app:potential_solutions}.

\paragraph{Spectral methods.}
Consider a spectral code working directly with the two potential functions $\Phi_{\ell m}$ and $\Psi_{\ell m}$ as defined in Sect.~\ref{methods_spectral} for the poloidal/toroidal decomposition.
The requirement that all components of the magnetic field be continuous (no current sheets at the surface) implies that the scalar potentials and their derivatives are continuous through the outer boundary. Therefore, the $\Bnabla \times \vec{B}= 0$ condition translates into
\beq
\Psi_{\ell m}=0 \, ,
\eeq
and the following differential equation for each radial function $\Phi_{\ell m}(r)$
\beq
(1-z) \frac{\partial^2 \Phi_{\ell m}}{\partial r^2} 
+ \frac{z}{r}\frac{\partial \Phi_{\ell m}}{\partial r}  
-\frac{\ell(\ell+1)}{r^2} \Phi_{\ell m} = 0\, ,
\label{ode2}
\eeq
where we assume the metric~(\ref{Schw}), and $z\equiv \frac{2 G M}{c^2 r}$. 
In this subsection, we explicitly reintroduce the relativistic corrections, as they will play an important role in the spin-down rate, discussed later in Sect.~\ref{GRspindown}.

We note that there is no $m-$dependence in the equation, so that the solution depends only on $\ell$ and we will omit the $m$ subindex hereafter.
 
In general, the family of solutions of Eq.~(\ref{ode2}) for any value of $\ell$
can be expressed in terms of generalized hypergeometric functions ($F([],[],z)$), 
also known as Barnes' extended hypergeometric functions, as follows:
\bear
\Phi_\ell
=  C_\ell ~r^{-\ell} ~F([\ell,\ell+2], [2+2\ell], z) + D_\ell ~r^{\ell+1} ~ F([1-\ell,-1-\ell], [-2\ell], z)\, , 
\label{outerPhi}
\ear
where $C_\ell$ and $D_\ell$ are arbitrary integration constants that correspond to the weight of each magnetic multipole $\ell$. Note that regularity at $r=\infty$ requires $D_\ell=0$ for each $\ell$. For any given value of $\ell$, one can also express the solution in closed analytical form. The explicit expressions for $\ell=1$ and $\ell=2$ are
\bear
\label{outerPhi1}
\Phi_1 &=& C_1 r^2 \left[ \ln(1-z) + z + \frac{z^2}{2}\right] \, ,
\\
\Phi_2 &=& C_2 r^3 \left[ (4-3z) \ln(1-z) + 4z - {z^2} - \frac{z^3}{6}\right]~. 
\ear
If we consider the Newtonian limit ($z\rightarrow0$), Eq.~(\ref{ode2}) simplifies to:
\beq  
 \frac{\partial^2 \Phi_{\ell}}{\partial r^2}   - \frac{\ell(\ell+1)}{r^2}\Phi_{\ell} = 0~.
\eeq 
The only physical solution (regular at infinity) of this equation is $\Phi_\ell=C_\ell \, r^{-\ell}$. Therefore, the requirement of continuity across the surface results in  
\beq 
\left. \frac{\partial \Phi_{\ell}}{\partial r}\right |_{r=R} = -\frac{\ell}{R}\Phi_\ell ~.
\label{OBC_Phi} 
\eeq 
In the relativistic case, we can implement Eq.~(\ref{outerPhi}) directly, or the most practical form, analogous to the Newtonian case:
\beq
\left.  \frac{\partial \Phi_{\ell}}{\partial r}\right |_{r=R} = -\frac{\ell}{R} f_\ell \Phi_\ell  \, ,
\label{relBC}  
\eeq
where the $f_\ell$'s are relativistic corrections that only depend on the value of $z$ at the star surface,
$z(r=R)$ (in the Newtonian limit all $f_\ell=1$), and can be evaluated numerically only once with the help of any algebraic manipulator and stored\footnote{See \citealt{radler2001} for an alternative form to evaluate $f_\ell$ based on the expansion in a series of powers of $1/r$.}.

\subsection{Force-free boundary conditions}\label{sec:forcefree}

In axial symmetry, the construction of force-free (FF) magnetospheres for (non-rotating) magnetars is a well-studied problem, even in the relativistic case (see, e.g., \citealt{kojima17} and references therein). In the context of magneto-thermal evolution, \citet{akgun18b} explored a method to impose such boundary conditions by solving the Grad-Shafranov equation, at each time step, to match the internal evolution of the star. Let us review their approach. 
Considering axial symmetry, the magnetic field can be written as follows: 
	\bear
	\vec{B} = \frac{(\partial P/\partial \theta)}{r^2 \sin\theta} {\hat{r}} - \frac{(\partial P/\partial r)}{r \sin\theta} {\hat\theta} + \frac{T}{r\sin\theta} {\hat\varphi} \, ,
	\label{mag_PT}
	\ear
where $P$ and $T$ are functions defining the poloidal and toroidal components, respectively (see more details in Appendix~\ref{app:formalism}).

The condition ($\vec{j} \times \vec{B} = 0$) implies that the electrical currents flow along magnetic surfaces, which are defined by constant $P$. Thus, the mathematical requirement of a vanishing azimuthal component of the local Lorentz force implies that the poloidal and toroidal functions must be functions of one another, say $T = T(P)$, that is, the poloidal and toroidal functions $P$ and $T$ are constant on the same magnetic surfaces\footnote{The magnetic flux through the area enclosed by the corresponding magnetic surface is $2\pi P$, and the current through the same area is $c T/2$. }.

From the definition of the current, one can arrive at the so-called Grad--Shafranov equation:
\bear
 \frac{\partial}{\partial r}\left(  \frac{\partial P}{\partial r}\ \right) +  \frac{\sin\theta}{r^2}  \frac{\partial}{\partial \theta}\left( \frac{1}{\sin\theta}  \frac{\partial P}{\partial \theta}\ \right) =
 - T(P) T'(P)   \ ,
\label{GS_eq}
\ear
where $T'(P) = dT/dP$. The current-free limit (potential solution) is simply recovered by taking the right hand side equal to zero.

In principle, there is an infinite family of external force-free solutions for a given radial magnetic field at the surface, because of the freedom to choose the functional form of $T(P)$. The main problem of this approach is how to continuously match the arbitrary field configuration, resulting from the evolution in the crust, while enforcing the force-free solution outside. In the crust, any line bundle marked by a given magnetic flux $P$ has in general different values of $T$ because, internally, the force-free condition does not hold. As discussed in \citet{akgun18b}, there is an intrinsic inconsistency in the possibly multi-valued function $T(P)$, if we strictly take it from the values at the surface ($r=R$). They address this problem by symmetrizing the numerical function $T(P)$, which is physically equivalent to allowing the propagation through the surface only of the modes compatible with solutions of the Grad-Shafranov equation. 

In Fig.~\ref{fig:twisted_taner}, we show a representative result, showing the evolution of an axisymmetric magnetospheric configuration physically connected to the interior. The initial model consists of both poloidal and toroidal dipolar components, with the latter extending beyond the surface. As the internal magnetic field evolves, the external magnetic field is consistently twisted, by the injection of magnetic helicity (i.e., currents) in the magnetosphere. 
Solutions are calculated at each time step until a critical point, where numerical solutions cannot be found anymore. The absence of a compatible solution physically means that the magnetosphere is expected to become unstable, possibly resulting in a global reconfiguration by opening of the twisted field lines and magnetic reconnection. 
Such reconfiguration, occurring on dynamical timescales (ms), is of extreme interest for the observed transient phenomenology of magnetars \citep{2011ASSP...21..247R}, but cannot be simulated with long-term evolution codes. These processes have been studied in detail in dedicated force-free electrodynamics simulations, in both the Newtonian and general relativistic (GR) cases \citep{parfrey13,carrasco19}.

\begin{figure}[t] \centering	\includegraphics[width=.9\textwidth]{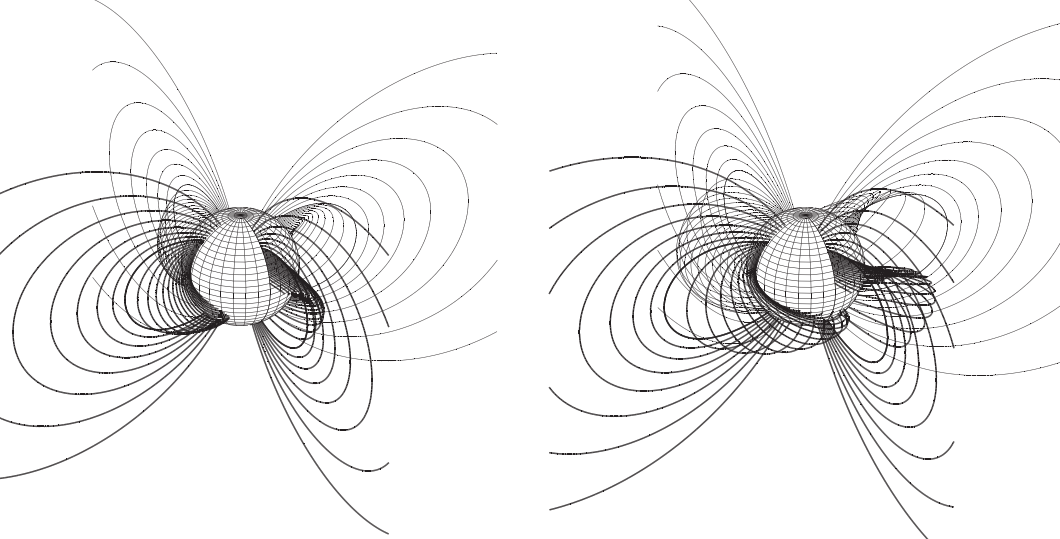}
\caption{Evolution of a twisted magnetosphere with coupling with the interior. The left and right panels show snapshots at $t=0$ and $t \sim 1.58$ kyr, the critical time when the magnetosphere of this particular model has stored the maximum possible twist. Image reproduced with permission from \citet{akgun17}, copyright by the author(s).}\label{fig:twisted_taner}
\end{figure}


\citet{Stefanou23a} presented a comprehensive study of force-free twisted magnetar magnetospheres with non-linear current distributions. Their work solves the force-free equations within a compactified spherical coordinate system, using the Grad–Rubin method. At the stellar surface, they applied suitable boundary conditions to prescribe both the current distribution and the magnetic field. The method’s accuracy is verified by reproducing a range of established analytical solutions and axisymmetric numerical results. Building on this validation, they explore fully 3D configurations with non-axisymmetric current patterns—for example, magnetic fields with localized twists that mimic surface hotspots. The study analyzes key physical quantities, including magnetic energy, helicity, and twist, and considers the implications for the magnetar’s energy budget, surface heating, and magnetic diffusion timescales, all in connection with possible observational signatures.

A particularly interesting model is a dipolar magnetic field combined with a localized surface current following a Gaussian profile, designed to reproduce the behavior of magnetospheres influenced by current-generating hotspots. They examine how the hotspot’s size and strength affect the magnetic energy, effective surface temperature, and magnetic diffusion timescale. The resulting temperature distributions and energy budgets align closely with observational inferences of magnetar hotspots, supporting the physical plausibility of the model.

In a recent paper \citep{Stefanou25}, building on the methodology of previous works in axisymmetry \citep{Stefanou23b,Urban23},
the authors employ a novel methodology based on PINNs to model pulsar and magnetar magnetospheres, spanning both axisymmetric and fully three-dimensional configurations for stars of varying compactness. The force-free equations are directly solved in the form
\begin{equation}
    \Bnabla \times \vec{B} = \alpha(\vec{r}) \vec{B}
\end{equation}
where $\alpha$ is an arbitrary, user-supplied function, associated with the strength of the twist (or equivalently, the ratio of toroidal to poloidal strengths of the magnetic field in the axisymmetric case) in the magnetospheric region.

Their framework successfully reproduces established axisymmetric solutions from the literature, including non-dipolar cases, while accurately capturing current sheet structures in the 2D rotating pulsar models. Models with surface current profiles designed to mimic the geometry of observed hotspots (Gaussian profiles for $\alpha$, with $\alpha_0$ denoting the maximum value of alpha at the center of the Gaussian) are imposed as boundary conditions at the star surface.
This analysis reveals that the lowest-energy solution branches allow only about 30\% more energy than current-free configurations in axisymmetric, globally twisted models. The excess energy available drops to about 5\% for fully three-dimensional cases with localized spots. In Figure \ref{fig:twisted_3D} we show samples of solutions with different
values of $\alpha_0$, the parameter that controls the intensity of the current. As its value increases, the twist of the lines threaded by currents becomes stronger.

These works highlight the promise of PINNs as an efficient and generalizable tool for simulating 3D magnetospheres (or other elliptic PDE problems), offering new future perspectives to investigate the magnetar phenomenology. The preliminary (axisymmetric) results presented in \citet{Urban23} demonstrate that this approach can be applied to the astrophysical problem of magnetic field evolution within a NS interior, coupled to a force-free magnetosphere. Using a PINN reduced the computational cost by more than an order of magnitude compared to a finite difference scheme applied to a similar case. 
These findings open the door to future 3D extensions of this or related problems, where implementing generalized boundary conditions is otherwise prohibitively expensive.

\begin{figure}[t]
	\centering
	\includegraphics[width=1.0\textwidth]{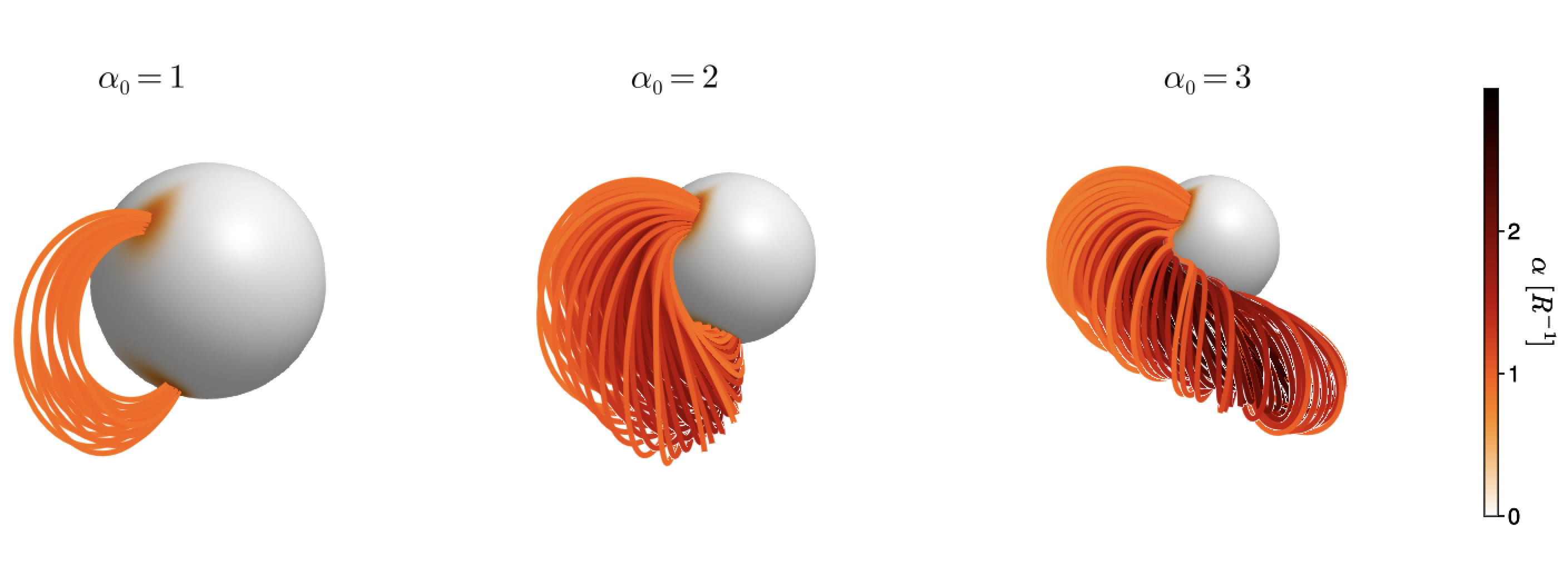}
	\caption{Illustrative examples of twisted magnetospheres with different values of $\alpha_0$. For clarity, only lines with $\alpha > 0.5 \alpha_0$ are shown. Figure courtesy of P.~Stefanou.}
	\label{fig:twisted_3D}
\end{figure}

\subsection{Dynamic force-free relaxation} \label{sec:extended_domains}

An alternative to imposing a precise mathematical boundary condition at the surface is to consider an extended domain, where we evolve at the same time all components of the field, but with physical coefficients that enforce the solution to meet the required conditions. Instead of imposing a boundary condition at the last numerical cell, this approach considers a generalized induction equation, where, at the surface, there is a sharp transition in the values of the pre-coefficients describing the physics ($\eta, f_H, f_a$). In the numerical GRMHD context, this approach has been successfully used to describe at the same time the resistive and ideal MHD inside and outside a NS \citep{palenzuela13}. 

The idea is that, since the magnetospheric timescales are many orders of magnitude shorter than the interior, the long-term evolution of the magnetosphere can be seen as a series of equilibrium states, attained immediately after every time step of the interior. Therefore, one can activate an artificial term that dynamically leads to the force-free solution. This approach is similar to the {\it magneto-frictional method} \citep{yang86,roumeliotis94}, as known in solar physics. The modified induction equation employed in the exterior of the star has a mathematical structure equivalent to an ambipolar term (as in Eq.~(\ref{ind_eq_final})), which forces currents to gradually align to magnetic field lines, without having to solve the elliptical Grad-Shafranov equation at every time step (which is numerically expensive). This also allows us to account for the transfer of helicity and provides a mechanism to continuously feed currents that twist the magnetosphere. The caveat is that the coupling coefficient (quantifying the ratio between the interior and exterior timescales) must be fine-tuned to prevent the exterior dynamics from being neither too fast (it would excessively limit the time step), nor too slow (it would not manage to relax to a force-free configuration and would cause non-negligible, unphysical feedback on the interior). In the NS long-term evolution scenario, such a strategy has only been explored preliminarily in the 3D Cartesian parallelized code used in \citet{vigano19}. More studies are needed to test the feasibility of this approach.

\subsection{Evolution of spin period and obliquity}
\label{sec:spin-down}

The magnetic field evolution also influences the rotational properties of the star. As NSs age, they lose angular momentum through electromagnetic torques, i.e. they spin down \citep{spitkovsky06,beskin13,philippov14}. The poloidal dipolar magnetic field dominates this process, since higher-order multipoles decay rapidly with distance. 
Since the period and its derivative are among the primary observables, it is relevant to examine the standard approximations and their limitations. The basic argument for quantitatively estimating the {\em dipolar component of the surface field at the magnetic pole}, $B_p$, involves equating the rotational energy losses, $I\Omega \dot\Omega$ (where $I$ is the moment of inertia of the interior component coupled to the magnetosphere, $\Omega$ is the spin angular velocity, and $\dot \Omega$ its time derivative), to the electromagnetic torque, which is
\begin{equation}\label{eq:lsd}
    L_{sd} = \frac{B_p^2R^6\Omega^4}{6c^3}f_\phi\, ,
\end{equation} where $R$ is the NS radius and $f_\phi$ is a factor $\sim {\cal O}(1)$ which depends on the inclination angle $\phi$, defined as the angle between the dipolar magnetic moment and the rotational axis. A widely used simplification consists in considering the analytical case for vacuum, in which case $f_\phi=\sin^2\phi$. For an orthogonal rotator, using a typical moment of inertia of $I=10^{45}$ g~cm$^2$ and a radius of $R=10$ km, one has:
\begin{equation} \label{eq:bp_vacuum}
    B_p^{\rm vac} \sim 6.4 \times 10^{19} \sqrt{ P[{\rm s}] \dot{P}} \text{ G}.
\end{equation}
where $P=2\pi/\Omega$ and $\dot P$ are the actual observables, the spin period and its time derivative measured by a distant observer. This estimate is ubiquitously used for its simple connection to the precisely measured timing observables. However, it does not consider the presence of plasma, which has to populate the magnetosphere \citep{goldreich69}.

A rotating plasma-filled magnetosphere has no trivial solution; therefore, numerical simulations are necessary. In a more general and realistic case, $f_\phi = \kappa_0 + \kappa_1\sin^2\phi$ \citep{spitkovsky06}, and the coupled evolution of spin period and inclination angle is governed by
\citep{philippov14}:
\begin{eqnarray}
&& \dot{P} = \beta \frac{B_p^2}{P} (\kappa_0 + \kappa_1 \sin^2{\phi})\, ,\label{eq:spindown} \\
&& \dot{\chi} = - \kappa_2 \beta \frac{B_p^2}{P^2} \sin{\phi} \cos{\phi}\, , \label{eq:alignment} 
\end{eqnarray}
where we have defined the auxiliary quantity
\begin{eqnarray}
\beta \equiv \frac{\pi^2 R^6}{I c^3} \, , 
\end{eqnarray}
and the coefficients $\kappa_0$, $\kappa_1$, $\kappa_2$ depend on the magnetosphere geometry and its physical conditions and determine the magnetospheric torque. In the classical (unphysical) vacuum dipole model $\kappa_0=0$, $\kappa_1=\kappa_2=2/3$, which incorrectly suggests that an aligned rotator ($\phi=0$) experiences no torque and would not spin down \citep[e.g.][but similar arguments are abundant]{johnston17}.
This is physically inaccurate, as realistic 3D numerical models of plasma-filled magnetospheres find $\kappa_0\sim \kappa_1\approx1$ \citep{spitkovsky06,philippov14}, with $\kappa_2$ ranging from 0 to 1.
Many groups achieve comparable results despite using different methods (force-free electrodynamics vs. particle-in-cell simulations) and physical components (resistive or purely force-free, relativistic or non-relativistic).

Moreover, \citet{philippov14} demonstrated that the alignment of the rotation and magnetic axes in a vacuum magnetosphere model occurs much faster (exponentially, with characteristic time $\tau_0 = \frac{P_0}{2 \beta B_0}$) compared to realistic plasma-filled magnetospheres (where the alignment angle decreases following a power-law). In a realistic case, variations in the inclination angle typically cause torque corrections of up to a factor of $\approx 2$ (similar to the uncertainty in the NS moment of inertia). Alignment cannot halt the star’s period evolution.

Conversely, the decay of the magnetic field can cause torque variations spanning multiple orders of magnitude, significantly impacting $P$ and $\dot{P}$. To model rotational evolution effectively, the key factor is the time evolution of $B_p$, provided by interior evolutionary models. The particular value of the initial period becomes irrelevant at later stages, provided it is sufficiently small, $P_0\ll P$. 
Another factor influencing rotational evolution is the time-dependent moment of inertia $I(t)$. While the overall neutron star structure remains stable, superfluidity can be a relevant factor. A superfluid component (e.g., neutrons in the core or inner crust) is only weakly coupled to the star’s rotation and does not contribute to $I$, which should only account for matter rigidly co-rotating with the magnetosphere.
Two other effects can alter the moment of inertia $I$: (1) the volume of the superfluid component may evolve, as the phase transition depends on density and temperature (as the star cools, the volume of the superfluid component gradually
increases); and (2) during glitches, the normal and superfluid components may temporarily couple, modifying $I$. These effects are challenging to model, requiring a two-fluid approach. Typically, these corrections are ignored by assuming a constant $I$ for the entire star in rigid rotation. For realistic stars, $I \sim 1.5 \times 10^{45}$~g~cm$^2$, with a $50\%$ uncertainty, yielding $\beta \sim 6 \times 10^{-40}$ s G$^{-2}$.

\subsubsection{General Relativistic effects}
\label{GRspindown}

An often-overlooked fact is that GR effects significantly enhance the spin-down luminosity $L_{sd}$ and overestimate the inferred values of the magnetic fields.

Several advanced GR force-free electrodynamics or particle-in-cell magnetospheric simulations (e.g., \citealt{ruiz14,philippov15,carrasco18})
indicate that the spin-down luminosity, measured through the Poynting flux at the light cylinder, exceeds that of Newtonian models, showing higher values of $\kappa_0$ and $\kappa_1$ \cite[see Table 2 in][]{petri16}. The physical reason for this is the larger fraction of open field lines as the compactness ratio $M/R$ increases,
with $M$ being the mass of the star and $R$ the areal radius.
Care must be taken when interpreting quantities that depend on the reference frame. The relativistic magnetospheric simulations mentioned above yield results expressed as the magnetic moment observed at infinity. Consequently, the reported slight increase in $L_{sd}$ with compactness must be interpreted in terms of the observable quantities in a specific frame.

As discussed by \citet{Rezzolla2004}, two additional GR effects suggest that $L_{sd}$ may increase even more with compactness. First, there is an effective amplification of the magnetic field strength near the pole due to spacetime curvature. Using the GR potential solutions Eqs.~(\ref{ode2},\ref{outerPhi},\ref{outerPhi1}), the dipolar field intensity, measured by an observer at the surface, $B_{p\star}$, is higher than the value seen by an observer at infinity, $B_{p0}$, by a factor $f_R$:
\begin{equation} \label{eq:relb}
    f_R := \frac{B_{p\star}}{B_{p0}} =  -\frac{3}{z^3}
    \left[\ln{\left(1 - z \right)} + z + \frac{z^2}{2} \right]>1,
\end{equation}
where $z = 2GM/c^2 R$ is the redshift factor introduced above. Similarly, the angular velocity of the star measured by a distant observer, $\Omega_0$, is lower than the value measured at the surface, $\Omega_\star$, by a factor $N_R$: 
\begin{equation}
  N_R := \frac{\Omega_0}{\Omega_*} = \sqrt{1 - z} < 1 \, ,
\end{equation}
Since the losses will depend on the local quantities, $L_{sd} \propto B_{p\star}^2 \Omega_*^4$, Eq.~(\ref{eq:lsd}), there is an effective amplification relative to the Newtonian luminosity \cite[see equation 150 in][]{Rezzolla2004}
\begin{equation} \label{eq:sdlum}
\frac{L_{sd}^{\rm GR}}{L_{sd}^{\rm N}} \equiv  \kappa = \frac{f_R^2}{N_R^4} ~.
\end{equation}
This correction scales sharply with compactness. For example, $\kappa = 4.2$ for $z = 0.34$ but $\kappa = 10.7$ for $z = 0.5$. Thus, the proper relativistic formula yields a substantial torque increase, potentially leading to a significant overestimation of the "measured" magnetic fields. 

We now examine the strength of the dipolar component, $B_{p\star}$, as observed in the local frame, as this is the physical quantity that simulations track.
The relativistic version of eqs. (\ref{eq:spindown}) and (\ref{eq:bp_vacuum}) is:  
\begin{equation} \label{eq:breduc}
\begin{aligned}
B_{p\star}^2 = \left(1 - \frac{z}{2} \right)^{7/2} \frac{3 c^3 P \dot{P} I}{2 \pi^2 R^6 \left(\kappa_0 + \kappa_1 \sin^2 \phi \right)}.
    \end{aligned}
\end{equation}
Using representative parameters 
($\phi = \pi/4$, $M=1.4~M_{\odot}$, $R = 12$~km, $I = 0.4~M R^2$, e.g.  \citealt{lattimer01,bejger02}),
\cite{Stefanou25} find that a 
GR-corrected inferred $B_p$  
reads
\begin{equation} \label{eq:bpgr}
    B_p^{\rm GR} \sim 1.6 \times 10^{19} \sqrt{ P{\rm [s]} \dot{P}} \text{ G},
\end{equation}
which is a factor 4 lower than the widely used Newtonian, Eq.~(\ref{eq:bp_vacuum}).

Note that these GR corrections affect the entire NS population, potentially introducing significant bias in the inferred magnetic fields. Additional effects may further skew the inferred fields in magnetars. On one hand, in a highly twisted magnetosphere, the Poynting flux can be further amplified as toroidal pressures expand field lines beyond the light cylinder \citep{parfrey13}. 
For instance, \citet{ng25} found that extreme twists could increase the spin-down luminosity by a factor of up to 16, which would further amplify this correction in some cases. On the other hand, particle winds produce a similar effect, particularly pronounced in magnetars \citep{tong13}. The combined impact of all effects likely results in a significant systematic overestimation of magnetic field strength when applying the classical Newtonian dipole model \citep{petri2019}.

\section{Magneto-thermal evolution of NSs}\label{sec:examples}

\subsection{Initial conditions and early evolution}
\label{sec: initial conditions}

Modeling the magneto-thermal evolution of NSs begins with a physically motivated initial model that defines the temperature and, crucially, the magnetic field configuration. 
Here, "initial" refers to the state just after the proto-NS cools and contracts to its final size due to neutrino transparency, approximately one minute after formation.
For the temperature evolution, the initial conditions are well-established: within hours to days after formation, the NS core becomes nearly isothermal, allowing the assumption of a uniform temperature. The precise value of the initial temperature, as long as it falls within the range $10^9$--$10^{10}$ K, influences only the early evolutionary stages (up to a few decades). This is because neutrino production, which scales non-linearly with temperature, acts as a self-regulating mechanism, causing the star to lose memory of its initial temperature quickly. Consequently, cooling curves starting from different initial temperatures $T_0$ converge rapidly to the same trajectory, provided $T_0$ is not unrealistically low.

Establishing the initial magnetic field configuration poses a significantly greater challenge. The complex dynamics of core-collapse supernovae result in highly intricate geometries for strong magnetic fields, leaving the question of the most probable realistic initial configurations unresolved. One possible approach assumes that the proto-NS remains in its hot, liquid phase long enough to achieve full MHD equilibrium. However, the possible equilibria are infinite, so that, in practice, MHD equilibrium-based initial conditions have been calculated only for a number of very simple geometries, often consisting of a dipole with the toroidal field contained in a torus in the equatorial region \citep{colaiuda2008, ciolfi2013}. These smooth solutions are characterized by having most of the currents in the core, thus rendering the crustal dynamics, including Ohmic dissipation and the related spin-down evolution, too slow to explain the observations. Moreover, it is unclear how nature could lead to such simple geometries, instead of redistributing the magnetic energy across a wide range of multipoles, as it is ubiquitously seen in astrophysics. Another approach, followed by most existing crustal evolution models, and heuristically driven by the need of having shorter dynamical timescales in the crust, consists of crustal-confined fields with an arbitrary degree of complexity. In these models, MHD equilibrium is usually not satisfied, the core is not magnetized and the electrical currents entirely circulate in the outer layers only. Note that qualitatively justifying the crust confinement by magnetic flux expulsion due to the Meissner effect (exhibited by type-I superconductors) is an argument that is in conflict with the type-II superconductivity (or a mix of types, see \citealt{wood22}) that protons are thought to display in these conditions.

In addition to geometry, the origin of ultra-strong magnetic fields remains an open question. Several mechanisms potentially active during the initial proto-NS phase have been proposed. Primarily, differential rotation can generate a large-scale toroidal magnetic field by twisting a weak pre-collapse field. Furthermore, the magnetorotational instability, which also depends on differential rotation, can exponentially amplify a weak seed field across a large-scale configuration. At saturation, this instability can sustain both poloidal and toroidal field components across various spatial scales \citep{mosta15,guilet2017,reboul2022}. Alternatively, compression and convection in the hot-bubble region between the proto-NS and stalled shock may play a role \citep{ober2015}. Another possibility is the Tayler--Spruit dynamo in a proto-NS spun up by fallback accretion; recent 3D MHD simulations \citep{Barrere25} show that a self-sustained dynamo emerges when the Brunt-V\"ais\"al\"a frequency exceeds the angular rotation frequency by a factor of four. Magnetic field amplification during NS–NS mergers has also gained attention \citep{ciolfi2019}, though the rarity of such events, and the likely outcome as a black hole rather than a NS, implies that this formation channel can account for only a tiny fraction of magnetars. Despite their differences, all these mechanisms involve the distribution of magnetic energy over a broad range of scales due to turbulence.

\begin{figure}
  \centering
\begin{minipage}{0.49\linewidth}
  \centering
\includegraphics[width=\linewidth]{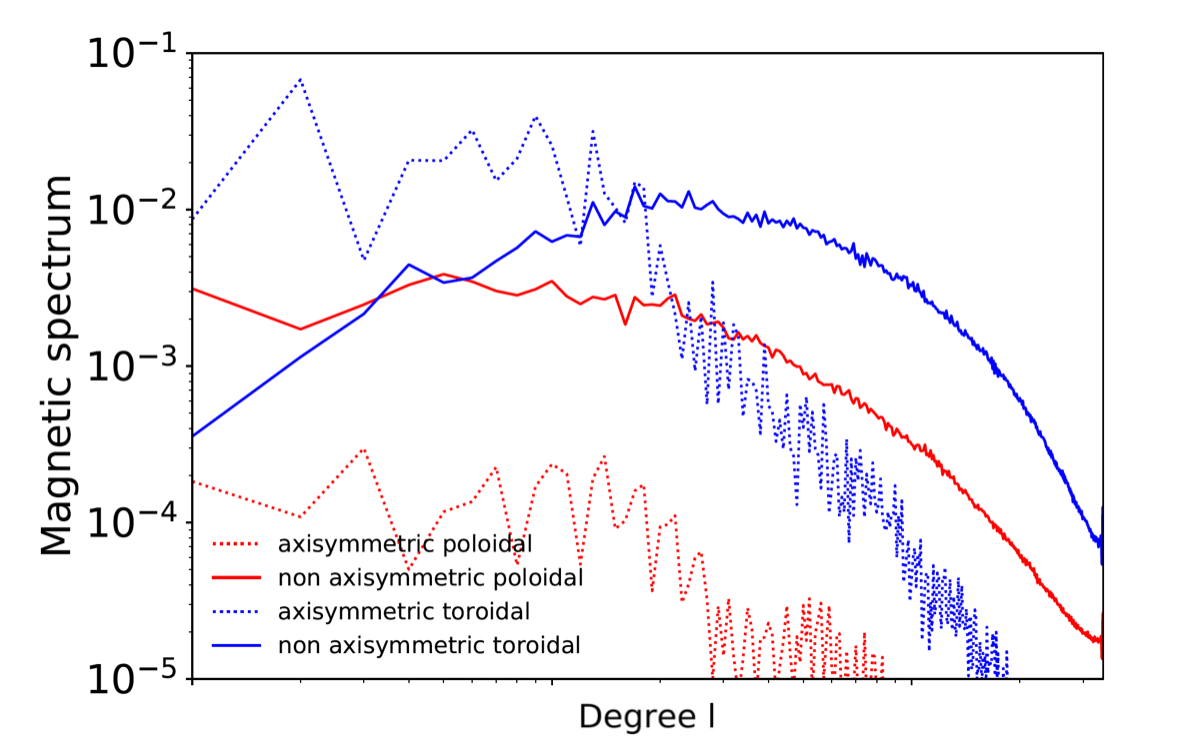}
\end{minipage}\hfill
\begin{minipage}{0.49\linewidth}
  \centering
  \includegraphics[width=\linewidth]{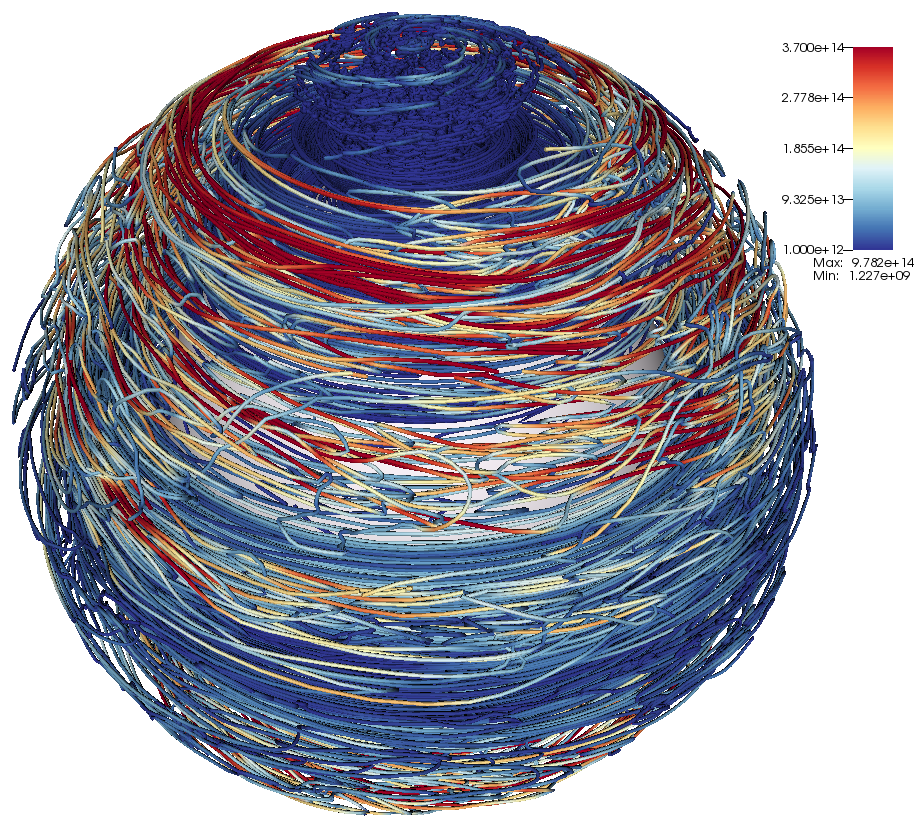}
\end{minipage}
  \caption{Representative magnetic configuration in a proto-NS, soon after birth, as obtained by dynamo simulations in a shell, with a background which mimics the typical differential rotation and thermodynamical properties observed in core-collapse simulations. Left: Different contribution to the volume-averaged magnetic-energy spectra. Shown are the poloidal (red) and toroidal (blue) components, each normalized to the total magnetic energy, as functions of the dimensionless multipole index $\ell$.
  Dotted (solid) lines denote axisymmetric (non-axisymmetric) contributions. Right: 3D visualization of magnetic field lines, with color indicating the magnetic field strength, in units of Gauss. Image reproduced with permission from \citet{reboul2022}, copyright by the author(s).
  }
  \label{fig:Reboul turbulence results}
\end{figure}

Only recently, simulations accounting for complex initial conditions have been carried on, although still confined to the crust \citep{gourgouliatos20,Igoshev21,Igoshev25,MATINS_MT,DehmanBrandenburg25,DehmanPons25}. 
Among these works, \citet{MATINS_MT} performed fully 3D magneto-thermal simulations of realistic NS crusts initialized with complex initial conditions inspired by the results of dynamo simulations having a proto-NS-like background (see Fig.~\ref{fig:Reboul turbulence results}). 
Following \citet{reboul2022}, the magnetic energy was initially stored in the toroidal component, especially the quadrupole ($\ell=2$), with the dipole contributing just a few percent. Interestingly, \citet{MATINS_MT} found that this energy distribution persists for hundreds of thousands of years, since the Hall term continuously taps energy from larger scales, whereas Ohmic dissipation removes energy from the small scales. Small-scale structures contribute noticeably to the stellar surface without dominating, and all scales decay gradually, preserving an approximately self-similar spectrum. As a result, the field remains tangled, and its complex structure does not disappear throughout the evolution. No evidence of an inverse cascade feeding back to amplify the dipole was found. A qualitatively similar evolution of the magnetic energy spectrum was observed in \citet{Igoshev25}, who used initial models from simulations of a Tayler–Spruit dynamo \citep{Barrere25}. Both works demonstrate that turbulent dynamo-generated fields at birth reproduce the expected properties of CCOs and low-field magnetars (relatively weak dipole, strong internal field in smaller scales). Notably, the thermal luminosities predicted by these simulations also agree with observations of such sources. However, they still fail to reproduce the classical magnetar picture (ultra-strong, dominant $\ell=1$ poloidal component at the surface). Thus, the origin of the strong, large-scale dipole required to explain magnetar spin-down remains uncertain (but see the discussion in Sect.~\ref{GRspindown}). 

\begin{figure}[t]
  \centering
\begin{minipage}{0.49\linewidth}
  \centering
\includegraphics[width=\linewidth,height=0.7\linewidth,keepaspectratio=false]{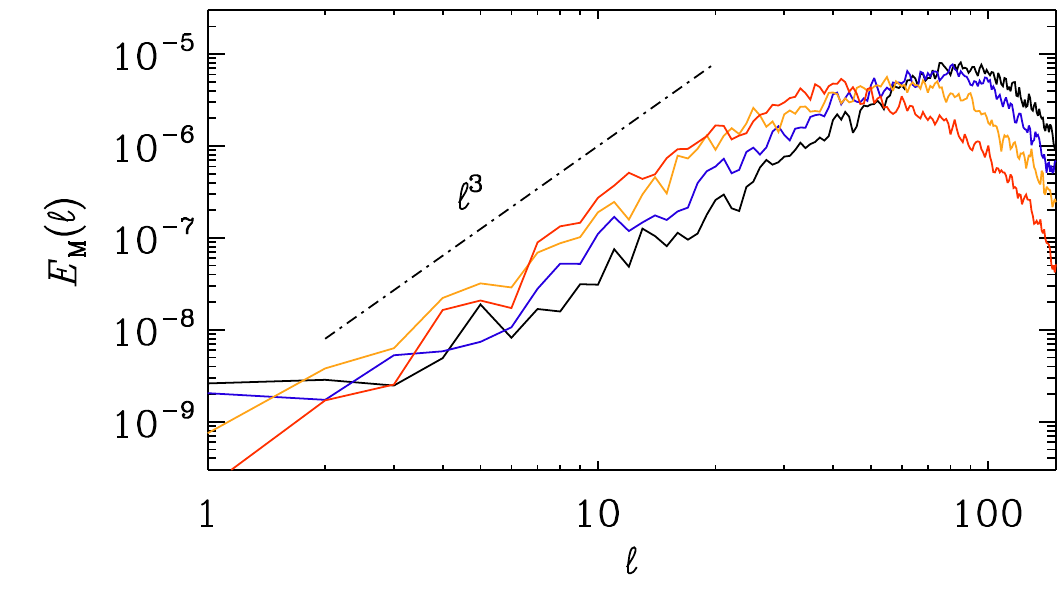}
\end{minipage}\hfill
\begin{minipage}{0.49\linewidth}
  \centering
  \includegraphics[width=0.8\linewidth]{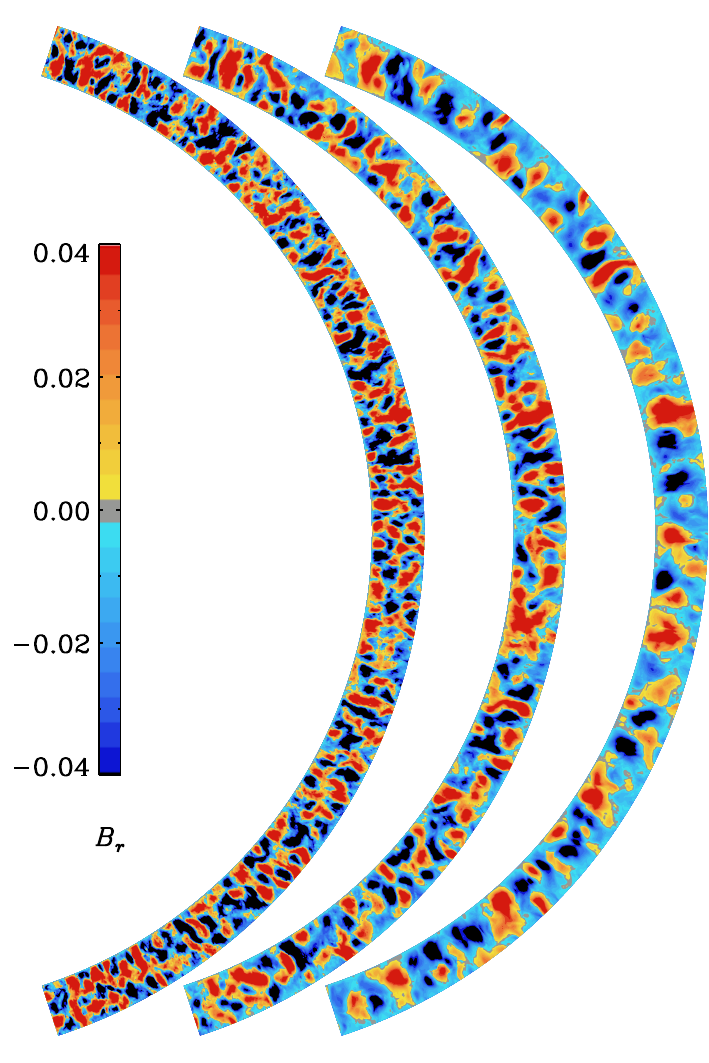}
\end{minipage}
  \caption{Evidence for inverse Hall cascade in magneto-thermal simulations with a non-zero initial net magnetic helicity. Left: Magnetic energy spectra at $0.11\,\tau_{Ohm}$ (black), $0.16\,\tau_{Ohm}$ (blue), $0.20 \,\tau_{Ohm}$ (yellow), and $0.26 \,\tau_{Ohm}$ (red). Right: Meridional slices of $B_r(r,\theta)$ for $\ell_0=200$, shown at $0.03 \,\tau_{Ohm}$, $0.11 \,\tau_{Ohm}$, and $0.16 \,\tau_{Ohm}$ (left to right). Times are given in units of $\tau_{Ohm} = \xi^2/\eta$, where $\xi$ is the characteristic length scale of the magnetic structures. Image reproduced with permission from \citet{DehmanBrandenburg25}, copyright by the author(s).
}
  \label{fig:IC results}
\end{figure}

An alternative scenario suggests that a newborn NS, initially permeated by small-scale turbulent magnetic structures, may reorganize its field into an ordered dipole. When the system possesses significant magnetic helicity, the nonlinear Hall term favors a direct rather than an inverse cascade (see Sect.~\ref{sec: Hall drift}). 
The first study of this process in NS crusts was performed in a box setup by \citet{brandenburg2020}, 
who demonstrated that the inverse cascade can shift the peak of the magnetic energy spectrum toward slightly larger scales, thereby amplifying the dipolar component to magnetar strengths. A more realistic study, incorporating the NS structure, the thin-crust aspect ratio, and detailed microphysics, showed that the Hall term, combined with initial non-zero net magnetic helicity, can trigger an inverse cascade. However, its efficiency is severely constrained by the extreme aspect ratio of the crust \citep{DehmanBrandenburg25}. 
In fact, the cascade is limited to multipoles $\ell \lesssim \mathcal{A}^{-1} \sim 30$, where $\mathcal{A} \approx 1/30$ denotes the crust aspect ratio (Fig.~\ref{fig:IC results}).
This occurs because nonlinear mode couplings halt the inverse cascade once its peak scale approaches the crust thickness. As a result, according to these first studies, the Hall-driven inverse cascade can transfer energy only into moderately low multipoles ($\ell \sim 10$–20), but it cannot generate the very large-scale dipole characteristic of magnetars.

\begin{figure}
    \centering
    \includegraphics[width=0.9\linewidth]{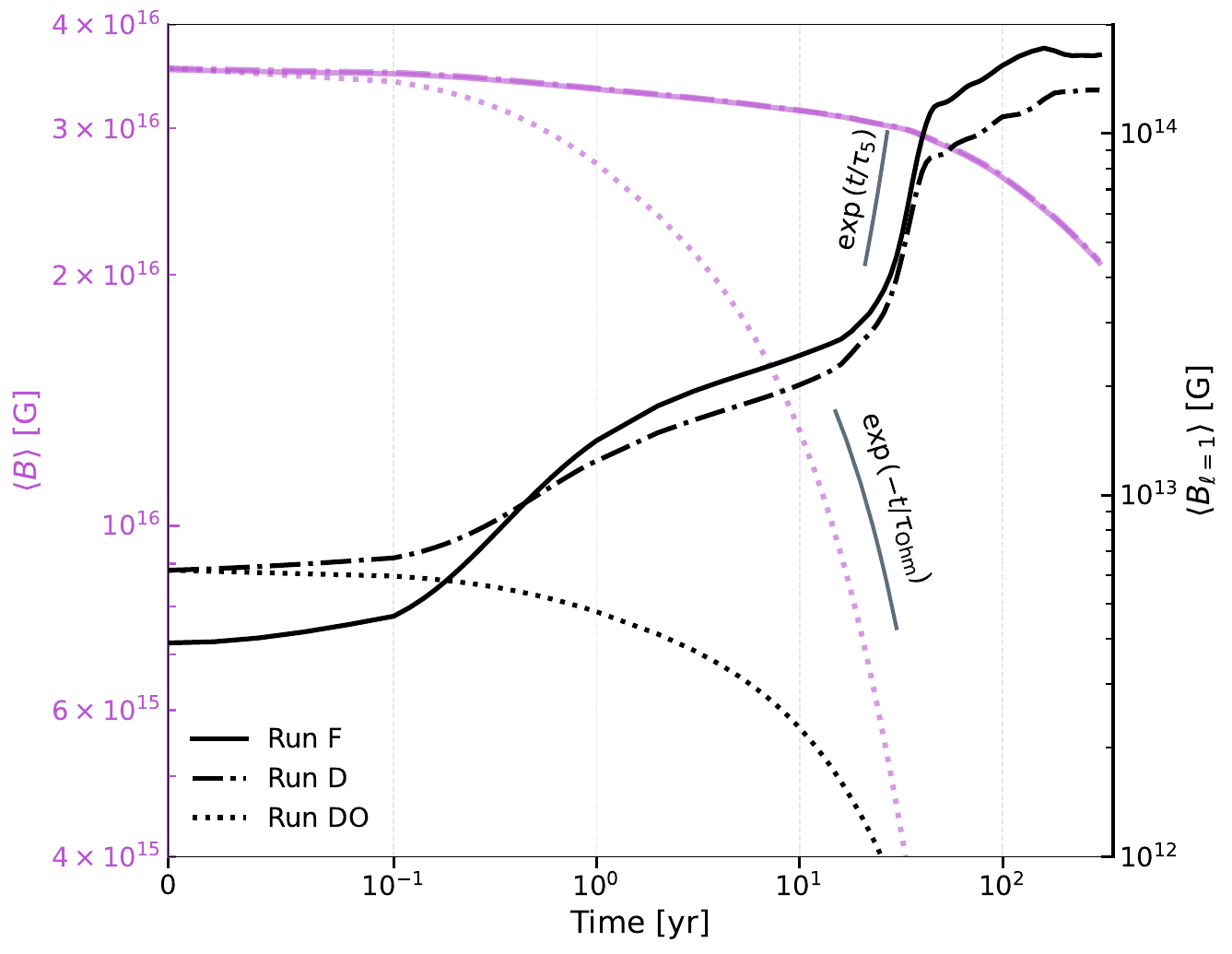}
    \caption{Time evolution of the average magnetic field (mauve, left axis; $4 \times 10^{15}$--$4 \times 10^{16}$ G) and dipolar field (black, right axis; $10^{12}$--$2 \times 10^{14}$ G) for representative simulations including CME. Solid and dash-dotted lines (Run~F and D, respectively) correspond to simulations with the CME active but different initial conditions, while dotted lines (Run~DO) show the purely Ohmic case (CME switched off). Gray lines are fits to the growth and decay phases, with $\tau_\mathrm{Ohm} \equiv 1/\eta k^2 \approx 20$–25 yr and $\tau_5 \equiv 1/\eta k k_5 \approx 5$--10 yr. Image reproduced with permission from \citet{DehmanPons25}, copyright by the author(s).} 
    \label{fig:CME result}
\end{figure}

Beyond the nonlinear Hall term, magnetic helicity conservation becomes especially important when the chiral term is included in the induction equation (see Sect.~\ref{sec: CME}). \citet{DehmanPons25} present 3D magneto-thermal simulations with \MATINS in which they show that, with the CME term, the dipolar component of the field can grow to magnetar strengths within 50--100 years after birth (see Fig.~\ref{fig:CME result}).
This work shows that a strong turbulent field with encoded  magnetic helicity can source and maintain a tiny chiral imbalance, with differences in chemical potentials of the order $\sim 10^{-11}$ MeV. The imbalance is sustained over some decades, acting as a catalyst and driving an efficient inverse-cascade-like mechanism. 

\subsection{Influence of boundary conditions on the long-term evolution}
\label{sec: impact of BC}

Despite their importance, boundary conditions often receive insufficient attention, even though different choices can significantly impact the interior evolution, affecting the interpretation of results and their alignment with observational data.
In the particular case of considering a crustal evolution of the magnetic fields, one must choose the boundary conditions at both the crust-core and crust-magnetosphere interfaces, and they have an impact.

\begin{figure}
	\includegraphics[width=.49\textwidth]{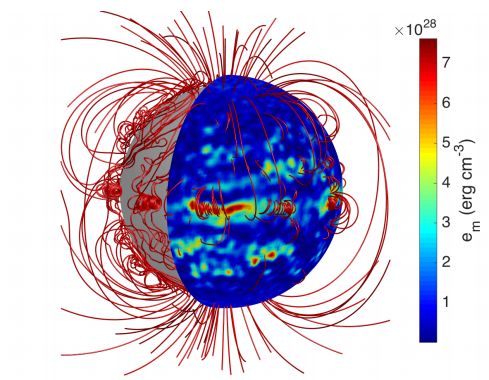}
	\includegraphics[width=.49\textwidth]{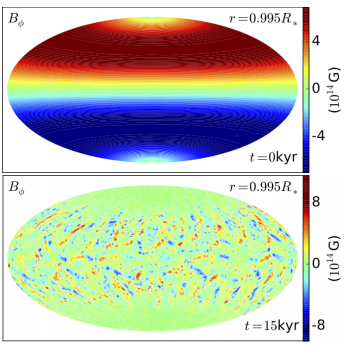}
	\caption{Left: Magnetic field lines and magnetic energy density maps on the star surface (in colors), at $t=15$ kyr, for an initial model consisting of an $l=1$ poloidal field, and $l=2$ toroidal field, plus a small non-axisymmetric perturbation. Right: Contour plot of the azimuthal component of the magnetic field at $r = 0.995R_\star$, with $R_\star$ being the star radius, for the same model. Images reproduced with permission from \citet{gourgouliatos16}, copyright by the author(s). }
	\label{fig:kostas_3d}
\end{figure}

A first example is the Hall instability. As initially proposed by \citet{RG2002} and later confirmed in 2D simulations \citep{Pons2010}, the occurrence of the instability is closely linked to the choice of boundary conditions, background magnetic field and the aspect ratio of the crust (similarly to the inverse cascade discussed in the previous subsection). The first 3D simulations of crustal-confined fields \citep{wood15,gourgouliatos16}, with an exterior boundary condition consisting of a general potential solution, reinforced this idea. The temperature was not included in the simulations, and the resistivity and density profiles were prescribed as analytical functions, fitted to mimic a realistic model at $T=10^8$ K \citep{cumming04}. 
These 3D studies show new dynamics and the creation of km-size magnetic structures persistent over long timescales. Even using initial axisymmetric conditions, the Hall instability breaks the symmetry and new 3D modes quickly grow, but the dominant growing modes are of the order of the crust thickness, as a result of the boundary conditions. This was confirmed in \citet{gourgouliatos19}. A typical model is shown in Fig.~\ref{fig:kostas_3d}. The surface field is highly irregular, with small regions in which the magnetic energy density exceeds by at least an order of magnitude the average surface value. By exploring many different initial models, \citet{gourgouliatos16} found that magnetic instabilities can efficiently transfer energy to small scales, which in turn enhances Ohmic heating and powers the persistent emission, confirming the 2D results. Similarly, \citet{gourgouliatos18} explored magnetic field configurations that lead to the formation of magnetic spots on the surface of NSs, extending previous 2D works \citep{geppert14}. They show how an ultra-strong initial toroidal component is essential for the generation of a single spot, possibly displaced from the dipole axis, which can survive on very long timescales. 

These simulations, however, adopted potential boundary conditions, which suppress helicity transfer into the magnetosphere. Different results could be expected with, for example, force-free boundary conditions. To date, only \citet{akgun18b} and \citet{Urban23} have carried out simulations that couple the interior evolution with a magnetospheric model including electric currents originated from the star internal evolution. Both works assume axial symmetry. They couple the interior evolution with a force-free magnetospheric configuration requires. \citet{akgun18b} solved the elliptic equation on an extended grid reaching far beyond the stellar surface to recover the correct asymptotic behavior (see Sect.~\ref{sec:forcefree}). This is computationally demanding, often requiring tens of thousands of iterations for each configuration prescribed by the interior evolution. Recently, this limitation has been partially mitigated through novel PINN-based approaches, which \citet{Urban23,Stefanou23b} showed to provide an efficient alternative.

\begin{figure}[ht]
    \centering
    \includegraphics[width=.49\textwidth]{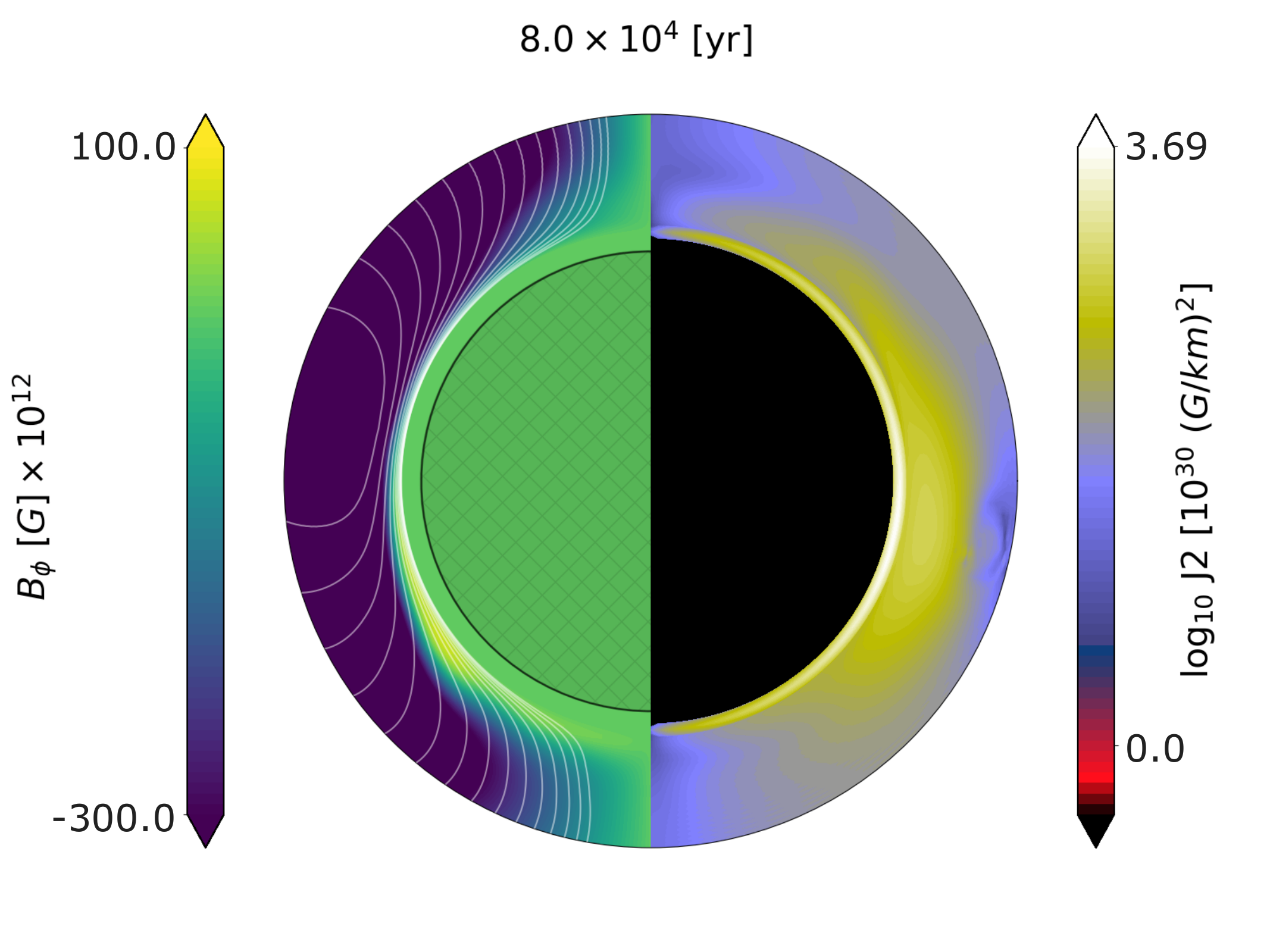}
    \includegraphics[width=.49\textwidth]{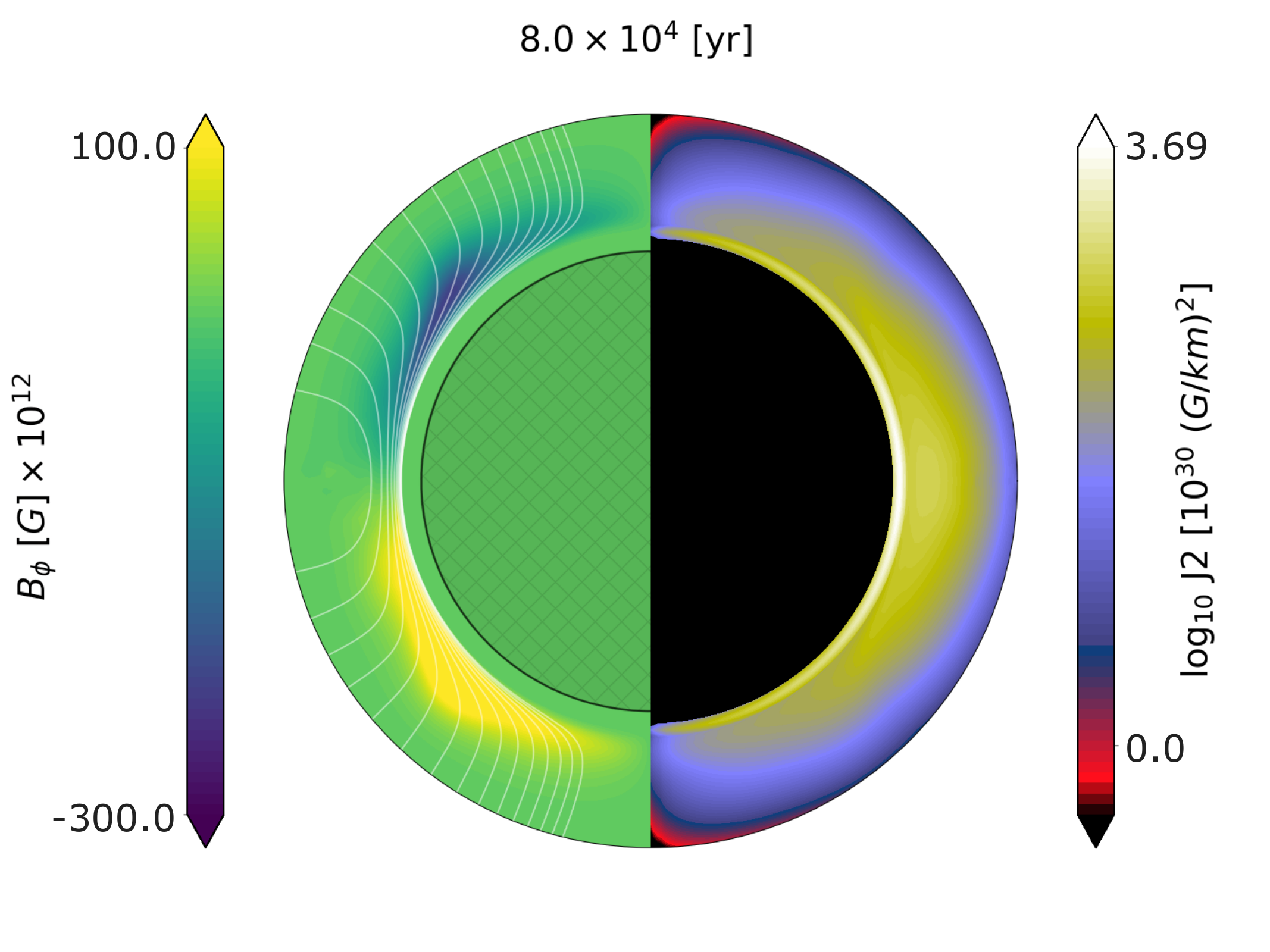}
    \caption{Snapshot of the magnetic field and electric current at $80$ kyr. The left hemisphere shows the meridional projection of the magnetic field lines (white) and the toroidal component (colors), while the right hemisphere displays the squared modulus of the electric current, $|J|^2$ (log scale). The crust is enlarged by a factor of 8 for clarity. Left panel: force-free boundary conditions. Right panel: vacuum boundary conditions.
    Image reproduced with permission from \citet{Urban23}, copyright by the author(s).}  
    \label{fig: B field PINN FF vs VAC}
\end{figure}

Figure~\ref{fig: B field PINN FF vs VAC} compares the results from crust-confined simulations adopting different boundary conditions: force-free (left panel), or potential (right). Both simulations use identical initial models: a force-free magnetic field with a poloidal surface strength of $3\times10^{14}$ G at the pole and a maximum toroidal field of $3\times10^{14}$ G. The outcomes at late times (80 kyr in the plot) differ significantly. With force-free boundary conditions, a stronger toroidal dipole forms near the surface, connected to the magnetosphere and pushing poloidal field lines, slightly shifted northward. In contrast, vacuum boundary conditions maintain approximate equatorial symmetry in the poloidal field, with the dominant toroidal component being quadrupolar and concentrated near the crust–core interface.
Current distributions also vary: vacuum boundary conditions suppress currents near the poles and surface, while force-free boundary conditions permit non-zero surface currents. The yellowish region in the left panel (northern mid-latitudes) shows significant current flowing into the magnetosphere. These differences have a large impact on the surface temperature, since currents are forced to pass through the highly dissipative envelope, as noted by \citet{akgun18b}.

\begin{figure}[h]
\includegraphics[width=\textwidth]{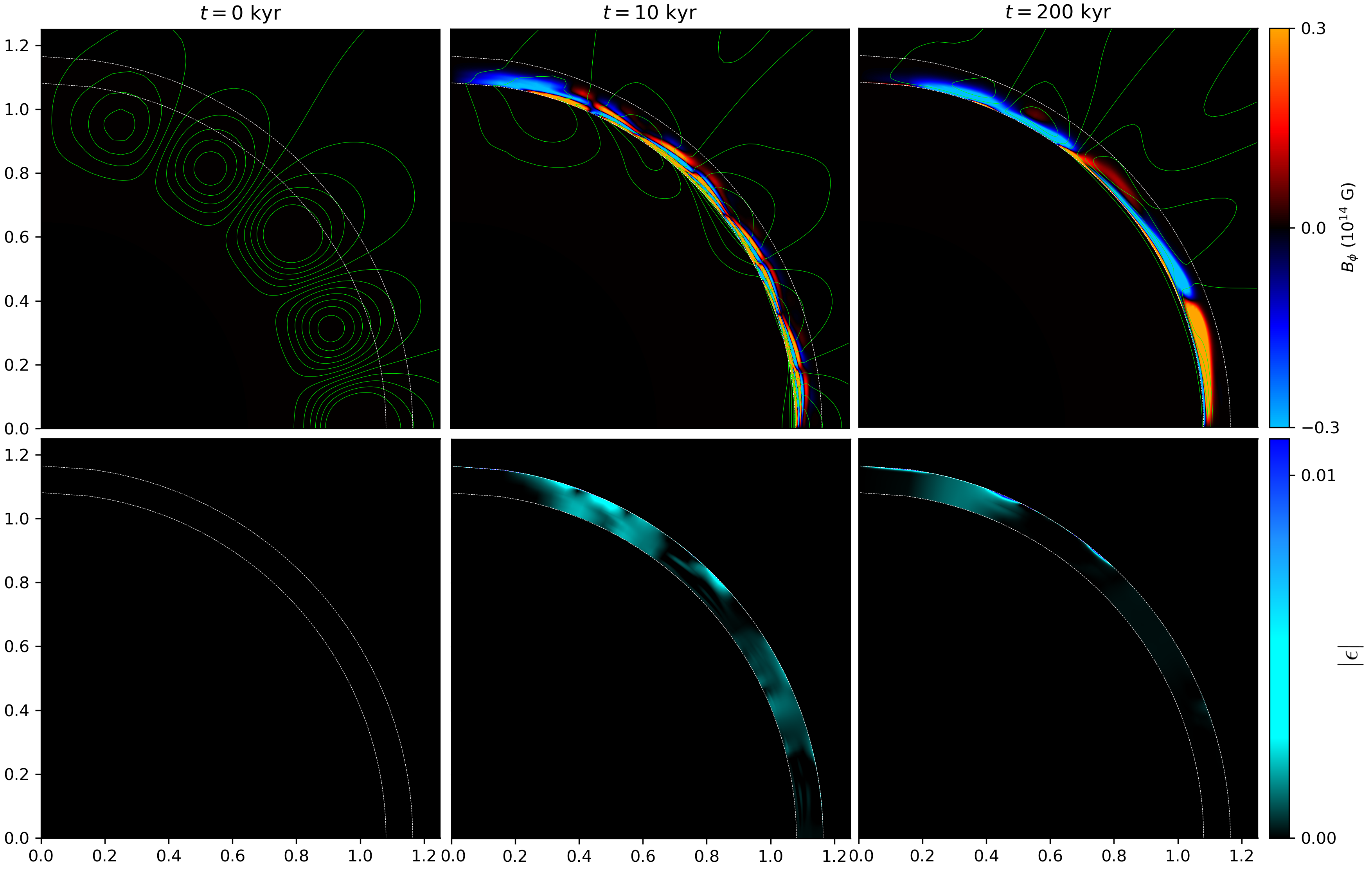}
\caption{Simulations involving a crust-core magnetic evolution coupling. Top panels: Snapshots of the magnetic field configuration at $t=0$~kyr, $t=20$~kyr, and $t=200$~kyr. Green curves show poloidal magnetic field lines, and color shows $B_\phi$ in units of $10^{14}$~G. Bottom panels: Snapshots of the von Mises strain $|\epsilon | = \sqrt{\frac{1}{2}\epsilon_{ij}\epsilon_{ij}}$ at the same times. The crust-core interface and the surface are indicated by the inner and outer dashed white curves, respectively. The axis show distance in units of $10^6$~cm. Image reproduced with permission from \citet{bransgrove25}, copyright by the author(s).
}
\label{fig: Hall wave}
\end{figure} 

A third example is connected to the treatment of the crust–core boundary. Most simulations ignore the core magnetic field and its influence on crustal magnetic-field evolution and impose a perfect-conductor (Meissner/type-I) boundary for simplicity. In a type-II superconducting core, however, magnetic flux threads the fluid as quantized tubes between the lower and upper critical fields, $H_{c1}$ and $H_{c2}$. For $B>H_{c2}$, superconductivity breaks down and the medium becomes a classic magnetized fluid. Importantly, even for $B<H_{c1}$, pre-existing flux tubes might persist due to pinning and slow drift; a true Meissner state may only be reached on longer, uncertain timescales. The efficiency of flux expulsion and the resulting core-crust coupling, and thus their impact on long-term crustal evolution, remain open issues.

Generally, the core evolution is anticipated to be slower than that of the crust due to its significantly higher conductivity, so the crustal-confined findings discussed earlier are expected to remain qualitatively valid. Nonetheless, more realistic inner boundary conditions that account for the magnetic field threading the core cannot be overlooked.

In a recent study, \citet{bransgrove25} model the evolution within a crust-like domain under the influence of Ohmic and Hall effects, introducing a novel inner boundary condition that accounts for a type-II superconducting core. Their approach simplifies the treatment by incorporating angular advection of magnetic field lines in the azimuthal direction at the inner boundary, while neglecting radial and meridional velocities at the crust-core interface.
The interior is approximated to be in hydromagnetic equilibrium, achieved through a relaxation method. Despite these simplifications, the study reveals significant new insights. They find that spin-down-driven advection can drive magnetic flux into the crust, generating strong interface currents and triggering Hall waves from the crust-core boundary (see Fig.~\ref{fig: Hall wave}).
With rapid initial rotation ($P \sim 10$ ms), the vortex–flux-tube coupling efficiently reorganizes core flux; by $\sim 10$ kyr much of the flux has been advected into the crust, amplifying an initial large-amplitude Hall pulse. Strong vortex–flux-tube interactions produce stronger interface currents and, consequently, stronger Hall waves, as the core is actively depleted of flux and approaches a Meissner-like state ($B=0$) on the spin-down timescale. The simulations also suggest that Hall waves might be sufficiently powerful to fracture the crust, potentially leading to starquakes that induce rotational glitches or other observable alterations in the spin-down properties. Additionally, these Hall waves interact with gradual magnetospheric changes, naturally resulting in braking indices $n \neq 3$ due to the time-dependent dipole moment \citep{pons12b,gourgouliatos15a}.


Thermal boundary conditions play a critical role in determining cooling timescales, as they govern energy losses through surface photon emission. Notably, significant differences arise when comparing non-magnetized envelope models \citep{gudmundsson83,potekhin97} with magnetized ones \citep{potekhin03,potekhin_rev15a}, for both light-element (hydrogen) and heavy-element (iron-like) compositions.
Light-element envelopes, commonly used for accreting sources, produce luminosities up to an order of magnitude higher than heavy-element envelopes during the neutrino-cooling phase. Due to the efficient surface photon losses, light-element envelope models cool down faster once the photon-cooling era begins, rendering the objects hardly detectable in terms of thermal $X$-rays ($L_X \lesssim 10^{32}$ erg/s), much before than the heavy-element models. This effect is further amplified by strong magnetic fields, 
as first shown by \citet{page1996} and more recently confirmed by \citet{Dehman23b}. They showed that the predicted thermal X-ray luminosity varies significantly based on multiple factors, particularly the envelope's properties (see Sect.~\ref{subsection:envelope} for details). Their findings are summarized in Fig.~\ref{fig: comparison different env}, which compares outcomes under different assumptions about the magnetic field, composition (iron versus light elements like hydrogen, typical in accreted envelopes), or the internal current distribution (core-threading field versus crustal-confined field).

\begin{figure}[h]
\includegraphics[width=\textwidth]{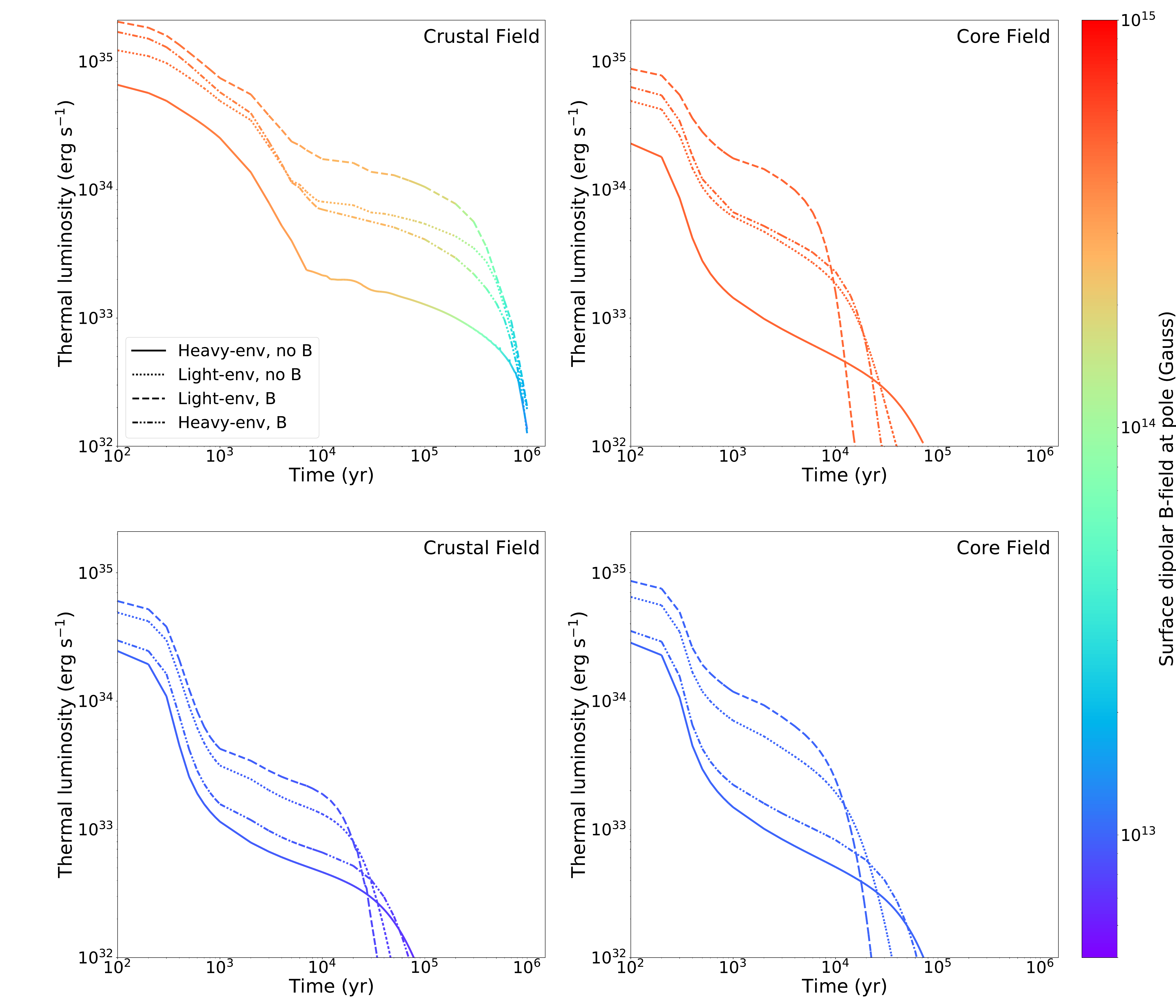}
\caption{Luminosity curves for four envelope models: non-magnetised heavy envelopes (solid, \citealt{gudmundsson83}), non-magnetised light envelopes (dots, \citealt{potekhin97}), magnetised light envelope (dashes, \citealt{potekhin03}), and magnetised heavy envelope (dot-dashes, \citealt{potekhin_rev15a}). Left panels show models with crust-confined magnetic fields, while right panels show those with core-dominated magnetic fields, in all cases consisting of initial large-scale components only. Results are presented for two initial polar field strengths: $B = 5 \times 10^{14}$ G (top panels) and $ B = 10^{13}$ G (bottom panels). Image reproduced with permission from \citet{Dehman23b}, copyright by the authors. }
\label{fig: comparison different env}
\end{figure}

\subsection{Late-time evolution.}
\label{subsec:3.1}
Beyond initial conditions, early-time ($t \lesssim 100$ yr) field reshaping and boundary-condition effects, the key question already mentioned in the previous section is arguably the location of the electric currents sustaining the magnetic field. Crustal-confined fields have been extensively studied and many works \citep{PonsGeppert2007,vigano12a,vigano13,gourgouliatos13,gourgouliatos14a,gourgouliatos14b,vigano21,gourgouliatos19,Degrandis20,Degrandis22,MATINS1,MATINS_MT} generally agree on the overall picture of the Hall-driven dynamics in these configurations. 

For typical field strengths of $10^{14}$ G, and starting from a predominantly poloidal dipolar field, we observe a stage dominated by the Hall drift (readjusting from initial conditions), which creates higher-order multipoles, followed by a quasi-stationary Ohmic stage. This structure, which has been called the Hall attractor \citep{gourgouliatos14b,bransgrove18},
is characterized by a nearly constant angular velocity of the "electron" fluid ($\Omega \approx j/e n_e r$) along each poloidal field line, and proportional to the magnetic flux. 
It is worth noting that Hall drift can significantly accelerate magnetic field dissipation by steadily channeling magnetic energy to smaller scales, where Ohmic dissipation is more efficient.

\begin{figure}[t]
	\centering
	\includegraphics[width=.32\textwidth]{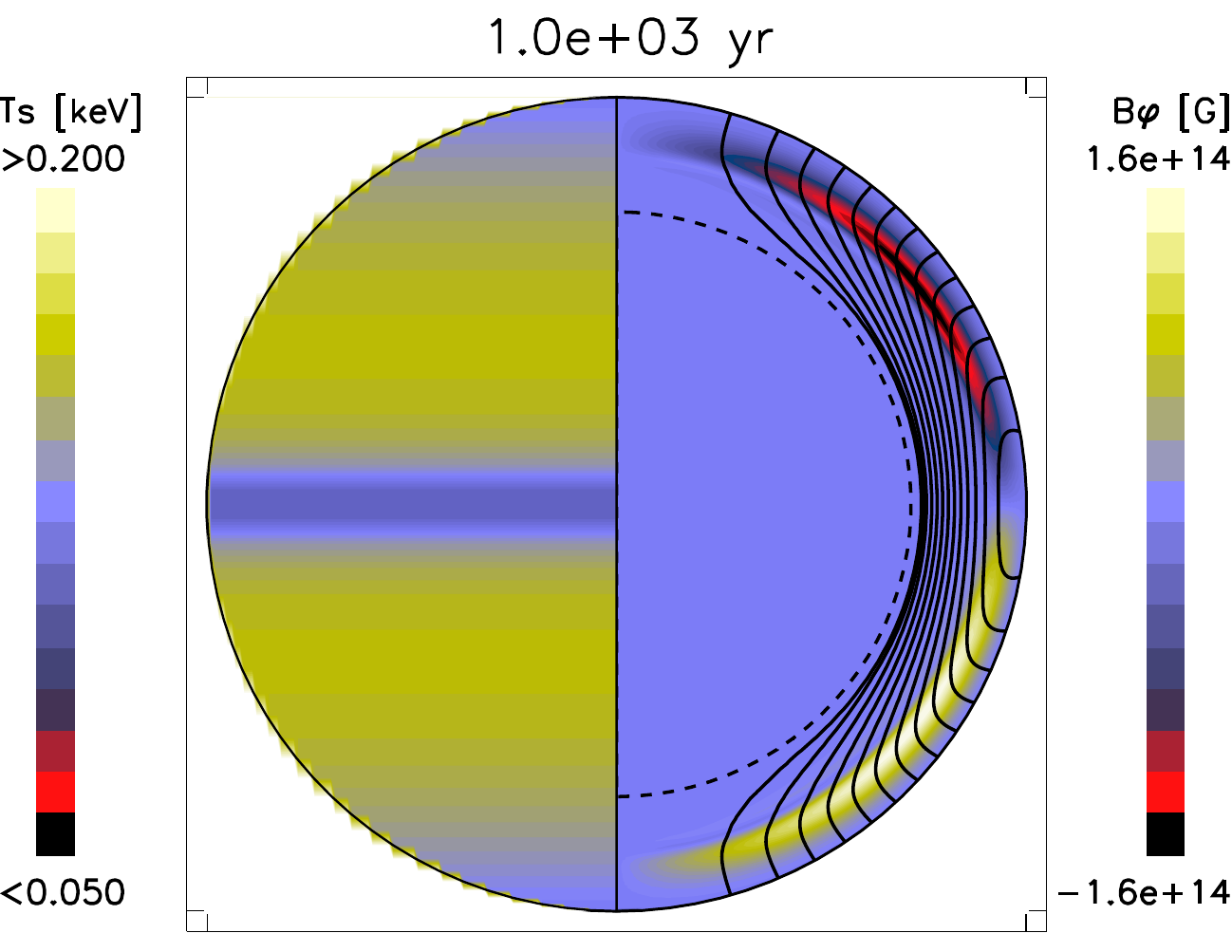}
	\includegraphics[width=.32\textwidth]{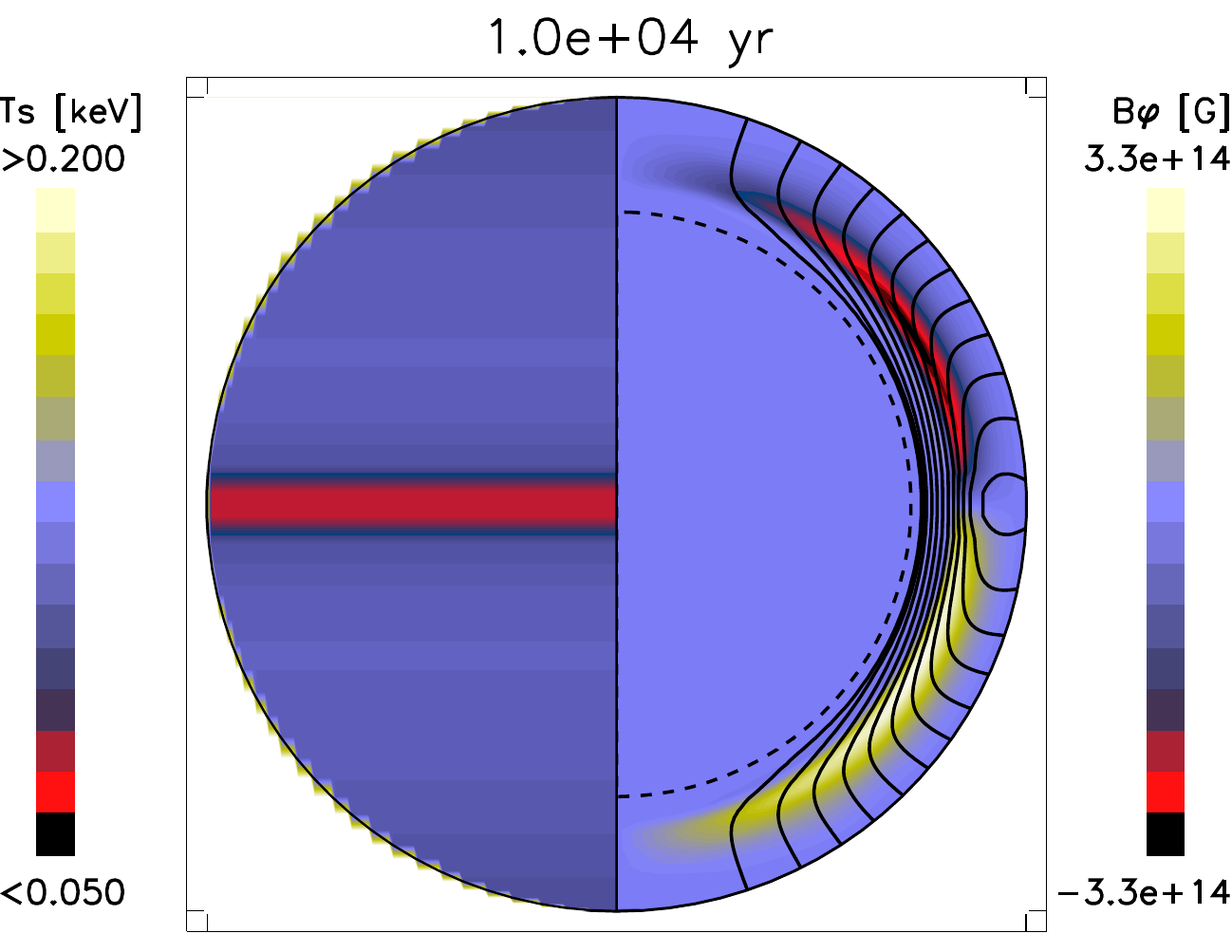}
	\includegraphics[width=.32\textwidth]{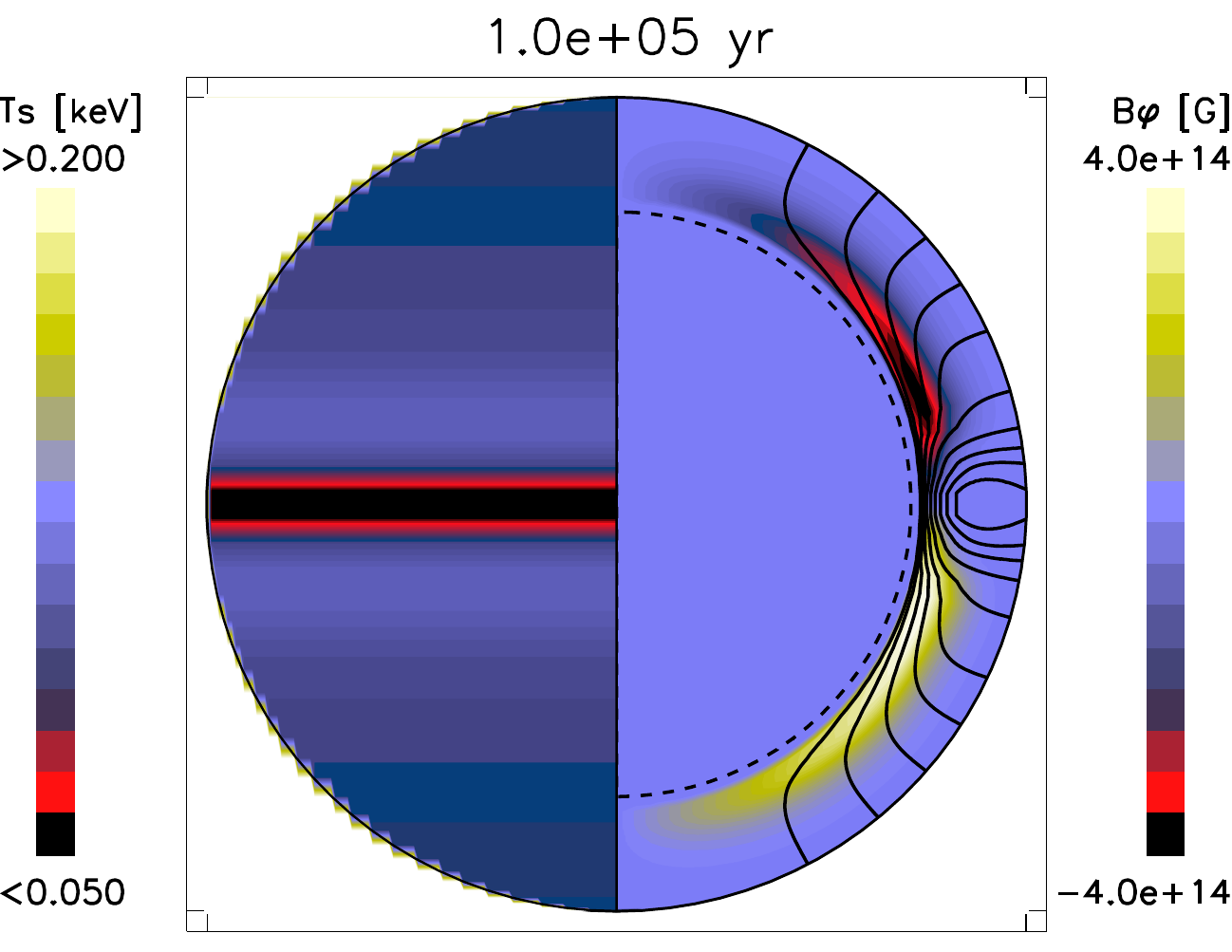}\\
	\vskip0.2cm
	\includegraphics[width=.32\textwidth]{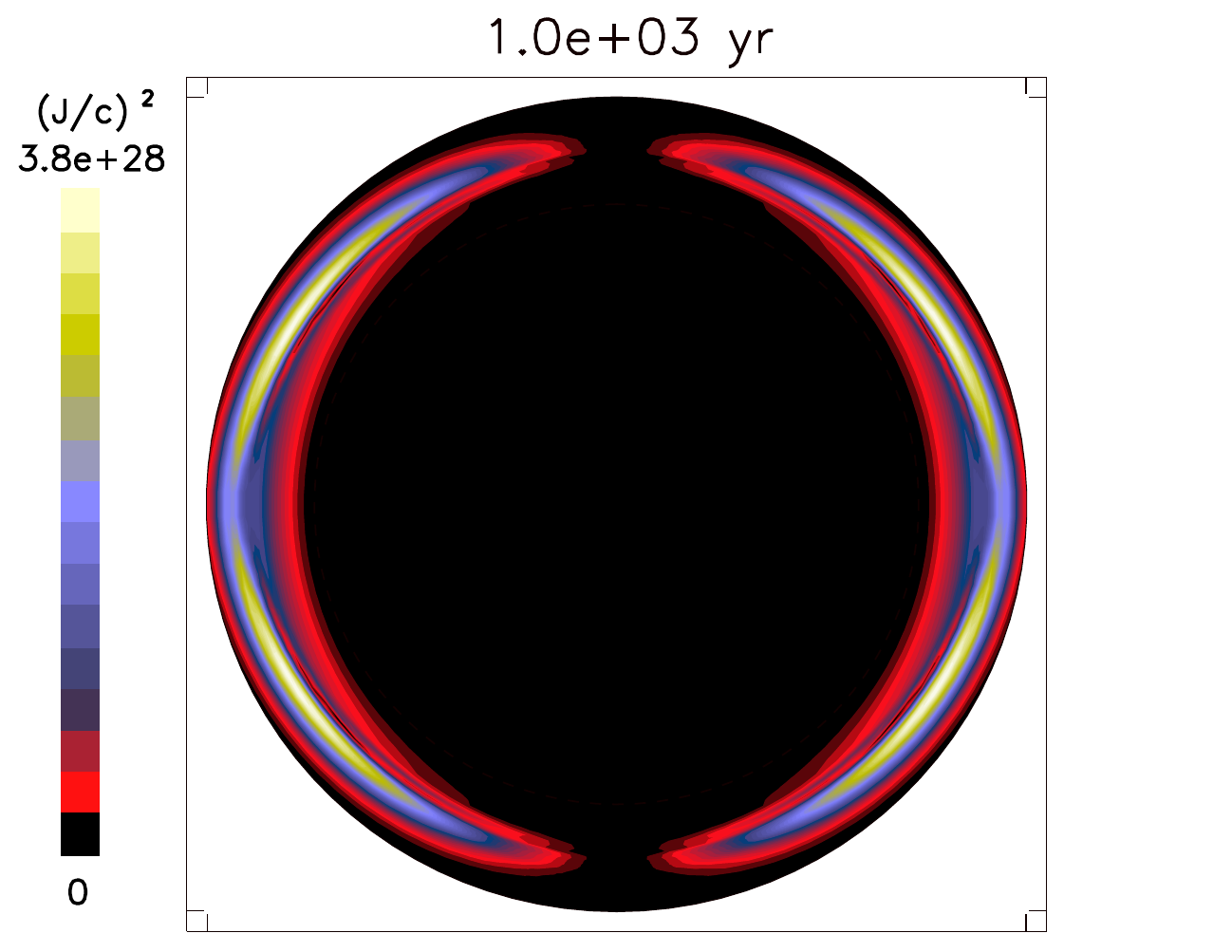}
	\includegraphics[width=.32\textwidth]{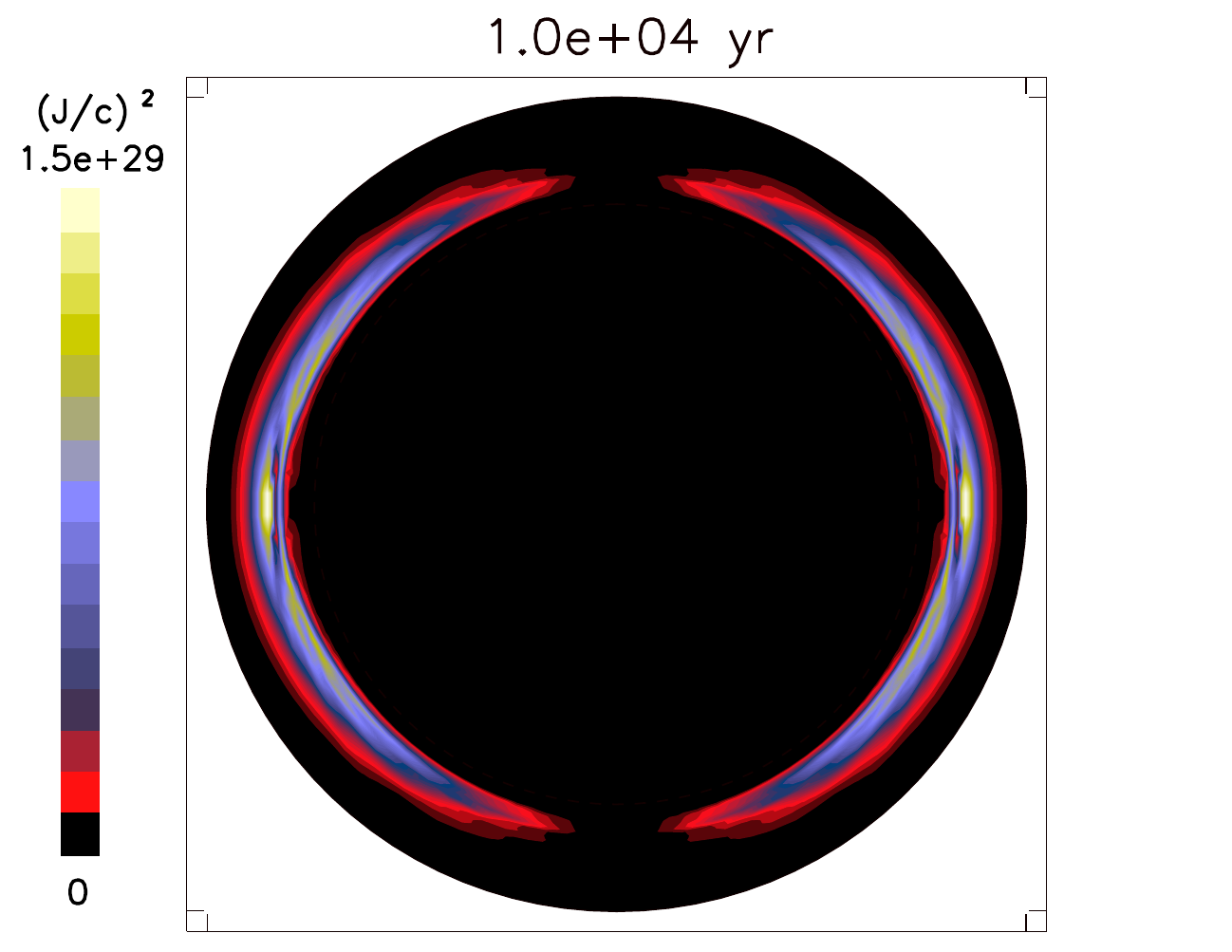}
	\includegraphics[width=.32\textwidth]{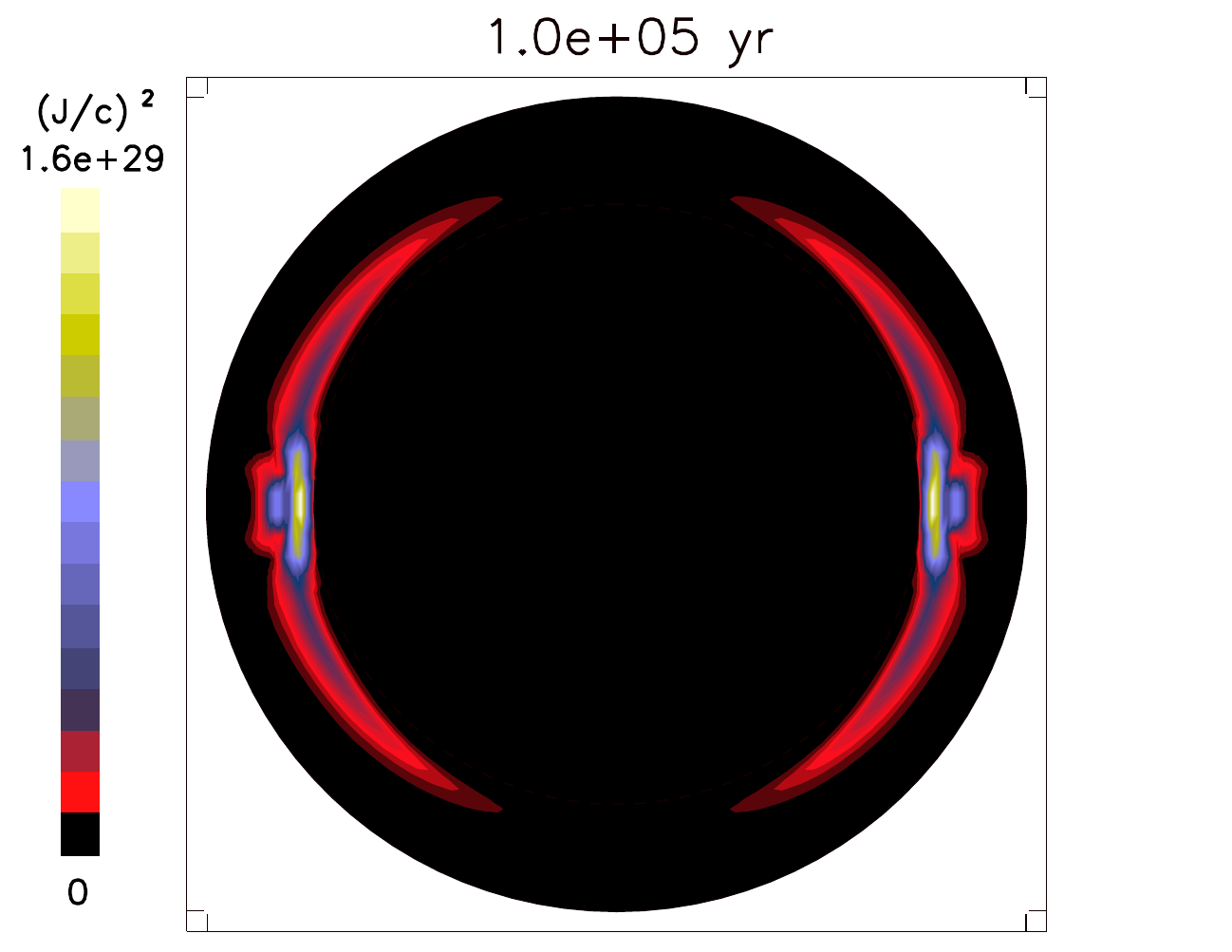}\\
	\vskip0.2cm
	\includegraphics[width=.32\textwidth]{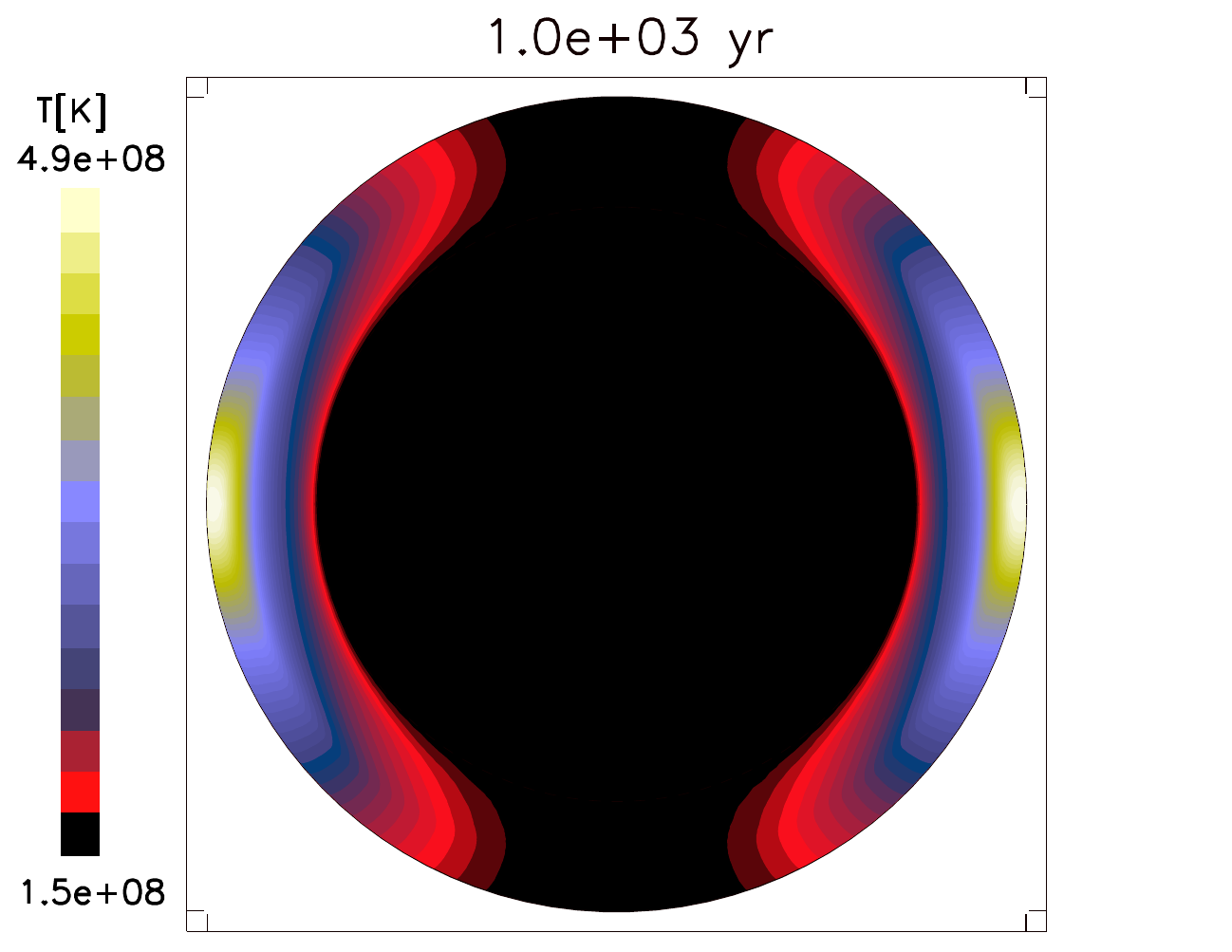}
	\includegraphics[width=.32\textwidth]{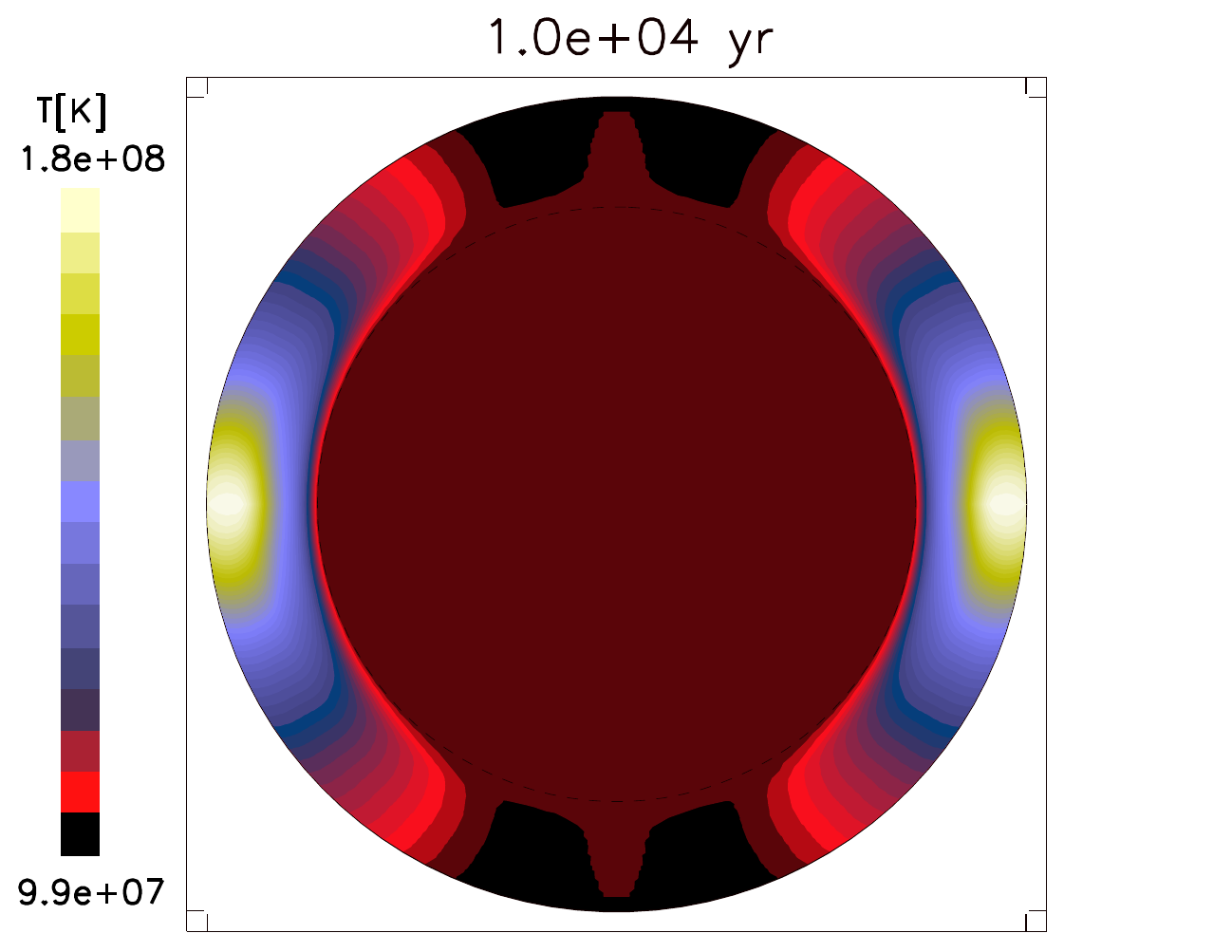}
	\includegraphics[width=.32\textwidth]{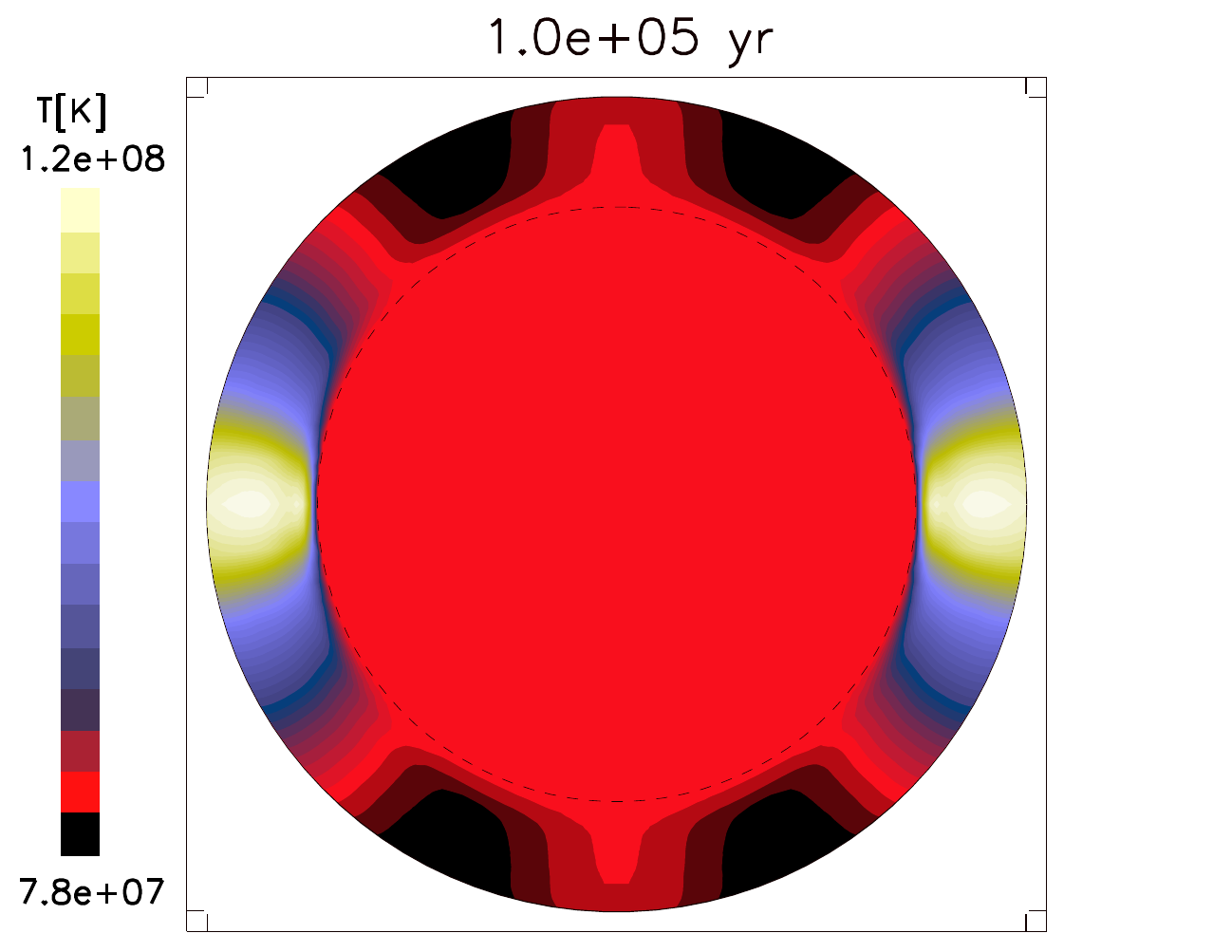}
	\caption{Snapshots of the magneto-thermal evolution of a NS model at $10^3, 10^4, 10^5$ yr, from left to right. Top panels: the left hemisphere shows in color scale the surface temperature, while the right hemisphere displays the magnetic configuration in the crust. Black lines are the projections of the poloidal field lines and the color scale indicates the toroidal magnetic field intensity (yellow: positive, red: negative). Middle panels: intensity of currents; the color scale indicates $J^2/c^2$, in units of $($G/km$)^2$. Bottom panels: temperature map inside the star. In all panels, the thickness of the crust has been enlarged by a factor of 4 for visualization purposes. Figure courtesy of \citet{vigano13}. Animations available in the supplementary material.}
	\label{fig:b14_evo}
\end{figure}

As an example, in Fig.~\ref{fig:b14_evo} we show three snapshots of the evolution of a simple crustal-confined axisymmetric model, initially a $\ell=1$ poloidal field with $B_p=10^{14}$ G (labeled as model A14 in \citealt{vigano13}). Such very simple initial configuration allows one to analyse and capture several general basic features characterizing the Hall-dominated dynamics. Let us recap the most important facts:
\begin{itemize}
\item The Hall term initially links the poloidal and toroidal magnetic field components, causing the rapid emergence of a toroidal field even if it starts at zero. Within approximately $10^3$ years, a quadrupolar toroidal magnetic field forms, reaching a strength comparable to the poloidal field, with $B_\varphi$ negative in the northern hemisphere and positive in the southern hemisphere.

\item Subsequently, the Hall drift dominates the evolution, driven by the toroidal magnetic field, which pulls currents deeper into the inner crust (as shown in the middle panels) and compresses magnetic field lines. The Hall term redistributes energy from the large-scale dipole to smaller scales, where higher-order multipoles become locally intense, potentially forming current sheets, particularly at the equator.

\item In regions with sufficiently small-scale structures, enhanced local ohmic dissipation counteracts the Hall drift, leading to a quasi-stationary state resembling the Hall attractor. After about $10^5$ years, the toroidal magnetic field is predominantly confined to the inner crust.

\item At this stage, most of the current flows near the crust/core interface, where magnetic energy dissipation is governed by the resistivity of this region. In this particular model, a highly resistive layer in the nuclear pasta region causes rapid magnetic field decay, directly affecting the observable rotational properties of X-ray pulsars \citep{pons13}.

\item Joule heating alters the temperature distribution. As shown in the bottom panels of Fig.~\ref{fig:b14_evo}, at $t=10^3$ years, the equator is approximately three times hotter than the poles due to the insulating effect of the strong magnetic field, as discussed in \S \ref{sec:T_anis}. Strong tangential components ($B_\theta$ and $B_\varphi$) insulate the surface from the interior. In a dipolar geometry, the magnetic field is nearly radial at the poles, maintaining thermal connection with the interior, while tangential field lines insulate the equatorial region. This creates a dual effect: if the core is warmer than the crust, the poles are hotter than the equator; however, if ohmic dissipation heats the equatorial region, the temperature distribution reverses, reflecting the poloidal magnetic field geometry that guides heat flow.

\end{itemize}

In the supplementary material, we provide the animations of two models with the same initial dipolar poloidal magnetic field with $B_p=10^{14}$ G and the same maximum intensity of the toroidal field, $B_{\rm tor}=10^{15}$ G, but differing in the geometry of the initial toroidal field ($\ell=1$ or $\ell=2$).


\begin{figure}
	\centering
	\includegraphics[width=.8\textwidth]{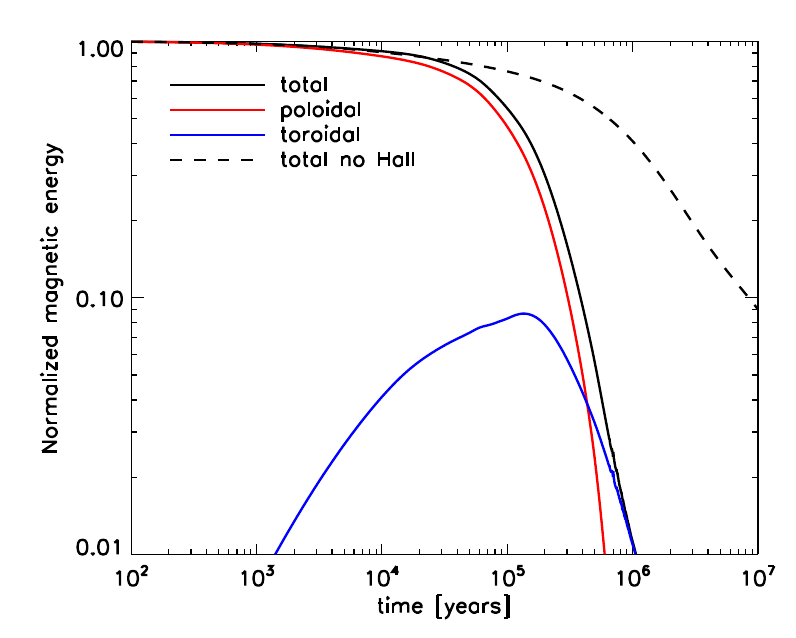}
	\caption{Magnetic energy in the crust (normalized to the initial value) as a function of time, for the same model of Fig.~\ref{fig:b14_evo}. The solid lines correspond respectively to the total magnetic energy (black), the energy in the poloidal component (red), and the energy in the toroidal component (blue). The dashed line shows the evolution of the same model when the Hall term is deactivated (only Ohmic dissipation). Image reproduced with permission from \citet{vigano13}, copyright by the authors.}
	\label{fig:entransf}
\end{figure}

In order to show more clearly the enhanced dissipation caused by the combined action of Hall and Ohmic terms, in Fig.~\ref{fig:entransf} we show the evolution of the total magnetic energy stored in each component, comparing the evolution of the previous model with another model with the same initial data but switching off the Hall term (purely resistive case). In this case, there is no creation of a toroidal magnetic field or smaller scales. When the Hall term is included, $\sim 99\%$ of the initial magnetic energy is dissipated in the first $\sim 10^6$ yr, compared to only the $60\%$ in the purely resistive case. At the same time, a $\sim 10\%$ of the initial energy is transferred to the toroidal component in $10^5$ yr, before it begins to decrease. Note that the poloidal magnetic field, after $10^5$ yr, is dissipated faster than the toroidal magnetic field. The poloidal magnetic field is supported by toroidal currents concentrated in the inner, equatorial regions of the crust. Here the resistivity is high for two reasons: the effect of the nuclear pasta phase, and the higher temperature (see right bottom panel of Fig.~\ref{fig:b14_evo}). Conversely, the toroidal magnetic field is supported by larger loops of poloidal currents that circulate in higher latitude and outer regions, where the resistivity is lower. As a result, at late times most of the magnetic energy is stored in the toroidal magnetic field. This example is very illustrative of the importance of knowing in detail the geometry of the field and the location of currents at different stages.

In 3D, details become even more relevant. Fig.~\ref{fig:3D_rendering} highlights the role of complex, multipolar magnetic structures close to the star surface in producing anisotropies in the temperature evolution. Such anisotropies have direct implications for the observable surface emission, potentially biasing cooling-age estimates and the interpretation of $X$-ray spectra, if oversimplified. This underscores the need for careful multi-dimensional treatments of the heat diffusion equation, rather than relying on one-dimensional cooling models \citep{Degrandis21,Igoshev21,Igoshev23b,MATINS_MT,MATINS2}.

\begin{figure}
    \centering
    \includegraphics[width = \textwidth]{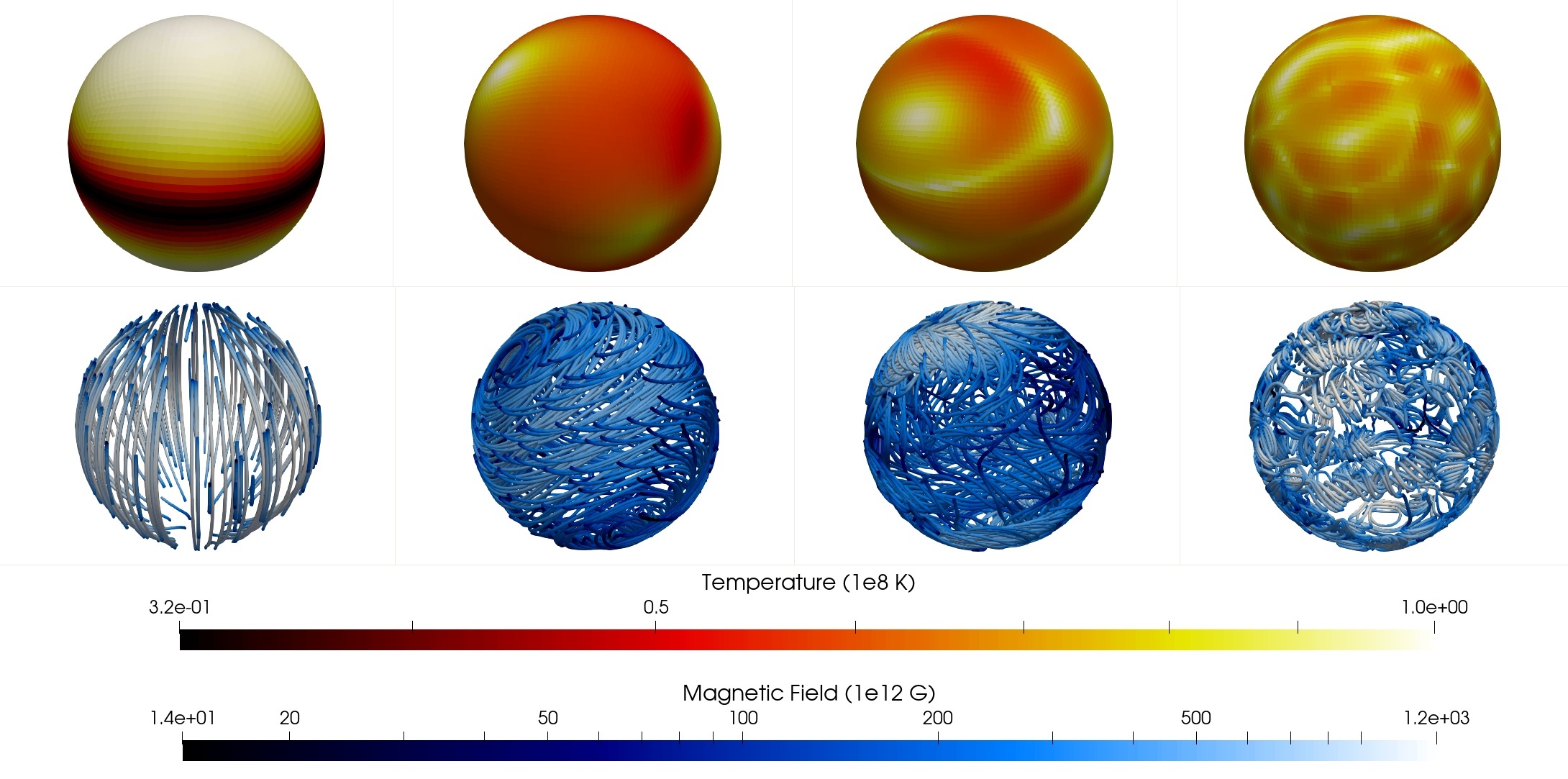}
    \caption{3D visualizations of different magnetic field configurations and their impact on the thermal surface distribution. Top: Temperature maps at the base of the envelope. Bottom: Magnetic field lines, with colors indicating field intensity (here not evolved). Simulations assumed a NS mass of $1.4 M_\odot$ with the SLy4 EoS (see Fig.~\ref{fig:ns_profile}). Image reproduced with permission from \citet{MATINS2}, copyright by the author(s).}
    \label{fig:3D_rendering}
\end{figure}

Core-threading configurations are less well understood due to the complex physics within the inner core of NSs. Two main distinctions exist between crustal-confined and core-threading configurations: first, the field curvature of large-scale components differs by about an order of magnitude, corresponding to the star’s size versus the crust’s thickness; second, with the two regions having significantly different conductivities (see Fig.~\ref{fig:cond_el}), the location of currents determines where Ohmic dissipation occurs and hence the timescale. As we have already observed in Fig.~\ref{fig: comparison different env}, the weaker Joule heating effect leads to core-threading models cooling much faster after the neutrino cooling era ($t \gtrsim 10^4$ yr). 
Thus, the observational appearance of a bright magnetar at late times hints for a consistent amount of electric currents in the crust. Note that the latter depends not only on how much the magnetic field penetrates in the core, but also on the complexity of the initial magnetic field: strong electrical currents can circulate for configurations like the ones discussed above, (e.g. \citealt{MATINS_MT}), extended or not to the core. 

Conversely, for lower field strengths (bottom panels of Fig.~\ref{fig: comparison different env}), crustal-confined and core-threading models show similar behavior with minimal differences, due to the little relevance of magnetic effects (Ohmic heating and transport anisotropy). This has important observational implications \citep{Marino24} that we discuss in the next subsection, where we compare observational data to different models.

\begin{figure}
	\centering
\includegraphics[width=0.8\textwidth,height=0.9\textheight,keepaspectratio=false]{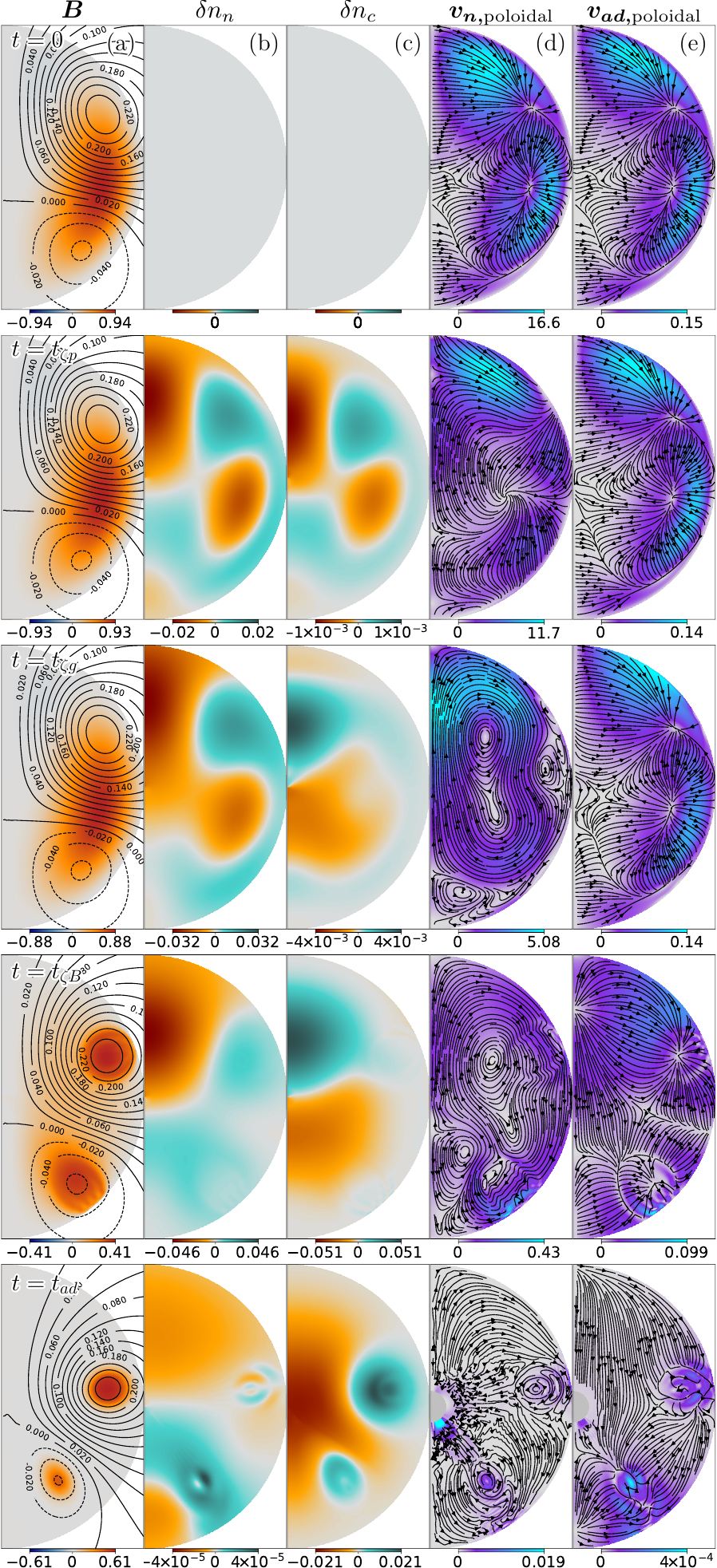}
\caption{Evolution of a mixed poloidal–toroidal core field under ambipolar diffusion. From left to right, columns show (a) Configuration of the magnetic field, where lines represent the poloidal magnetic field 
and colors the toroidal potential;
(b) and (c) density perturbations $\delta n_n$ and $\delta n_c$, respectively, both normalized to $n_{c0}$; (d) and (e) 
poloidal component of the neutron velocity, $\boldsymbol{v}_n$, and ambipolar diffusion velocity, $\boldsymbol{v}_{ad}$, where arrows represent the direction and colors the magnitude normalized to $R/t_0$. Rows correspond to different times: $t=0, \, t_{\zeta p}, \, t_{\zeta g}, \, t_{\zeta B}$, and $t_{ad}$. Image reproduced with permission from \citet{castillo20}, copyright by the authors. 
}
	\label{fig:castillo20}
\end{figure}

Prior to concluding this section, some more remarks on ambipolar diffusion in the core are in order. Despite some limitations, 2D
\citep{castillo17,passamonti17b,castillo20,vigano21,castillo2025,moraga2025} and 3D \citep{Igoshev23a} studies demonstrate that, under some circumstances, ambipolar diffusion can drive the long-term reorganization and decay of magnetic fields in NS cores. As an illustrative example, Fig.~\ref{fig:castillo20} shows the results of a two-fluid simulation of an initial mixed large-scale-only poloidal–toroidal core field under ambipolar diffusion, assuming axial symmetry. The figure summarizes some of the expected key features, occurring at different characteristic timescales, from shorter to longer: the timescale for propagation of sound waves $t_{\zeta p}$ (characteristic of p-modes), the timescale associated with the Brunt-V\"ais\"al\"a frequency
$t_{\zeta g}$ (characteristic of g-modes), the Alfv\'en crossing time, $t_{\zeta B}$ ,  
and the ambipolar diffusion timescale $t_{ad}$, given by $$t_{ad}\sim (0.3\text{–}3)\times10^{3}\left(\frac{10^{15}\,\mathrm{G}}{B}\right)^{2}
\left(\frac{T}{10^{8}\,\mathrm{K}}\right)^{2}
\left(\frac{L}{1\,\mathrm{km}}\right)^{2}\ \mathrm{yr},
$$ 
(see Fig.~\ref{fig:castillo20} for detailed definitions of the rest of characteristic timescales).

The simulation shows how the interplay of magnetic field and the two-fluid (neutral and charged fluids) dynamics drives the NS core through the following sequence of quasi-equilibria.
Following a brief relaxation phase (where $t_{\zeta p}$, $t_{\zeta g}$, and $t_{\zeta B}$ represent rapid dynamical timescales on the order of fractions of a second), the system exhibits small density perturbations in its two components and establishes a nearly steady velocity field, approaching a twisted-torus equilibrium where magnetic, pressure, and buoyancy forces are almost balanced. This quasi-equilibrium is non-barotropic, as neutrons and charged particles contribute differently to the force balance \citep{castillo20}. By $t \approx t_{\zeta B}$, the toroidal magnetic field has been fully restructured, primarily persisting within closed poloidal loops. Density perturbations transition from being correlated with the magnetic field to uncorrelated, and the poloidal force imbalance diminishes, as evidenced by the reduced amplitude of velocities. From this point, ambipolar diffusion, that operates on a much longer timescale becomes the driver. From $t_{\zeta B} \to t_{ad}$, the magnetic force pushes charged particles relative to neutrons, transporting flux and reducing $|\delta n_n|$ while $|\delta n_c|$ grows until charged-particle gradients alone balance the field (note the scales at the basis of each panel). The last row of the figure shows both signatures: small $|\boldsymbol{v}_{ad}|$ compared with earlier times and suppressed neutron perturbations, consistent with a transition toward Grad--Shafranov like equilibria supported mainly by the charged fluid. Although this evolution is slow, it occurs more rapidly than in scenarios with stationary neutrons.

However, significant uncertainties remain. Realistic cores are expected to be superfluid and superconducting, which should at the very least alter the couplings between superfluid neutrons and superconducting protons (thus modifying the time-scales for the processes simulated in these studies) and even more importantly, modify the governing induction equation \citep{glampedakis11b,Graber2015,kantor18}. We anticipate that this would be one very active line of research in the incoming years.

\subsection{Comparison with observations}

Magneto-thermal simulations of isolated NSs are particularly valuable as they enable direct comparison with observational data, notably quiescent thermal X-ray luminosities, as well as timing properties $P$ and $\dot{P}$. A key example is the work of \citet{vigano13}, who consistently re-analysed data from 40 isolated, thermally emitting NSs and showed that their phenomenological diversity can be explained by varying only the initial magnetic field, NS mass, and envelope composition. More recently, \citet{potekhin20} conducted a complementary survey, presenting estimated ages, surface temperatures, and thermal luminosities of middle-aged NSs with relatively weak to moderately strong magnetic fields. Their comparison with theory demonstrated that the agreement between observational data and theoretical cooling curves improves significantly when models assume weak neutron superfluidity in the stellar core.

In recent years, increasing attention has turned to the fast cooling scenario \citep{Mendes22,Marino24}. This interest was reinforced by the careful monitoring of three sources with well-determined ages that appear significantly colder by nearly an order of magnitude than other objects of comparable youth. These include two standard radio pulsars, \psra ($P=70$ ms, $B_p = 7\times10^{12}$ G, age $=841$ yr; \citealt{Kothes2013}) and \psrb ($P=490$ ms, $B_p \sim 2\times10^{13}$ G, age $
\sim 7700$ yr; \citealt{YarUyaniker2004}), and one CCO \velajr (age $\sim 2500$-5000 yr; \citealt{Allen2015}).

The secular cooling of NSs is influenced by the EoS, mass, magnetic field, and composition of the envelope, with the last three factors varying from star to star. By measuring the surface temperatures of numerous objects across a wide age range, NS cooling models (and consequently, the EoS) can be effectively constrained \citep{Page2004}.
Cooling curves are generally categorized into standard (minimal) and enhanced regimes. The surface temperatures of observed neutron stars typically align with standard cooling models \citep{Page2004, potekhin_rev15a}, although evidence of enhanced cooling has been noted for decades, with the Vela pulsar serving as a prime example. However, uncertainties in spectral energy distributions, precise ages, and accurate distances have hindered robust constraints on the equation of state (EoS) in this context.

A detailed study of the three above-mentioned exceptionally cold, young, and nearby NSs was presented in \citet{Marino24}. Reconciling theoretical models with these observations requires the inclusion of enhanced cooling processes, which in turn provides constraints on the NS EoS. A large set of simulations explored three representative EoSs spanning different cooling channels: SLy4 \citep{douchin01}, which forbids enhanced cooling; BSK24 \citep{pearson2018}; and GM1A \citep{gusakov2014}, allowing for fast cooling. Simulations covered three masses (1.4, 1.6, and 1.8 M$_\odot$) with moderate magnetic fields ($\lesssim 7\times10^{13}$ G) and an iron envelope, to prevent high luminosities that would be incompatible with these sources (see Fig.~\ref{fig: comparison different env}).
The results, shown in Fig.~\ref{fig:fast cooling}, show that several scenarios fail to reproduce the faint thermal luminosities of the three cold sources. In particular, with the SLy4 EoS (orange curves), the sharp luminosity drop cannot be obtained for any mass or magnetic field configuration. By contrast, in the GM1A case with hyperons (blue/green curves), cooling can proceed rapidly enough to match the data. Similarly, for the BSK24 EoS, massive stars ($M \geq 1.6$ M$_{\odot}$) activate nucleon direct Urca, producing enhanced cooling tracks consistent with the observations. These results provide compelling evidence of enhanced cooling, showing that only EoSs (and compositions) allowing fast cooling within the first few thousand years can reproduce the observed thermal emission from this sample \citep{Marino24}. It also reinforces the early suggestion by \citet{aguilera08a} that DUrca cooling may be masked by strong magnetic fields in other NSs, potentially leading to misidentification. Importantly, the EoS should explain both exceptionally bright objects, such as magnetars, and extremely faint sources at young ages. 

\begin{figure}
    \centering
    \includegraphics[width=\linewidth]{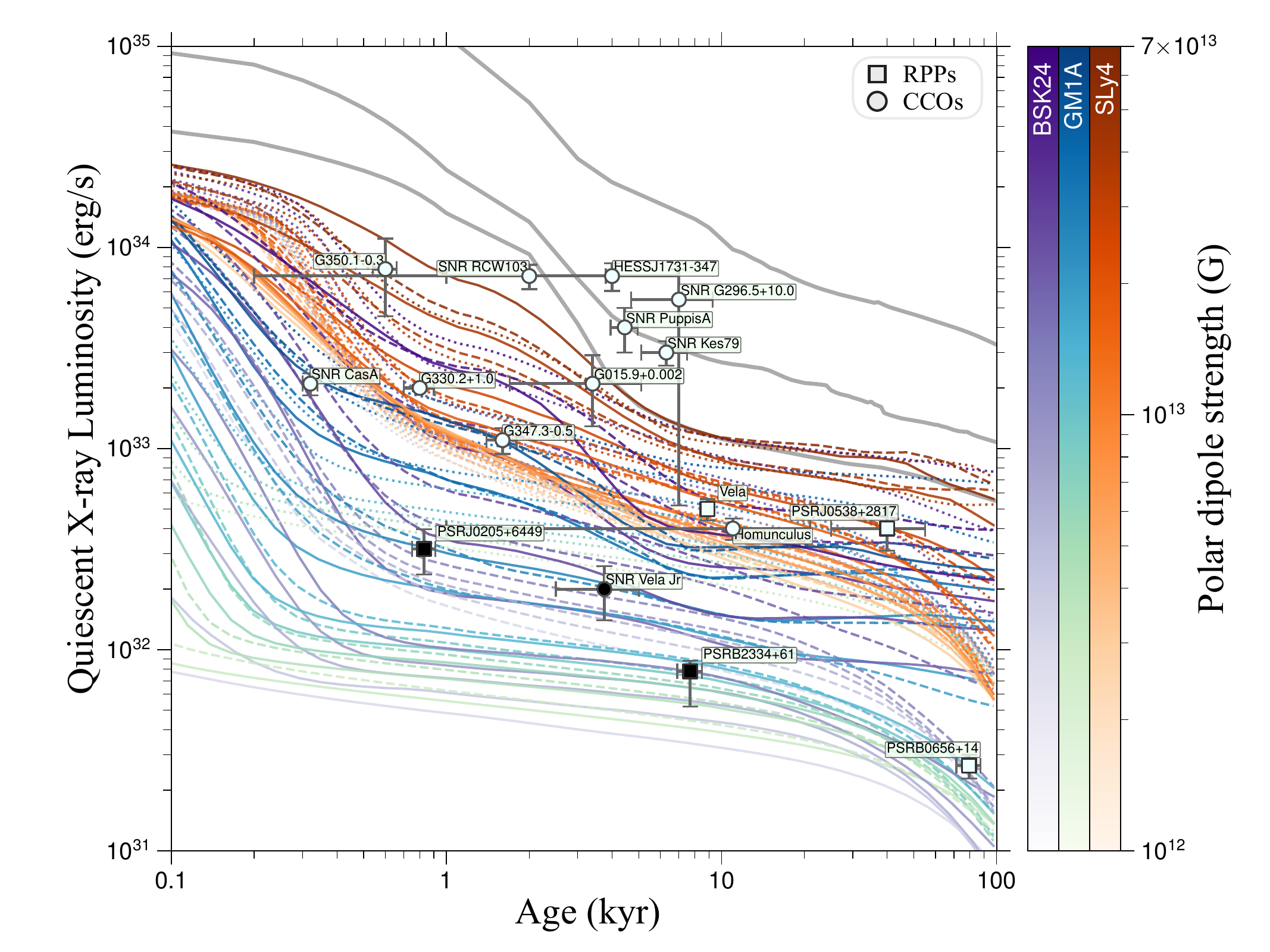}
    \caption{Comparison between observational data and theoretical cooling curves. Standard rotation-powered pulsars are shown as squares and CCOs as circles. The 81 theoretical cooling curves used in our analysis are shown for three EoSs: SLy4 (orange), BSk24 (violet), and GM1A (blue). We consider three masses: $1.4\,M_\odot$ (dots), $1.6\,M_\odot$ (dashed), and $1.8\,M_\odot$ (solid). We explore nine initial polar surface dipolar fields from $B_p=1\times10^{12}$ to $7\times10^{13}$ G. For comparison only, we also plot three gray curves for stronger fields with an initial  surface polar value of $10^{14}$, $3\times10^{14}$, and $10^{15}$ G (BSk24, $1.6\,M_\odot$). In all cases, fields are crust-confined and initially purely large scale. Image adapted from \citet{Marino24}.}
    \label{fig:fast cooling}
\end{figure}

\begin{figure}
	\centering
	\includegraphics[width=\textwidth]{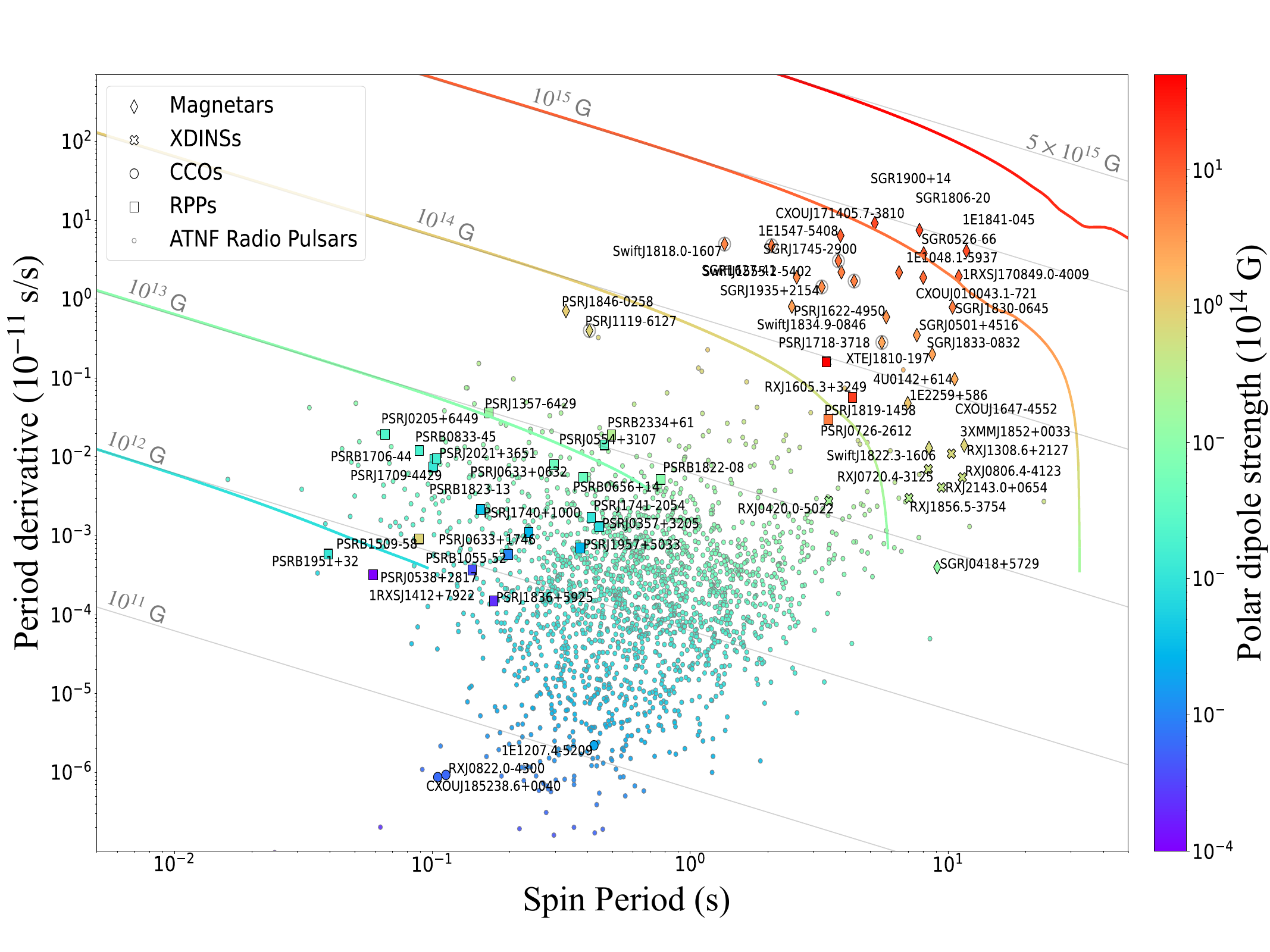}
	\caption{Evolutionary tracks in the $P-\dot{P}$ diagram, computed with the vacuum spin-down formula (Sect.~\ref{sec:spin-down}) for a $1.6 \, M_\odot$, R=12.5 km NS (BSk24 EoS) with initial polar fields $B_p^{0}=10^{12},\, 10^{13},\, 10^{14}, \, 10^{15},\,  5\times 10^{15}$ G, evolved under Hall drift and Ohmic dissipation.
    Solid gray lines show the tracks followed without considering magnetic field decay. Labels: magnetar-like sources (diamonds), nearby X-ray-dim isolated NSs (XDINSs; asterisks), central compact objects (CCOs; circles), rotation-powered pulsars (RPPs; squares), and ATNF radio pulsars (dots). The color bar shows the surface dipolar field at the pole in units of $10^{14}$ G.}
\label{fig:ppdot}
\end{figure}

Finally, we turn our attention to the rotational evolution of NSs. In Fig.~\ref{fig:ppdot} we present evolutionary tracks in the $P-\dot{P}$ diagram for a NS of 1.6 $M_\odot$ with varying initial magnetic field strengths, using a crustal-confined magnetic field configuration. Thin gray lines represent trajectories without field evolution, assuming a constant magnetic field, and appearing as straight lines in the diagram. 
In contrast, solid lines incorporating realistic field evolution deviate significantly from these models. Initially, the tracks coincide, as 
$B_p$ retains its initial value for early times ($t\lesssim 10^3-10^5$ yr).
Over time, the field dissipates faster than the spin period evolves, causing the tracks to bend downward at nearly constant $P$. This behavior is suggested as the primary cause of the observed period clustering in isolated X-ray pulsars \citep{pons13}. The limiting period depends mainly on the initial magnetic field and crust-core interface resistivity. The significant differences between constant and realistic magnetic field models highlight the need to account for the interplay between temperature, magnetic, and rotational evolution.

\section{Future prospects}\label{sec:conclusions}

The future of research in NS evolution (in particular, transient phenomena) is set to advance through enhanced survey capabilities and innovative instrumentation. In parallel, numerical advancements are critical: while 3D simulations have already become available, their application to NSs with realistic microphysics remains undeveloped, particularly for modeling small-scale hotspots linked to X-ray spectra and localized magnetic structures. Realistic boundary conditions at the surface of the star, moving beyond simple potential/vacuum solutions to include twisted magnetospheres, are essential for understanding interior dynamics and localized heating, since currents through the envelope potentially cause high temperatures. Additionally, the evolution of the core is complicated by physical processes involving superfluid neutrons and superconducting protons, marking a high-priority task to include a consistent theoretical description to advance our understanding of these extreme astrophysical objects.

\begin{acknowledgements}
We acknowledge support from the Conselleria d'Educació, Cultura, Universitats i Ocupació de la Generalitat Valenciana, through grant CIPROM/2022/13, and from the AEI grant PID2021-127495NB-I00.
CD acknowledges support from the Ministerio de Ciencia, Innovación y Universidades, co-funded by the Agencia Estatal de Investigación, the Unión Europea (FSE+), and the Universidad de Alicante. Her contract is part of the fellowship JDC2023-052227-I, funded by MCIU/AEI/10.13039/501100011033 and the FSE+. DV is supported by the European Research Council (ERC) under the European Union’s Horizon 2020 research and innovation program (ERC Starting Grant ”IMAGINE” No. 948582), and by the  “Maria de Maeztu” award to the Institut de Ciències de l’Espai (CEX2020-001058-M).
\end{acknowledgements}

\section*{Declarations}
{\small
\noindent
\textbf{Conflict of interest} The authors declare no conflict of interest.
}


\appendix
\normalsize

\section{Poloidal-toroidal decomposition of the magnetic field}\label{app:formalism}

Any three-dimensional, solenoidal vector field $\vec{B}$, can be expressed in terms of  its {\em poloidal} and {\em toroidal} components
\begin{equation} 
\vec{B}=\vec{B}_{\rm pol} + \vec{B}_{\rm tor}~.
\end{equation} 
In the literature, one can find different formalisms and notations to describe the two components. In this appendix we go through some of the ideas of the mathematical formalism and compare the most common notations.

Adopting the notation of \citet{geppert91}, the magnetic field can be written in terms of two scalar functions $\Phi (\vec{r},t)$ and $\Psi (\vec{r},t)$ (analogous to the stream functions in hydrodynamics) as follows:
\begin{eqnarray}\label{eq:def_decomposition_app}
&& \vec{B}_{\rm pol}=\vec{\nabla}\times(\vec{\nabla}\times\Phi\vec{k})\, ,\\
&& \vec{B}_{\rm tor}=\vec{\nabla}\times\Psi\vec{k}\, ,
\end{eqnarray}
where $\vec{k}$ is an arbitrary vector. This decomposition is particularly useful in situations where $\vec{k}$ is taken to be normal to one of the physical boundaries. Therefore, for a spherical domain, and using spherical coordinates $(r,\theta,\varphi)$, a suitable choice  is $\vec{k}=\vec{r}$. In this case, $\vec{\nabla}\times\vec{r}=0$, and
we can write:
\begin{eqnarray}\label{eq:def_pot_vect_app}
&& \vec{B}_{\rm pol}=\vec{\nabla}\times(\vec{\nabla}\Phi\times\vec{r}) = -\vec{r} ~\nabla^2 \Phi + \Bnabla 
\left( \frac{\partial (r\Phi)}{\partial r}\right)\, ,\\
&& \vec{B}_{\rm tor}=\vec{\nabla}\Psi\times\vec{r}~.
\end{eqnarray}
Generally speaking, the radial component of the magnetic field is included in the poloidal part, while the $\theta$ and $\varphi$ components are shared between poloidal and toroidal components. In axial symmetry, $\Phi=\Phi(r,\theta)$ and $\Psi=\Psi(r,\theta)$, the expressions are further simplified: the toroidal magnetic field is directed along the azimuthal direction $\hat{\varphi}$. In this case the potential vector is purely azimuthal and given by $\vec{A_\varphi} = - \vec{r} \times \Bnabla \Phi \, ,$ and the poloidal field can be directly derived from $\vec{B}_{\rm pol} = \Bnabla \times \vec{A_\varphi}~.$ 

Alternatively, another common notation expresses the magnetic field in terms of two other scalar functions, $P$ and $\Theta$ as:
\begin{equation}\label{eq:mf_clebsch_app}
\vec{B}=\vec{\nabla} P \times \vec{\nabla} \Theta~.
\end{equation}
In axial symmetry, and with the choice $\Theta=\varphi-\xi (r,\theta)$, the \textit{magnetic flux function} $P(r,\theta)$ is related to the $\varphi-$component of the vector potential by
\begin{equation}\label{eq:gamma_aphi_app}
P(r,\theta)= r\sin\theta ~ A_\varphi(r,\theta) \, ,
\end{equation}
and the poloidal and toroidal components are 
\begin{eqnarray}\label{eq:def_poloidal_app}
&& \vec{B}_{\rm pol}=\frac{\vec{\nabla} P(r,\theta) \times \hat{\varphi}}{r \sin\theta}\, ,\\
&& \vec{B}_{\rm tor}=(\vec{\nabla} \xi)_{\rm pol} \times (\vec{\nabla}P)_{\rm pol} \equiv \frac{T}{r\sin\theta}\hat{\varphi}\, ,
\end{eqnarray}
where we have introduced the scalar stream function $T$ used, for instance, in \citet{akgun17} and following works (in the force-free case, $T$ is a function of $P$, see Sect.~\ref{sec:forcefree}). The conversion between the two formalisms in axial symmetry is shown in Table~\ref{tab:formalism}.

\begin{table}[t]
		\caption{Comparison between different notations in axial symmetry. \citet{Pons2009} used the same notation as \citet{geppert91}, and in \citet{gourgouliatos16} $\Phi$ and $\Psi$ are denominated 
		by $V_p$ and $V_t$, respectively.}
	\begin{center}
		\begin{tabular}{l r r r}
			\toprule
			{\bf Formalisms}  &  \citet{akgun17} & \citet{kojima17} & \citet{geppert91} \\
			\midrule
			Poloidal function 	& $P(r,\theta)$		& $G(r,\theta)$	     & $\Phi(r,\theta)$\\
			Toroidal function 	& $T(r,\theta)$           & $S(r,\theta)$    &   $\Psi(r,\theta)$\\
			Toroidal potential vector $A_\varphi$         & $P(r,\theta)/r \sin\theta$	   & $G(r,\theta)/\varpi$		    	& $-\partial_\theta\Phi$ \\
			Magnetic flux		& $2\pi P$			& $2\pi G$				& $-2\pi r\sin\theta\partial_\theta \Phi$\\
			Poloidal magnetic field $\vec{B}_{\rm pol}$ 	& $(\vec{\nabla}P\times\hat{\varphi})/r\sin\theta$  & $(\vec{\nabla}G\times\hat{\varphi})/\varpi$	& $\vec{\nabla}\times(\vec{\nabla}\Phi\times\vec{r})$ \\
			Toroidal magnetic field $\vec{B}_{\rm tor}$ 	& $(T/r\sin\theta)~\hat{\varphi}$    & $(S/\varpi)~\hat{\varphi}$ 	& $\vec{\nabla}\Psi\times\vec{r} $ \\
			\bottomrule
			\label{tab:formalism}
		\end{tabular}
	\end{center}
\end{table}

\section{Potential solutions with Green's method}\label{app:potential_solutions}
For potential configurations, we can express the potential magnetic field in terms of the magnetostatic potential $\chi_m$, so that
\begin{eqnarray}
&& \vec{B}=\vec{\nabla}\chi_m\, , \label{eq:magnetostatic_potential}\\ 
&& \nabla^2 \chi_m=0~.\label{eq:laplace_potential}
\end{eqnarray}
The second Green's identity, applied to a volume enclosed by a surface $S$, relates the magnetostatic potential $\chi_m$ with a {\em Green's function} $G$ (see Eq.~(1.42) of \citealt{jackson91}):

\begin{equation}\label{eq:green_psi1}
2\pi \chi_m(\vec{r}) = -\int_S \frac{\partial G}{\partial n'}(\vec{r},\vec{r}') \chi_m(\vec{r}') {\rm d}S' + \int_S G(\vec{r},\vec{r}')\frac{\partial \chi_m}{\partial n'}(\vec{r}'){\rm d}S'\, ,
\end{equation}
where $\hat{n}'$ is the normal to the surface. Comparing with the electrostatic problem, we see that no volume integral is present, because $\dive\vec{B}\equiv \nabla^2\chi_m=0$. Note also that the factor $2\pi$ appears instead of the canonical $4\pi$, because inside the star Eq.~(\ref{eq:laplace_potential}) does not hold, thus $2\pi$ is the solid angle seen from the surface. The Green's function has to satisfy $\nabla'^2 G(\vec{r},\vec{r}')=-2\pi \delta(\vec{r}-\vec{r}')$. The functional form of $G$ is gauge dependent: given a Green's function $G$, any function $F(\vec{r},\vec{r}')$ which satisfied $\nabla'^2 F=0$ can be used to build a new Green's function $\tilde{G}=G+F$. The boundary conditions determine which gauge is more appropriate for a specific problem.

In our case the volume is the outer space, $S$ is a spherical boundary of radius $R$ (e.g., the surface of the star), and $\hat{n}'=-\hat{r}'$. We face a {\em von Neumann boundary condition problem}, because we know the form of the radial magnetic field
\begin{equation}
B_r(R,\theta) \equiv \frac{\partial\chi_m}{\partial r}(R,\theta)~.
\end{equation}
In order to reconstruct the form of
\begin{equation}
B_\theta(R,\theta)\equiv \frac{1}{R}\frac{\partial\chi_m}{\partial\theta}(R,\theta)\, ,
\end{equation}
we have to solve the following integral equation for $\chi_m$:

\begin{eqnarray}\label{eq:green_psi2}
2\pi \chi_m(\vec{r}) & = & R^2\left\{\int_0^\pi\int_0^{2\pi}  \frac{\partial G}{\partial r'}(\vec{r},\vec{r}') \chi_m(R,\theta')\sin\theta' {\rm d} \varphi' {\rm d} \theta' + \right.\nonumber\\
&& \left. - \int_0^\pi\int_0^{2\pi} G(\vec{r},\vec{r}') B_r(\theta')\sin\theta' {\rm d} \varphi' {\rm d} \theta'\right\}~.
\end{eqnarray}
So far, we have not specified the Green's function. In our case, the simplest Green's function is:

\begin{eqnarray}\label{eq:green0}
&& G(\vec{r},\vec{r}')=\frac{1}{|\vec{r}-\vec{r}'|}= [(r\sin\theta\cos\varphi - r'\sin\theta'\cos\varphi')^2+\nonumber\\
&& + (r\sin\theta\sin\varphi - r'\sin\theta'\sin\varphi')^2+(r\cos\theta - r'\cos\theta')^2]^{-1/2}~.
\end{eqnarray}
In axial symmetry, we can set $\varphi=0$, to obtain

\begin{equation}\label{eq:green1}
G(\vec{r},\vec{r}')=[(r\sin\theta - r'\sin\theta'\cos\varphi')^2+(r'\sin\theta'\sin\varphi')^2+(r\cos\theta - r'\cos\theta')^2]^{-1/2}~.
\end{equation}
We can evaluate $G$ and its radial derivative at $r=r'=R$
\begin{eqnarray}
&& G(R,\theta,\theta',\varphi')=\frac{1}{\sqrt{2}R}\left[1-\cos(\theta-\theta')+2\sin\theta\sin\theta'\sin^2\left(\frac{\varphi'}{2}\right)\right]^{-1/2}\, , \label{green_eps0} \\
&& \derparn{r'}{G}(R,\theta,\theta',\varphi') \rightarrow -\frac{G}{2R}~. \label{dergreen_eps0}
\end{eqnarray}
Casting the two formulas above in Eq.~(\ref{eq:green_psi2}), we note that the following integral appears in the two right-hand side terms:

\begin{equation}
f(\theta,\theta')\equiv \sin\theta'\int_0^{2\pi}RG(R,\theta,\theta',\varphi'){\rm d} \varphi'~.
\end{equation}
As $G$ depends on $\varphi'$ via $\sin^2(\varphi'/2)$, we can change the integration limits to $[0,\pi/2]$, and $\varphi'\rightarrow 2\varphi'$, therefore

\begin{equation}\label{eq:green_f}
f(\theta,\theta')=\sqrt{8}\sin\theta'\int_0^{\pi/2} [1-\cos(\theta-\theta')+2\sin\theta\sin\theta'\sin^2\varphi']^{-1/2}{\rm d} \varphi'~.
\end{equation}
Casting Eq.~(\ref{eq:green_f}) in Eq.~(\ref{eq:green_psi2}), and substituting $\chi_m(\theta)=R\int_0^\theta B_\theta(R,\theta') \de\theta'$, we have

\begin{equation}\label{eq:greeneps0_final}
4\pi\int_0^\theta B_\theta(\theta') {\rm d} \theta' + \int_0^\pi  B_\theta(\theta')\left[\int_{\theta'}^{\pi}f(\theta,\theta'')\de\theta''\right]\de\theta' = -2\int_0^\pi B_r(\theta')f(\theta,\theta')\de\theta'~.
\end{equation}
In Eq.~(\ref{eq:green_f}), if $\theta=\theta'$, then $f(\theta,\theta')\rightarrow 2\int_0^{\pi/2}(\sin\varphi')^{-1}{\rm d} \varphi'$, which is not integrable because of the singularity in $\varphi'=0$ (corresponding to $\vec{r}=\vec{r}'$). However, in both terms where it appears, the function $f(\theta,\theta')$ is integrated in $\theta'$, and both terms of the equation are integrable.

For numerical purposes, we can express Eq.~(\ref{eq:greeneps0_final}) in matrix form, introducing $f_{ij}=f(\theta_i,\theta'_j)$ evaluated on two grids with vectors $\theta_i,\theta'_j$, with $m$ steps $\Delta\theta$. The coefficients of the matrix $f_{ij}$ are purely geometrical, therefore they are evaluated only once, at the beginning. The grid $\theta_i$ coincides with the locations of $B_r(R,\theta)$, while the resolution of the grid $\theta'_j$ is $M$ times the resolution of the grid $\theta_i$ ($M\gtrsim 5$) to improve the accuracy of the integral function $f_{ij}$ near the singularities $\theta_i\rightarrow\theta_j$. The resolution of the grid of $\varphi'_k$ barely affects the result, provided that it avoids the singularities $\varphi'=0,\pi/2$. We typically use $M=10$ and $n_\varphi'=1000$. The calculation of the factors $f_{ij}$ is performed just once and stored. The matrix form is:

\begin{equation}\label{eq:greeneps0_matrix}
\sum_{j=1}^m [4\pi\delta_{ij} + f_{ij}\Delta\theta] \chi_m(\theta_j)= \sum_{j=1}^m [-2f_{ij}\Delta\theta]B_r(\theta_j), \qquad i=1,m~.
\end{equation}
From this, we obtain $B_\theta$ by taking the finite difference derivative of $\chi_m(\theta)$.

\phantomsection
\addcontentsline{toc}{section}{References}
\bibliographystyle{spbasic-FS}      
\bibliography{9-references}   


\end{document}